\newif\iffigs\figstrue
\documentclass[12pt]{article}
\usepackage{latexsym}
   \input{epsf}
\def\mycaptions#1{%
\refstepcounter{figure}
\begin{center}
\hskip 1pt\vskip -0.6cm
\small {\bf Fig.\hskip -3pt \arabic{figure}}: {\sl #1}
\null\hskip 1pt\vskip -0.2cm
\end{center}}
\def\mycaptionl#1{%
\refstepcounter{figure}
\begin{center}
\hskip 1pt\vskip -0.6cm
\begin{minipage}{16.5cm}
\small {\bf Fig. \hskip -3pt\arabic{figure}}: {\sl #1}
\end{minipage}
\null\hskip 1pt\vskip -0.2cm
\end{center}}
\def\remark#1{\vskip 5pt\noindent {\sl #1}~}
\newcommand{\eqn}[1]{(\ref{#1})}
\newcommand{\ft}[2]{{\textstyle\frac{#1}{#2}}}
\newsavebox{\uuunit}
\sbox{\uuunit}
    {\setlength{\unitlength}{0.825em}
     \begin{picture}(0.6,0.7)
        \thinlines
        \put(0,0){\line(1,0){0.5}}
        \put(0.15,0){\line(0,1){0.7}}
        \put(0.35,0){\line(0,1){0.8}}
       \multiput(0.3,0.8)(-0.04,-0.02){12}{\rule{0.5pt}{0.5pt}}
     \end {picture}}
\newcommand {\Cbar}
    {\mathord{\setlength{\unitlength}{1em}
     \begin{picture}(0.6,0.7)(-0.1,0)
        \put(-0.1,0){\rm C}
        \thicklines
        \put(0.2,0.05){\line(0,1){0.55}}
     \end {picture}}}
\def\sch#1#2{\{#1;#2\}}
\newcommand {\unity}{\mathord{\!\usebox{\uuunit}}}
\newcommand  {\Rbar} {{\mbox{\rm$\mbox{I}\!\mbox{R}$}}}
\newcommand{\Ka}{K\"ahler}

\def\Im{{\rm Im ~}}

\csname @addtoreset\endcsname{equation}{section}
\def\dop{{\rm d}\hskip -1pt}

\def\ii{{\rm i}}
\def\dd{{\rm d}}

\font\cmss=cmss10 \font\cmsss=cmss10 at 7pt

\def\inbar{\vrule height1.5ex width.4pt depth0pt}
\def\bfzero{\relax\,\hbox{$\inbar\kern-.3em{\rm 0}$}}
\newcommand{\bfone}{\unity}
\def\dop{{\rm d}\hskip -1pt}
\def\real{{\rm Re}\hskip 1pt}

\def\ii{{\rm i}}

\def\sch#1#2{\{#1;#2\}}
\def\IG{\relax\,\hbox{$\inbar\kern-.3em{\rm G}$}}
\def\IB{\relax{\rm I\kern-.18em B}}
\def\ID{\relax{\rm I\kern-.18em D}}
\def\IL{\relax{\rm I\kern-.18em L}}
\def\IF{\relax{\rm I\kern-.18em F}}
\def\IH{\relax{\rm I\kern-.18em H}}
\def\II{\relax{\rm I\kern-.17em I}}
\def\IN{\relax{\rm I\kern-.18em N}}
\def\IP{\relax{\rm I\kern-.18em P}}
\def\IQ{\relax\,\hbox{$\inbar\kern-.3em{\rm Q}$}}
\def\IR{\relax{\rm I\kern-.18em R}}

\def\ZZ{\relax\ifmmode\mathchoice
{\hbox{\cmss Z\kern-.4em Z}}{\hbox{\cmss Z\kern-.4em Z}}
{\lower.9pt\hbox{\cmsss Z\kern-.4em Z}}
{\lower1.2pt\hbox{\cmsss Z\kern-.4em Z}}\else{\cmss Z\kern-.4em
Z}\fi}
\begin{document}
\begin{titlepage}
\begin{flushright} KUL-TF-98/19\\  IFUM 616/FT \\ hep-th/9803228
\end{flushright}
\vfill
\begin{center}
{\Large\bf The rigid limit in Special K\"ahler geometry }    \\
\vskip 0.3cm
{\large {\it  From  $K3$-fibrations to Special Riemann surfaces:
a detailed case study.}}
\vskip 0.3cm
{\large {\sl }}
\vskip 10.mm
{\bf   Marco Bill\'o$^1$,  Frederik Denef $^{1,+}$, \\[2mm]
Pietro Fr\`e$^2$,
Igor Pesando$^2$, Walter Troost$^{1,*}$,\\[2mm]
Antoine Van Proeyen$^{1,\dagger}$ and  Daniela Zanon$^3$ } \\
\vfill
{\small
$^1$ Instituut voor theoretische fysica, \\
Katholieke Universiteit Leuven, B-3001 Leuven, Belgium\\
\vspace{6pt}
 $^2$  Dipartimento di Fisica Teorica dell' Universit\`a,\\
                 via P. Giuria 1,
                I-10125 Torino, Italy\\
\vspace{6pt}
 $^3$ Dipartimento di Fisica dell'Universit\`a di Milano and\\
INFN, Sezione di Milano, via Celoria 16,
I-20133 Milano, Italy. }
\end{center}
\vfill

\begin{center}
{\bf ABSTRACT}
\end{center}
\begin{quote}
The limiting procedure of special K\"ahler manifolds to their rigid limit is
studied for moduli spaces of Calabi--Yau manifolds
in the neighbourhood of certain singularities.  In two examples
we consider all the periods in and around the rigid limit, identifying the
non-trivial ones in the limit as periods of a meromorphic form on the relevant
Riemann surfaces.
We show how the K\"ahler potential of the special K\"ahler manifold
reduces to that of a rigid special K\"ahler manifold.
We make extensive use of the structure of these Calabi--Yau
manifolds as $K3$ fibrations, which is useful to obtain the
periods even before the $K3$ degenerates to an ALE manifold in the limit.
We study various methods to calculate the periods and their properties. The
development of these methods is an important step to obtaining exact results
from
supergravity on Calabi--Yau manifolds.
\vskip 2.mm
 \hrule width 5.cm
\vskip 2.mm
{\small
Work supported by
the European Commission TMR programme ERBFMRX-CT96-0045,
in which D.Z. is associated to U. Torino.\\
\noindent $^+$ Aspirant FWO, Belgium \\
\noindent $^*$ Onderzoeksleider FWO, Belgium \\
\noindent $^\dagger$ Onderzoeksdirecteur FWO, Belgium }
\end{quote}
\end{titlepage}

\section{Introduction}
String theory has proven to be a quite valuable tool in obtaining exact
results for non-trivial supersymmetric quantum field theories.
In particular, the relation of type~II theories on Calabi--Yau manifolds
to $N=2$ supersymmetric $D=4$ theories has been studied extensively
\cite{KKLMV,KLMVW}, see also \cite{revLerche,Klemmreview} and
references therein.
In many instances, special K\"{a}hler geometry \cite{DWVP}
and its rigid version \cite{PKTN2}, and, in particular,
the extraction of the rigid limit, play a key role.
Special geometry is the geometric structure defined by
the two derivative actions of the scalars in vector multiplets of
$N=2$, $D=4$ theories. The highly constrained structure
of the special geometry makes it possible to find
exact analytical expressions for many low-energy effective actions
\cite{SeiWit}.
\par
There are two types of special K\"{a}hler geometry: `rigid' and
`local'. Local special geometry applies to local supersymmetry, i.e.\ supergravity
and strings, while rigid special geometry applies to
rigid supersymmetry, i.e.\ $N=2$ supersymmetric gauge theory in flat
spacetime. In spite of the obvious similarities, their precise relation
is not completely straightforward. In fact, even in the context of
supergravity it is not completely obvious how to decouple gravitons,
gravitinos and their graviphoton supersymmetrically.
One particular limiting procedure
was discussed in \cite{ItalianN2}.
In the present paper we want to identify the essential steps which are
involved in this limiting procedure. We want to embed the rigid theory
in the supergravity theory. This also implies the
calculation of parts of the action which decouple from the field theory
degrees of freedom. The detailed explanation of the methods involved
can contribute towards the calculation of the next-to-leading terms in the
expansion of a phenomenologically important class
of supergravity actions, namely those which are obtained from
compactifications on $K3$ fibred Calabi--Yau manifolds.
\par
Type~IIB string theory compactified on a Calabi--Yau (CY) manifold leads to an
$N=2$ theory in four dimensions. In this
case the (local) special geometry of the vector multiplet
moduli space is given by the classical geometry of the CY complex structure
moduli \cite{Seiberg88} (see \cite{revKlemmTheisen} for a review),
and it is known to receive no quantum corrections.
Therefore, by going to a rigid limit of this classical moduli space
and identifying the corresponding rigid low-energy quantum theory
(usually a field theory)
one should obtain an exact solution for the two-derivative low-energy
effective action \cite{KKLMV,KLMVW}.
The identification of the rigid theory is conceptually the most
non-trivial part of this programme, and indeed it was solved only
after the discovery
of heterotic/type~II duality \cite{2ndqms}
(suggesting non-Abelian gauge theories)
and $D$-branes as solitonic states \cite{Polchinski}
(providing the `missing' massive gauge vector multiplets).
The relevant rigid special geometry often turns out to be naturally
associated to the rigid special geometry of a certain class of Riemann
surfaces, reproducing and extending the Seiberg--Witten (SW) solution of $N=2$
quantum Yang--Mills theory. Furthermore, within this framework
many features of quantum field theory
have a beautiful geometrical interpretation and this provides
quite elegant solutions to problems which would be hard to tackle with
 ordinary field theory techniques, like for example the existence
and stability of BPS states \cite{KLMVW, revLerche,Klemmreview}.
\par
By now, a very large class of $N=2$, $d=4$ quantum field theories (and
even more exotic theories) can be `engineered' and solved
geometrically in this way. The usual procedure \cite{GE, GE-review} is
to find a local IIA model which in the rigid limit produces the field
theory to be solved; to map this IIA theory into an equivalent IIB
theory using local mirror symmetry and finally to solve this IIB
theory exactly (in the low-energy field theory limit) using classical
geometry. One argues that the restriction to local models and local
mirror symmetry---where `local' means that one only considers a
certain small region of the Calabi--Yau manifold---is allowed roughly
because the relevant (light brane) degrees of freedom are all
localized well inside that region.
 \par
An alternative, but not unrelated, approach consists in
making use of $M$-theory brane configurations in flat space \cite{M-sol}.
\par
In this paper we study in detail the rigid limit on the type~IIB side,
without assuming {\it a priori\/} that we can restrict ourselves to the local
considerations mentioned above.
In \cite{KLMVW} it was explained how
one finds the periods of a Riemann surface and the associated rigid special
\Ka\ structure of its moduli space by expanding the Calabi--Yau 3-fold around
singular points and reducing it to an ALE-fibration. This idea
is also instrumental for the present work, and
many of the results obtained here were already contained in those papers.
However, the limiting theory
does not teach us how the rigid limit is approached in supergravity.
For this we study in detail the behaviour  of the
full \Ka\ potential and its rigid limit. Hence we will keep
{\it all\/} Calabi--Yau periods (which are the building blocks of
local special \Ka\ geometry), not just those that stay finite
in the ALE limit. This may also be useful for a possible
continuation of the present work, namely a computation of the next-to-leading
terms in the expansion of supergravity.
\par
We will study the rigid limit very explicitly in two examples. The
corresponding Calabi--Yau surfaces are $K3$-fibrations,
which is a generic feature of type~II
Calabi--Yau compactifications dual to heterotic strings
\cite{KLM,AspinwallLouis}.
We obtain exact expressions for all the periods using the $K3$-fibration
structure. Only afterwards will the expansion be made, and the reduction of
special \Ka\ geometry from local to rigid will be demonstrated explicitly.
We will see how certain periods decouple from the \Ka\ potential
to lowest order, and how the
latter reduces from its general $Sp(2n+2)$ structure (where $n$ is the complex
dimension of the special \Ka\ moduli space) to the $Sp(2r)$ structure (where
$r$ is the complex dimension of the rigid special \Ka\ moduli space, with,
in our examples, $r=n-1$).
For one of the examples we consider, we obtain the Seiberg--Witten
Riemann surfaces corresponding to the different rigid limits as
degenerating branches of a genus-5 Riemann surface, defined for {\it
all\/} values of the Calabi--Yau moduli. The same genus-5 surface gives
rise to a genus-2 surface in the $SU(3)$ rigid limit and to the
genus-1 surfaces for the $SU(2)$ and $SU(2)\times SU(2)$ limits.
\par
In section~\ref{ss:ing} we present some of the essential ingredients.
First we give a definition of special geometry, the one which is most
useful for our purpose \cite{PietroSKGlect,whatskg}.
Then we review how moduli spaces of Calabi--Yau and Riemann
surfaces realize special \Ka\ geometry,
and we point out the difference
between rigid and local special geometry. Then we explain how
we expect to make the transition between the two, i.e.\ how to perform
the rigid limit.
\par
Section~\ref{ss:ALEfromCY} is devoted to a description of CY surfaces
realized as $K3$ fibrations in weighted
projective spaces. We illustrate some
aspects of $K3$ manifolds, in particular, the distinction between
cycles in the Picard and in the transcendental lattice
\cite{GrifHar,Aspinwall}. As the former
have trivial periods, the latter are the relevant ones.
Then we describe a class of $K3$ fibrations in which our two examples will
fit and we explain how the CY periods can be computed by taking advantage
of this $K3$ approach.
\par
In section~\ref{ss:exa} we introduce the two examples which we
will study in detail. In particular, we analyse the structure of the moduli
spaces, using a parametrization which allows us to keep
the gauge invariances
related to redefinitions of the coordinates on the manifolds unfixed.
Then we discuss the singularity structures
 and recall how one can expand the manifolds
around these singularities in such a way that they reduce locally to
fibrations of ALE spaces.

\par
In section~\ref{ss:pmiex1} we present a detailed study of the first example.
We construct cycles and periods on the CY manifold in terms of the
corresponding ones on the $K3$ fibre.
The aim is to obtain the periods in a basis with known intersection matrix,
which is necessary to calculate the \Ka\ potential.
Although not strictly necessary for our discussion, we derive exact
expressions of the periods since these quantities would be useful
if we were to consider an expansion
of the full supergravity action in the vicinity of its rigid limit.
In order to obtain closed expressions of the periods in an
integral basis we have found it convenient to combine two approaches:
the use of the Picard--Fuchs differential equations \cite{PFeqns} and the
explicit
integration over one particular cycle in a certain region of moduli
space. Analytic continuation then leads to the construction of all the
other periods.
\par
In section~\ref{ss:pmiex2} the analysis is extended to the second example.
Since in this case
the CY manifold can be viewed as a double fibration, first a $K3$
and then a torus fibration, the CY periods are constructed
taking advantage of this feature. We obtain exact expressions for
the periods and determine monodromy matrices and the intersection
matrix.
\par
Finally, in section~\ref{ss:rigidlimitend} the limiting procedure is applied
and the rigid limit is described. We show that the \Ka\
potential of (local) special geometry indeed reduces to that of rigid
special geometry. We show how the scheme, set out in
section~\ref{ss:ing}, emerges from the limiting behaviour of the CY
periods. We finally compare the result in the limit with the periods
of the Seiberg--Witten Riemann surfaces.
\par
Section~\ref{ss:conclusions} contains a summary of the results
obtained in our work and some conclusive remarks. We also make some
connections to the $M$ theory 2/5-brane picture, albeit only at the
formal level.
\par
In the appendices we have collected some additional material.
Appendix~\ref{app:k3fibremodulispace}
presents the structure of the duality and monodromy groups from the point
of view of general group-theoretical considerations.
Appendix~\ref{ss:permonK3} contains instead the relevant calculations
for obtaining the Picard--Fuchs equations and their solutions in the
two examples considered in the main text. These are derived using a
method inspired by toric geometry,
which amounts in practice to simple  manipulations of Griffith's
representation of the periods. Finally, appendix~\ref{ss:z8c} gives
an alternative way to define cycles in the first example, which makes
contact with \cite{candmirror1}, and allows us to obtain a compact
expression of the intersection matrix.

\section{Ingredients}\label{ss:ing}
\subsection{Special geometry}
The first construction of general
couplings of vector multiplets to $N=2$ supergravity \cite{DWVP} used
conformal tensor calculus.
The geometry of the scalars, `special \Ka\ geometry', was
described in terms of `special coordinates'. The construction
started from a `prepotential' which corresponds to the function of
the vector superfields which is the superspace density. It was
realized in \cite{CdAF} that coordinate-independent formulations
would be more useful. Meanwhile it was recognized that a special \Ka\
geometry occurs in the moduli space of Calabi--Yau (CY) manifolds, and in
\cite{special} an attempt was made to give definitions of the special
geometry which are more useful from that point of view. Crucial in
these formulations is the symplectic group, $Sp(2(n+1))$ for the
coupling of $n$ physical vector multiplets, which is the group of
dualities of the vectors. This structure was already apparent in
the conformal tensor calculus in which vector multiplets realize this
$Sp(2(n+1))$ group. For defining the special geometry,
the symplectic structure is a more
natural starting point than
the prepotential. This point of view is supported by the observation
that some actions cannot be obtained directly from a prepotential
\cite{f0art} (in that case they are dual to an action obtained from a
prepotential \cite{whatskg}). In the conformal tensor calculus
these theories can be obtained from defining the symplectic vector of
superconformal multiplets \cite{symplconf}, rather than
from a superconformal invariant action. Finally,
the symplectic definition which we will
describe below is also most useful for the
application in moduli spaces.
\par
First we summarize the relevant geometric concepts, both for
the rigid\footnote{Rigid special geometry was first found in
\cite{PKTN2}.} and for the local case. We consider
symplectic vectors $V(u)$ (rigid), respectively $v(z)$ (local)
which are holomorphic functions of $r$, (respectively $n$)
complex scalars $\{u^i\}$ (respectively $\{z^\alpha \}$).
These are
$2r$-vectors for the rigid case (in correspondence with the electric
and magnetic field strengths), and $2(n+1)$-vectors in the local
case (because in that case there is also the graviphoton). A
symplectic inner product is defined as
\begin{equation}
\langle V,W \rangle=V^T\, Q^{-1}\, W\ ;\qquad
\langle v,w \rangle=v^T\, q^{-1}\, w\ , \label{symplinner}
\end{equation}
where $Q$ (and $q$) is a real, invertible, antisymmetric matrix
(we wrote $Q^{-1}$
in \eqn{symplinner} in view of the meaning which $Q$ will get in the
moduli space realizations). Canonically, $Q$ (or $q$) is
\begin{equation}
Q=\pmatrix{0&\unity \cr -\unity &0}\ , \label{canQ}
\end{equation}
but we will use also other symplectic metrics.
\par
The \Ka\ potential is, for the rigid and local manifold, respectively,
\begin{equation}
K(u,{\bar u})=\ii\langle V(u),\bar V(\bar u)\rangle\ ;     \qquad
{\cal K}(z,{\bar z})=-\log(-\ii\langle v(z),\bar v(\bar z)\rangle)\ .
\label{Kahlerpotentials}
\end{equation}
In the rigid case there is a rigid invariance $V\rightarrow e^{i\theta}V$,
but in the local case there is a symmetry with a holomorphic
dependence: $v(z)\rightarrow e^{f(z)}v(z)$. This gives a \Ka\
transformation
${\cal K}(z,\bar z)\rightarrow {\cal K}(z,\bar z) - f(z) -\bar f(\bar z)$.
Because of this local symmetry we have to introduce covariant
derivatives
\begin{equation}
{\cal D}_{\bar \alpha}v=\partial _{\bar \alpha}v= 0\ ;\qquad
{\cal D}_\alpha v=
\partial_\alpha v +\left( \partial_\alpha {\cal K}\right) v\ .
\label{specgeomcovder}
\end{equation}
There also exists a more symmetrical formulation, using as symplectic
sections $e^{{\cal K}/2}v$, but we will restrict here to one formulation.
 In both cases we
still need one more constraint (leading to the `almost always'
existence of a prepotential), which is for rigid (respectively local)
supersymmetry:
\begin{equation}
\langle \partial_i V,\partial_j V\rangle =0\ ;  \qquad
\langle {\cal D}_\alpha v,{\cal D}_\beta v\rangle =0\ .
\label{specgeom_conditions}
\end{equation}
There are further global requirements, such as that in
overlaps of regions the symplectic vectors should be related by
$ISp(2r,\Rbar)$, respectively $Sp(2(n+1),\Rbar)\times \Cbar$ transformations,
and the Hodge--\Ka\ nature of the local manifold, but for an exact
formulation we refer to \cite{PietroSKGlect,whatskg}.

\subsection{Special \Ka\ geometry in moduli spaces}\label{ss:spKamod}
Although the concept of special \Ka\ geometry includes cases that are
not necessarily related to moduli spaces of surfaces, and although
the rigid limit can also be addressed for these cases, in the present
paper we focus on such moduli spaces.
\par
Hence,
we recall how special geometry (of both kinds) is realized in the
moduli spaces of certain complex manifolds.
We consider first a CY manifold. It has $n=h^{21}$ complex structure
moduli to be identified with the complex scalars $z^\alpha$. The $h^{11}$
\Ka\ moduli do not play any role in this paper: in $N=2$
supergravity they are associated with the scalars in hypermultiplets,
which we do not consider here.
\par
There are $2(n+1)$ cohomologically different 3-cycles $c_\Lambda$,
and their intersection matrix will be identified with the symplectic metric
$q_{\Lambda\Sigma}=c_\Lambda\cap  c_\Sigma$. Canonically these are the
$A$ and $B$ cycles with intersection matrix \eqn{canQ}. One
identifies  $v$ with the `period' vector formed by integration of the $(3,0)$
form over the $2(n+1)$ three-cycles:
\begin{equation}
v_\Lambda=\int_{c_\Lambda}\Omega^{(3,0)}\ ;\qquad
{\cal D}_\alpha v_\Lambda=\int_{c_\Lambda}\Omega^{(2,1)}_{(\alpha)}\ .
\end{equation}
With a $(3,0)$ form which is holomorphic in the moduli
space, we satisfy all requirements for {\it local\/} special geometry
automatically \cite{FerStro}, e.g.\
\begin{equation}
\langle{\cal D}_\alpha v,{\cal D}_\beta v \rangle = \int_{c_\Lambda}
\Omega^{(2,1)}_{(\alpha)}
\cdot q^{\Lambda\Sigma}\cdot  \int_{c_\Sigma}\Omega^{(2,1)}_{(\beta)}= -
\int_{CY}\Omega^{(2,1)}_{(\alpha)}  \wedge \Omega^{(2,1)}_{(\beta)}= 0\ .
\end{equation}
{\em Rigid\/} special \Ka\ geometry is realized in moduli spaces of
Riemann surfaces (RS).
A RS of genus $g$ has $g$ holomorphic $(1,0)$ forms.
Now, in general, we need a family of Riemann surfaces with $r$ complex
moduli $u^i$, such
that one can isolate $r$ $(1, 0)$-forms which are the derivatives of a
meromorphic 1-form $\lambda$ up to a total derivative:
\begin{equation}
\gamma_i=\partial_i\lambda +d\eta_i\
;\qquad\alpha=1,\ldots ,r\leq g\ .
\end{equation}
Then one should also identify $2r$ 1-cycles $c_A$ forming a
complete basis for the cycles over which the integrals of $\lambda$
are non-zero. We  identify the symplectic vector of rigid special geometry
as
\begin{equation}
V_A=\int_{c_A}  \lambda   \ .
\end{equation}
As follows from the above explanation, this construction of rigid
special geometry is much less straightforward than the construction of
local special geometry in the CY moduli space, since except for genus~1
it is not clear {\it a priori\/} how to find the subspace of moduli space that
has the required special property.

\subsection{The rigid limit}\label{ss:rigidlimit}
Let us start from a manifold with local special geometry.
We expect, in some limit, to find the structure of {\it rigid\/}
special geometry; we will call this the {\it rigid limit}.
We will investigate the rigid limit at the level of the \Ka\ potential.
One may think of the construction as in section~\ref{ss:spKamod},
but the treatment here is more general.
The scalar fields  $\{z^\alpha\}$,
with $\alpha=1,\ldots ,n$, are moduli of this manifold, and we
consider a region in moduli space
where they are expanded in terms of a small expansion parameter
$\epsilon$. This expansion is made keeping some (functions of the) moduli
$\{ u ^i\}$ with $i=1,\ldots ,r$ fixed:
\begin{equation}
z^\alpha =z^\alpha(\epsilon ,u ^i  )\ .    \label{zepsilonlambda}
\end{equation}
An  example of such a procedure was given in section~10 of \cite{ItalianN2}
in which case $r=n$, while in the examples which we will consider below
$r=n-1$, and \eqn{zepsilonlambda} can be considered as a local
coordinate change from $z^\alpha$ to $\{\epsilon ,u ^i\}$.
We now imagine that the period vector
(or more generally the symplectic vector) is the sum of
pieces with different $\epsilon$ dependence, of the form
\begin{equation}
v=v_0 (\epsilon)+\epsilon^a v_1(u )+v_2(\epsilon,u  )\ .
\label{rigidexpv}
\end{equation}
The dominant piece in the $\epsilon\rightarrow 0$ limit
is imagined to be $v_0$, a piece independent of the surviving moduli.
$v_1$ is a vector such that the derivatives with respect to the
moduli form a matrix of rank $r$.
The pieces $v_1$ and $v_2$ are assumed to contribute
to the local \Ka\ potential ${\cal K}$ of \eqn{Kahlerpotentials}
in a progressively less dominant manner as follows:
\begin{equation}
\langle v,\bar v\rangle =\ii M^2 (\epsilon)+ |\epsilon|^{2a}
\langle  v_1(u ),\bar  v_1(\bar u )\rangle
+R(\epsilon,u ,\bar u )\ ,  \label{expvbarv}
\end{equation}
where
\begin{equation}
\ii M^2(\epsilon)= \langle v_0(\epsilon) ,\bar v_0(\epsilon) \rangle \ ,
\end{equation}
\begin{equation}
\frac{|\epsilon|^{2a}}
{M^2 }\stackrel{\epsilon\rightarrow 0}{\longrightarrow } 0\ ;\qquad
\frac{R}{|\epsilon|^{2a}}
\stackrel{\epsilon\rightarrow 0}{\longrightarrow } 0 \ .
\end{equation}
and $a$ is some real number (which could be normalized
to 1 by choosing $\epsilon$ appropriately).
For this one needs that
$\Im \langle v_0(\epsilon),\bar v_1(\epsilon) \rangle={\rm o} \big(|\epsilon|^{2a}
\big)$,
and  similar conditions for $v_2$.
If this structure is realized, the local \Ka\ potential ${\cal K}$
of
\eqn{Kahlerpotentials} can be written as
\begin{equation}
{\cal K} = - \log (M^2)+\ii\frac{|\epsilon|^{2a}}{M^2}
\langle  v_1(u ),\bar  v_1(\bar u )\rangle +\ldots =
- \log (M^2)+\frac{|\epsilon|^{2a}}{M^2}
K(v_1(u),\bar   v_1(\bar u ))+\ldots
\ .
\label{Kahler_expansion}
\end{equation}
This indicates that the local \Ka\ potential thus reduces to
the \Ka\ potential for a rigid special geometry
up to an additive constant and an overall multiplicative renormalization,
see also \eqn{Kahlerpotentials}
with the identification of $V(z)$ with $v_1(u )$.
\par
To confirm this interpretation,  we also have to
consider the covariant derivative in \eqn{specgeomcovder}.
The expansion \eqn{rigidexpv} implies for the derivatives
with respect to the surviving moduli
\begin{equation}
D_u  v= \epsilon^a\partial_u  v_1 + \ldots \ ,
\end{equation}
where the dots indicate subleading terms that can be neglected in the limit.
This implies that the first expression of \eqn{specgeom_conditions}
is implied by the second, in this limit.
The constraints of local special \Ka\ thus reduce to those of
rigid special \Ka\ geometry.
\par
Relating  $\epsilon$ to the inverse Planck
mass, in this way one should obtain the limit of the theory when
gravitational interactions are switched off.
One of our tasks is therefore to show how to obtain the structure
anticipated in \eqn{rigidexpv} and \eqn{expvbarv}.
In \cite{ItalianN2} it was proposed that
$v_0$ is a constant vector, orthogonal to $v_1(z^\alpha)$ (of rank $2n$),
and $v_2$ is of order $\epsilon^{3a}$. The various objects appearing
in rigid and local special geometry were then related.  $\epsilon$ is
in our case one of the fields of the supergravity model which is
fixed in the rigid theory. We will have to demonstrate
that the structure required for \eqn{expvbarv}, like the
approximate orthogonality of $v_0$ and $v_1$, really holds.
\par
In previous work \cite{KLMVW}, the
$\epsilon\rightarrow 0$ limit was taken by considering not the full CY
space, but only its structure around the specific singularity in moduli
space.
The CY surfaces treated there are $K3$ fibrations,
and in the rigid limit they  degenerate into fibrations of an
asymptotically locally Euclidean (ALE) space, with a corresponding
set of forms and cycles. One expects that in the rigid limit only the
periods corresponding to these forms and cycles are relevant.
The particular ALE space then determines the Seiberg--Witten auxiliary
surface for the corresponding Yang--Mills theory.
We will not make use of this degeneration, but rather work on the full
CY space. Nevertheless, inspired by \cite{KLMVW}, we will
also consider CY manifolds that are $K3$ fibrations. Therefore in
section~\ref{ss:K3-generalities}
we discuss some common properties of the particular CY and
$K3$ surfaces we need in the following.

\subsection{Monodromies and the Picard--Lefshetz formula} \label{ss:monodrPL}
In the course of the programme sketched above, we will often use monodromy
matrices. Let us state explicitly our convention about monodromy and
intersection matrices.
\par
Consider a set of functions $U_i(\mu)$
(e.g.\ a basis of solutions to a differential equation, to fix the ideas),
with a cut running from $\mu=0$. Consider the analytic continuation of
$U_i(\mu)$,
following $\mu\rightarrow e^{\ii\phi}\mu$ from $\phi=0$ to $\phi=2\pi$, with
$|\mu|$ small
(which we will indicate by $\mu\rightarrow e^{2\pi\ii}\mu$). After
crossing the cut, these continuations, $U'_i$,
will be re-expressed in the $U_i$ are still solutions of the
same differential equations, and can be expressed  in the $U_i$
basis by $U'_i=M_i{}^j  U_j$; $M_i{}^j$ is the monodromy
matrix around $\mu=0$.
We remark that the monodromy is thus connected to the crossing of the cut.
If more than one cut runs from $\mu=0$, the monodromy
matrix can depend on the base point,
i.e.\ the value of $\mu$ where we start.
 This dependence is in practice only on the order in which
different cuts to the singular points are crossed during the loop.
\par
If we cross two cuts, the first one related to the monodromy
matrix $M$, and the second one to $N$, then the combined monodromy is
ordered as $M\,N$, because the monodromies are always defined with respect to
the original basis $U_i$.
With these conventions we have for $I$ an intersection matrix, $M$
a monodromy matrix, and $U^{\prime}=SU$ a change of basis:
\begin{equation}
M\,I\,M^T=I\ ;\qquad I'=S\, I\, S^T\ ;\qquad M'=S\, M\, S^{-1} \label{cbIM}
\end{equation}
for the new intersection and monodromy matrices $I'$ and $M'$.

A useful ingredient is the Picard--Lefshetz formula. This says
that for a cycle $C$ going around a
singularity where $\nu$ is the vanishing cycle
\begin{equation}
\label{sr35}
C \to C - s^N (-1)^{N(N+1)/2}\left(\nu \cdot C\right)\, \nu\ ,
\label{Picard-Lefshetz}
\end{equation}
where $N$ is the complex dimension of the manifold, $s =\pm 1$,
the conventional intersection number $X\cdot Y$ of
the real axis $X$ and the imaginary axis $Y$ in the complex plane. We used
$s=+1$. Further, the dot product means the intersection number.
 Note that for $K3$, as for any manifold with $N=2$ mod 4, this
 implies $\nu\cdot \nu=-2$. For CY and $K3$ we can thus use
\begin{equation}
C \to C + (C\cdot \nu)\, \nu\ . \label{Picard-LefshetzK3CY}
\end{equation}
\section{$K3$ fibrations, singularities and periods}
\label{ss:ALEfromCY}
\subsection{Description of the CY and $K3$ in weighted projective spaces}
\label{ss:K3-generalities}
The complex manifolds we shall be concerned with in this paper are all
described as hypersurfaces in a {\it weighted projective ambient space\/};
they correspond to loci of a complex {\it polynomial equation\/}
$W(x)=0$ with $(x_1,\ldots x_{N+2}) \sim (\lambda^{w_1} x_1 \ldots
\lambda^{w_{N+2}} x_{N+2})$, i.e.\ with
$x\in \IP^{w_1,\ldots w_{N+2}}$,
and with $W(\lambda x) = \lambda^d W(x)$,
if they are $N$ dimensional.
The first Chern class of such a manifold vanishes provided
the sum of the weights equals the homogeneity degree of $W$, i.e.\ if
$\sum w_i = d$; if this is the case, the hypersurface is named a CY
manifold if three dimensional, it is $K3$ if two dimensional and the torus if
one dimensional.

We adopt the standard  notation to denote such manifolds:
for example, $X_{24}[1,1,2,8,12]$ is a CY space defined by $W(x)=0$ with
$x\in \IP^{1,1,2,8,12}$ and $W(\lambda x) = \lambda^{24} W(x)$.
 \par
A general quasi-homogeneous polynomial $W$ in weighted projected space
is specified by many coefficients, 335 in the  example above.
Coordinate transformations in the ambient weighted projective space,
however, induce `gauge
transformations' of these coefficients, numbering 92 in the present case.
Fixing the gauge reduces the number
of true parameters, to 243 in the example.
\par
We will consider surfaces for which a  discrete
group of additional {\it global identifications\/} is imposed  on the ambient
space. This group  we will call $G$ for CY and $G'$ for $K3$.
For instance, for $X^*_{24}[1,1,2,8,12]$,
which is one of those we will treat extensively, we impose, with respect to
its unasterisked version%
\footnote{The * in the
notation in facts indicates the mirror manifold, for which a
particular set of global identifications is necessary, as we will do
in our examples. }, additional identifications
\begin{eqnarray}
&& x_j \simeq \exp(n_j \frac{2 \pi i}{24}) \, x_j~,\nonumber\\
&&(n_1,n_2,n_3,n_4,n_5)= m_1 (1,-1,0,0,0) +m_2(-1,-1,2,0,0) \ ,
\label{identif}\end{eqnarray}
where $m_i\in \ZZ$ mod 24. Using a shorthand
notation, the group $G$ in (\ref{identif}) is generated by
the rescalings $(1,-1,0,0,0)_{24}$ and $(-1,-1,2,0,0)_{24}$.
Notice that these rescalings are defined mod $(1,1,2,8,12)_{24}$ because of
the projective identifications.
The defining polynomial $W$ is then restricted to a sum of those monomials
that are invariant under these identifications. In this way
the number of parameters is reduced to 11 and the number of gauge
invariances to 8. We thus remain with three true complex structure moduli. The
manifold turns out to be the dual of the one without the identifications, and
we have $h^{21}=3$ and $h^{11}=243$.
\par
For some purposes it is useful to keep (at least some of) the gauge
invariances, rather then fixing them, as we will see.
\par
The relevant forms  both for local as for rigid special geometry
 can be represented using the Griffiths residue theorem \cite{griffiths,LSW}.
We will use this representation  for the holomorphic forms
$\Omega^{(N,0)}$, with $N=3$ for the CY $(3,0)$ form
and $N=2$ for the $(2,0)$ form in $K3$. In that case the theorem states that
\begin{equation}
\Omega^{(N,0)}=\frac{|G|}{(2\pi\ii)^{N+1}}\int \frac{\omega}{W}\ ,
\label{griffiths30}
\end{equation}
where $|G|$ is the number of elements in the group $G$ (or $G'$ for the $K3$).
the normalization is chosen for later convenience and
the integral is over a 1-cycle that encircles the surface $W=0$
in the ambient space $\IP^{w_1,\ldots w_{N+2}}$, of which $\omega$ is the
volume form \cite{Morrison}:
\begin{equation}
\label{volform}
\omega = w_1 x^1 dx^2\ldots dx^{N+2} - w_2 x^2 dx^1\ldots dx^{N+2} + \ldots
+ (-)^{N+1} w_{N+2} x^{N+2} dx^1\ldots dx^{N+1}~.
\end{equation}
The relevant periods are obtained by
performing integrals of these forms over integral 3-cycles on the CY surface.

The expression (\ref{griffiths30}) makes it clear that periods of
$\Omega^{(N,0)}$ over $N$-cycles depend holomorphically on the moduli
(not on their complex conjugates)
appearing in the polynomial $W$. This remains true if the normalization
of $\Omega^{(N,0)}$ is redefined by any holomorphic function of the moduli.
 This freedom leads in the local \Ka\
 potential to a \Ka\ transformation.
 To obtain the expansion of the periods into the form
 of (\ref{expvbarv}) it may be necessary to perform such a \Ka\ transformation.
\subsection{$K3$ and its Picard and transcendental lattices}
\label{ss:K3Pictransc}
We will encounter $K3$ manifolds as fibres in a suitable description of
the Calabi--Yau manifolds. It is thus perhaps useful to recall some properties
of $K3$, also in comparison with those of the CY 3-folds themselves.
The number of homology 2-cycles (and of cohomology 2-forms)
of $K3$ is $b^2=22$: beside $\Omega^{(2,0)}$ and $\bar\Omega^{(0,2)}$ there are
$h^{1,1}=20$ elements in $H^{1,1}$.
\par
Just as for the CYs we are interested in the periods of the holomorphic
form $\Omega^{(3,0)}$, for $K3$ we will need the periods of $\Omega^{(2,0)}$.
 Yet, for their moduli spaces
  it is important to stress
 the following three important differences
with respect to those of the Calabi--Yau 3-folds:
 \begin{enumerate}
\item
In contrast to what happens for a CY, where the number of 3-cycles
along which we calculate the periods just equals the
third Betti number: $b^3 =2+2 \, h^{(2,1)}$, for $K3$
the periods of $\Omega^{(2,0)}$ along
a subset of 2-cycles, named `algebraic' (see below), vanish.
Therefore in later sections,
when discussing the 2-cycles for certain realizations of $K3$, we will
consider exclusively the remaining ones, called `transcendental'.
\item In the Calabi--Yau case the calculation of the periods is instrumental
for the very determination of the special K\"ahler metric
$g_{\alpha\bar \beta}(z,{\bar z})$ on moduli space. In the $K3$ case the
determination of the periods, for a fixed algebraic representation of $K3$,
corresponds only to the solution of a
uniformization problem, i.e.\ establishing the relation between
a canonical and a non-canonical coordinate system. The metric on
the complex structure moduli space is actually known {\it a priori\/} as this
space
is the homogeneous non-compact coset manifold
$O(2,n)/O(2)\times O(n)$, where $n$ is the number of transcendental cycles.
\item
The moduli-space of a CY manifold is the direct product
of its \Ka\ and complex structure moduli spaces.
In contrast, the $K3$ moduli space,
a homogeneous coset manifold modded by a discrete subgroup,
has no such direct product structure.
\end{enumerate}

Let us now be more precise, and consider a specific algebraic
realization $\Sigma_{K3}$ of the $K3$ manifold.  The proper language for
distinguishing between `{\em complex structure 2--cycles}' and
`{\em K\"ahler
class  2--cycles}' relative to such a realization, is by making use of
the concept of the Picard lattice and its orthogonal complement,
the transcendental lattice (see e.g.\ \cite{Aspinwall}).
The {\it algebraic\/} 2-cycles are those that can be {\it holomorphically\/}
embedded in $\Sigma_{K3}$, and the integral of $\Omega^{(2,0)}$ on them
clearly vanish. The Picard lattice ${\rm Pic} \left( \Sigma_{K3}\right)$
is the sublattice of $H^2 \left(K3,\ZZ \right)$
spanned by their Poincar\'e duals. They  have a purely
$H^{1,1}\left( \Sigma_{K3}\right)$
representative, and we can equivalently define
\begin{equation}
{\rm Pic} \left( \Sigma_{K3}\right)\equiv
H^{1,1}\left( \Sigma_{K3}\right) \cap H^2
\left(K3,\ZZ \right) \ .
\label{picardo}
\end{equation}
Also with this formulation it is clear that the periods of $\Omega^{(2,0)}$
along cycles dual to the 2-forms in the Picard
lattice are always zero.
Fixing a certain lattice to be the Picard
lattice puts a constraint on the allowed complex structure deformations, whose
number is reduced to $n\equiv 20-\rho(\Sigma_{K3})$, where
$\rho(\Sigma_{K3})$, the `Picard number', is the
rank of the Picard lattice.
A completely generic $K3$ has Picard number zero. For an algebraic $K3$
however, $\rho(\Sigma_{K3})\geq 1$ and the lattice has
signature $(1,\rho(\Sigma_{K3})-1)$. Indeed,
the Kodaira--Spencer theorem ensures that in a projective algebraic manifold
the \Ka\  class is proportional to an integer class.
The generic quartic has $\rho(\Sigma_3)=1$. Here
the curve $x_i=0$, with $x_i$ any
homogeneous coordinate on $\IP^3$, is algebraic. Hence there are 19 complex
structure
deformations. The purely Fermat quartic has $\rho(\Sigma_{K3})= 20$.
The two examples which we will consider have  Picard number
19 and 18, respectively. The orthogonal complement of ${\rm Pic} \left( \Sigma_{K3}\right) $
in $ H^2\left(K3,\ZZ \right) $ is called the
{\it transcendental lattice\/}
\begin{equation}
\Lambda^{\rm tr}=\left\{\hat  C \in H^2(K3,\ZZ) | \hat C \cdot Pic =
0 \right\}\ ,
\end{equation}
the dot product being determined by the intersection of the dual cycles.
This lattice has rank
$22-\rho(\Sigma_{K3})$ and
signature $ \left( 2,20-\rho(\Sigma_{K3})\right)$.
Denote by $\{\hat c_A\}$ for $A=1,\ldots 22$ a basis of $H^2(K3,\ZZ)$, by
$\{\hat c_a\}$ for $a=1,\ldots \rho(\Sigma_{K3})$ a basis of ${\rm Pic}$,
and $\{\hat c_I\}$ for $I=\rho(\Sigma_{K3})+1,\ldots 22$ a basis of
the transcendental lattice.
We thus have
\begin{eqnarray}
\hat c_a&=& {\cal G}_a{}^A\hat c_A\qquad\mbox{with }{\cal G}_a{}^A\in\ZZ
 \nonumber\\
\hat c_I&=& {\cal G}_I{}^A\hat c_A\qquad\mbox{with }\left\{
\begin{array}{ll}
{\cal G}_I{}^A\in\ZZ\\
{\cal I}_{Ia}\equiv {\cal G}_I{}^A {\cal I}_{AB}{\cal G}_a{}^B=0
\end{array}\right. \ ,
\end{eqnarray}
with ${\rm rank~}({\cal G}_a{}^A) =\rho(\Sigma_{K3}) $.
The total transformation of basis for the 22 two-cycles is thus
determined by the matrix
\begin{equation}
\pmatrix{{\cal G}_a{}^A\cr {\cal G}_I{}^A} \ .     \label{matcalG}
\end{equation}
If this matrix is of determinant 1, then
$H^2(K3,\ZZ) = {\rm Pic}\oplus \Lambda^{\rm tr}$,
but in general the $b^2$ vectors $\{c_a,c_I\}$,
do {\em not\/} form a lattice basis of $H^2(K3,\ZZ)$.
\par
Since the periods of the $(2,0)$-form over the algebraic cycles (the Poincar\'e
duals of the forms in ${\rm Pic}$) vanish,
the complex structure moduli space at fixed Picard
lattice is (over)parametrized by the periods along the transcendental cycles
(dual to elements of $\Lambda^{\rm tr}$).
\par
Consider the general relation
\begin{eqnarray}
\int_{c_A} \lambda\ {\cal I}^{AB}\int_{c_B}\omega = \int_{K3}
\lambda\wedge \omega\ ,
\end{eqnarray}
for any 2-forms $\lambda$ and $\omega$ and with $c_A$ a complete basis
of cycles with ${\cal I}^{AB}$ the inverse of the complete intersection
matrix. This relation remains true for any change of basis, even if
not integral. Therefore, using the transformation (\ref{matcalG}),
this equation reduces for $\lambda=\omega=\Omega^{(2,0)}$
to a sum over the cycles $\hat c_I$ forming a basis of the $\Lambda^{\rm
tr}$. Defining the periods
\begin{equation}
\theta_I = \int_{\hat c_I} \Omega ^{(2,0)}\ ,   \label{transcper}
\end{equation}
we obtain
\begin{equation}
 \theta_I \,{\cal I}^{IJ} \, \theta_J =\int_{K3} \Omega^{(2,0)}
 \wedge \Omega^{(2,0)} \, = \, 0\ ,
 \label{interfono}
\end{equation}
where the symmetric matrix ${\cal I}^{IJ}$, of signature
$(2,20-\rho(\Sigma_{K3}))$, is the inverse of the
intersection matrix of the transcendental 2-cycles:
\begin{equation}
 {\cal I}_{IJ} \,  \equiv \, {\hat c}_{I} \, \cap \,{\hat c}_{J}\ .
\label{intersezione2}
\end{equation}
\par
In the next section we shall introduce a class of CY manifolds that admits
a $K3$ fibration. This class includes the two examples that
will subsequently be treated in detail. We will indicate
how the fibration can be used to reduce the calculation of the CY periods
to that of the $K3$ periods, and discuss some strategies to carry out this
computation.
\subsection{A class of $K3$ fibrations}
\label{ss:classK3}
We consider Calabi--Yau manifolds of the form\footnote{Here we write
$X_{2K}$ for a manifold which has at least some global identifications
as we will explain. In practice we will use the one with all identifications
such that it is $X^*_{2K}[1,1,2k_3,2k_4,2k_5]$.}
\begin{equation}
X_{2K}[1,1,2k_3,2k_4,2k_5]\qquad\mbox{with }\ K=1+k_3+k_4+k_5\ ,
\label{genCY}
\end{equation}
that is, zero-loci of a quasi-homogeneous polynomial
$W(x)=\lambda^{2K}W(\lambda x)$ in the weighted projective
space $\IP^{1,1,2k_3,2k_4,2k_5}$, where a point of coordinates
$(x_1,x_2,x_3,x_4,x_5)$ is identified with $
(\lambda x_1,\lambda x_2,\lambda^{2k_3} x_3,\lambda^{2k_4} x_4,
\lambda^{2k_5} x_5)$.
The ambient space is subject to a group of discrete
identifications $G$ of the form
\begin{equation}
G= \ZZ_K  \times G'\ , \label{GisZKGp}
\end{equation}
where the first factor corresponds to the identifications
\begin{equation}
x_1\sim e^{\frac{2\pi i}{2K}} x_1\ ;\qquad
x_2\sim e^{\frac{-2\pi i}{2K}} x_2\
\label{genident}
\end{equation}
(note that the $K{\rm th}$ power of (\ref{genident}) is a projective
transformation) and $G'$ leaves invariant the ratio $x_1/x_2$.
The most general polynomial of degree $2K$ in the $x$'s
invariant under (\ref{genident}) can be written (up to a rescaling
of $x_1/x_2$) as
\begin{equation}
W=\frac{B}{2K}\left( x_1^{2K} +x_2^{2K}\right) +\hat W(x_1x_2,
x_3,x_4,x_5;\psi_i)-\frac{1}{K}\psi_s (x_1x_2)^K\ ,
\label{genpol}
\end{equation}
where $B,~\psi_i,~\psi_s$ are moduli and $\hat{W}$ is
a polynomial of weight $K$ in the projective space $\IP^{1,k_3,k_4,k_5}$.
The $K3$-fibration of the CY manifold we are considering is
exhibited introducing new coordinates $x_0$ and $\zeta$,
invariant under (\ref{genident}),
by
\begin{equation}
x_1=\zeta^{1/(2K)}\sqrt{x_0}  \ ;\qquad
x_2=\zeta^{-1/(2K)}\sqrt{x_0}\ .
\label{x1x2x0z}
\end{equation}
In these coordinates the potential $W$ becomes
\begin{equation}
W= B'\frac{1}{K}x^K_0 + \hat W(x_0, x_3,x_4,x_5;\psi_i)\ ,
\label{defgenK3}
\end{equation}
where we have introduced
\begin{equation}
B'=\frac12\left(B\zeta +\frac{B}{\zeta}-2\psi_s\right)\ .
\label{B'}
\end{equation}
The polynomial $W$ in (\ref{defgenK3}) then describes a $K3$ manifold
of type $X_K[1,k_3,k_4,k_5]$ (subject to a discrete identification group
$G'$), with one of the moduli, namely  $B'$, varying from point to point
in the base as in (\ref{B'}).

Note that the `gauge invariances' in the moduli space
of both the CY and the $K3$ always include the projective transformation
\begin{equation}
B'\rightarrow \lambda^K B'\ ; \qquad
\psi_i\rightarrow\psi_i\lambda^{a_i} \ ;\qquad \lambda\in \Cbar_0\ ,
\label{projmoduli}
\end{equation}
where the weights $a_i$ are determined  so as to reabsorb the coordinate
rescaling $x_0\to \lambda^{-1} x_0$:
\begin{equation}
W(x_0, x_3,x_4,x_5;B',\psi_i)=
W(\ft1\lambda x_0, x_3,x_4,x_5;B'\lambda^K,\psi_i\lambda^{a_i})\ .
\label{defai}
\end{equation}
\subsection{Using the $K3$ fibration for the periods}
\label{ss:K3fperiods}
In the calculation of periods and monodromy
matrices for Calabi--Yau 3-folds one may take full advantage
of the fibration structure of the
manifold. In the expression for the $(3,0)$-form, (\ref{griffiths30}), we can
make use of the relation between the volume forms
\begin{eqnarray}
\omega_{CY}&=& \frac{1}{K} \omega_{K3}\,\frac{d\zeta}{\zeta}\ ;
\qquad \Omega^{(3,0)} =\Omega^{(2,0)}\,\frac{d\zeta}{2\pi\ii\,\zeta}
\nonumber\\
\omega_{K3}&=&x_0\,  dx_3\, dx_4\, dx_5 -dx_0
\left(k_3 x_3\, dx_4\, dx_5  +
k_4 x_4\, dx_5\, dx_3  +k_5 x_5\, dx_3\, dx_4\right) \ .
\label{omegaCYK3}
\end{eqnarray}
The factor $K$ is cancelled in the relation between the forms due to
(\ref{griffiths30}) and  (\ref{GisZKGp}).
To make full use of this factorized form, it is convenient to
also describe the relevant CY 3-cycles in a factorized form:
this allows us to integrate first
over 2-cycles in the $K3$ manifold, thus giving periods of the $K3$,
leaving the $\zeta$ integral to the end.
The structure of this last integration will eventually
lead to the auxiliary  Riemann surfaces
of \cite{SeiWit}, and the associated special \Ka\ geometry.
Let us therefore first consider the structure of the 3-cycles.
The homology cycles $C$
of the Calabi--Yau manifold  can be constructed as fibrations%
\footnote{In the following, for simplicity, we write just
$C = \gamma\times c$, but we understand a fibration.}
of homology cycles $c$ of the $K3$--fibre over (possibly {\em open\/})
paths $\gamma$ in the base manifold $\IP^1$; that is,
there is a fibration $f:C \to \gamma $ such that for every point
$p \in \gamma$, the fibre $f^{-1}(p)$ is a $K3$ 2-cycle $c$.
The intersection of two 3-cycles $C_1, C_2$ is then
\begin{equation}
\label{fibint}
C_1 \cdot C_2 = \sum_{p \in \gamma_1 \cap \gamma_2}
s(p) \; {f_1}^{-1}(p) \cdot f_2^{-1}(p),
\end{equation}
where $s(p)$ is the sign of the intersection in $p$ of the paths
$\gamma_1$ and $\gamma_2$.
This simply means that we have to add (with the appropriate sign)
the $K3$ intersections of the fibres (which are 2-cycles) above
the intersection points in the base\footnote{Note that, in general,
$f_1^{-1}(p) \cdot f_2^{-1}(p)$ will be $p$-dependent (due
to monodromies). The sum factorizes only if the CY cycles have
a true global direct product structure.}.

Consider a particular non-trivial 2-cycle in the $K3$ manifold.
This cycle may nevertheless vanish for a value of the moduli
that makes the $K3$ singular. For a fixed Calabi--Yau manifold
this singularity corresponds to a certain value of $\zeta$. Now
one can make two constructions
\begin{itemize}
\item One can consider a
path between two (singular) points in the $\zeta$-base space where the same
$K3$ cycle vanishes. The $K3$ 2-cycle along this path gives a closed
3-cycle in the CY manifold.
\item One can consider a non-trivial loop in the base space
transporting a cycle that has trivial monodromy around that loop.
\end{itemize}
\subsection{Strategies for obtaining $K3$ periods}\label{ss:stratK3per}
In view of this fibration structure we will first address the calculation of
periods, monodromies and intersection matrix for the $K3$ manifold,
leaving the dependence of the moduli on the $\IP^1$ base coordinate
$\zeta$ for later.
\par
The most straightforward approach to the computation of the
$K3$ periods would be to explicitly integrate the (2,0) form
over an integral basis of 2-cycles\footnote{We will further consider only the
transcendental cycles, and will therefore just talk about these as
`the cycles'.}.
This is, however, not always practical. In the first example, in fact,
we will use a variation on this method \cite{candmirror1}.
We explicitly integrate  over one specific cycle,
which is easy to parametrize, in the neighbourhood of the
`large complex structure' point in moduli space; this gives the so-called
`fundamental period'. Analytic continuation  of this period,
as a function of the modulus, outside the original neighbourhood
reveals cuts. Associated monodromy transformations change it into
(integer) linear combinations of a basis of integral periods.
An approach which is complementary in this respect is the use of
the {\it Picard--Fuchs equations\/} satisfied by
the periods of the $(2,0)$ form.
Using (\ref{interfono}) the intersection matrix is
obtained modulo a multiplicative constant.
In turn, while a generic basis of solutions of the PF equation is a basis
for the $K3$ periods, it does not necessarily correspond to integrals
over {\em integer\/} cycles:
it is not an {\em integral basis}, and the explicit expression
of the period over the fundamental cycle is needed in order to be able to
define such a basis.
\par
Finally one can try to apply the strategy used for the Calabi--Yau
periods to the $K3$ periods themselves. In the second example this
works because the $K3$ is
itself a {\em torus fibration.} The forms and cycles can be decomposed in
forms and cycles on the torus, fibred over a second $\IP^1$ base space.
For that example this procedure has the advantage that the starting point is
the torus, where much more is known explicitly
for  cycles, periods and the intersection matrix.
\section{Introducing the examples: \\
fibrations and singularities} \label{ss:exa}
In this section, we turn to the explicit description
of the two examples of CY manifolds that we study in detail,
the $K3$ fibrations that we use in this study,
and their singularity structures.
We use the notation introduced in subsection~\ref{ss:classK3}.
\subsection{Description and fibration of $X^*_{8}[1,1,2,2,2]$}
\label{ss:descrCY2}
The first example is the Calabi--Yau manifold in the class (\ref{genCY})
with $k_3=k_4=k_5=1$.
The group $G$ contains the identifications
\begin{equation}
x_j \simeq \exp(n_j \frac{2 \pi i}{8}) \, x_j
\label{equivalentie2}
\end{equation}
with
\begin{eqnarray}
(n_1,n_2,n_3,n_4,n_5)&=&m_0 (1,1,2,2,2)+m_1(1,-1,0,0,0)\nonumber\\ &&+
m_2 (0,0,2,-2,0)+m_3(0,0,2,0,-2) \ ,
\label{identif2}
\end{eqnarray}
where $m_1,m_2,m_3\in \ZZ$. The transformation parametrized with  $m_0$
is the identifying relation for the weighted projective space in which our
CY is embedded. If $m_0$ is an integer, (\ref{equivalentie2})
leaves the polynomial invariant. The identification $m_1$
is the transformation (\ref{genident}). The group $G$ is thus
$\ZZ_4^3$ as in (\ref{GisZKGp}) with $G'=\ZZ_4^2$.

The most general polynomial of
degree 8 in the variables $x_1,\ldots,x_5$, invariant under (\ref{identif2}),
is (up to rescalings of the $x_i$)
\begin{equation}
W^{(1)}=\frac{b_1}{8} x_1^8 + \frac{b_2}{8} x_2^8 + \frac{b_3}{4}x_3^4
+ \frac{b_4}{4}x_4^4 + \frac{b_5}{4}x_5^4
- \psi_0 \, x_1 x_2 x_3 x_4 x_5 - \frac{1}{4} \psi_s \, (x_1 x_2)^4\ .
\label{w11222g}
\end{equation}
There are still 5 gauge invariances, induced by rescaling the $x_i$,
such that there remain 2 independent moduli.
We could take as invariants, for example,
$\psi_s^2/(b_1 b_2)$ and $\psi_0^4/(\psi_s b_3 b_4 b_5)$.
This would be valid in a patch where $\psi_s b_1 b_2 b_3 b_4 b_5\ne 0$.
We will regularly make such choices without further comments.
For example, in the form of (\ref{genpol}), the gauge invariance
has already been used\footnote{Although $b_1=0$ will be of interest for
our work, this is not a limitation:
see footnote~\ref{b1=0} on page~\pageref{b1=0}.}
also  to put $b_1=b_2=B$.
It is sometimes useful to postpone such choices, notably for
the discussion of singularities (section~\ref{Singuls}), and also for the
derivation of Picard--Fuchs equations (section~\ref{ss:permonK3}).
\par
We continue with the $K3$ fibration, following the general pattern of
section~\ref{ss:classK3}: introducing $x_1 x_2 =x_0$ leads to (\ref{defgenK3}),
which in this case becomes
\begin{equation}
W^{(1)}=\frac{B'}{4}x_0^4+ \frac{b_3}{4}x_3^4+ \frac{b_4}{4}x_4^4
+ \frac{b_5}{4}x_5^4-\psi_0 x_0x_3 x_4 x_5\ .
\label{K3-I}
\end{equation}
The remaining gauge invariances are the rescalings of the $x_i$.
Therefore the moduli space of this fibre is in fact 1-complex dimensional.
The truly invariant variable is
\begin{equation}
z=  - \frac{ B' b_3 b_4 b_5}{\psi_0^4} \ .\label{defz}
\end{equation}
(with minus signs for later convenience). Mostly we will  partially gauge
fix the rescalings by putting
$b_3=b_4=b_5=1$, such that the polynomial is
\begin{equation}
W^{(1)}(x;B',\psi_0)=\ft14\left(B' x_0^4+ x_3^4+ x_4^4+x_5^4
\right) -\psi_0 x_0x_3x_4x_5
\label{K31111}
\end{equation}
with  $B'$ given in (\ref{B'}).
The moduli space has the projective transformations given in general in
(\ref{projmoduli}) (with $K=4$, $a_0=1$).
The identifications in
(\ref{identif2}) give rise to corresponding identifications
on the $K3$ manifold
\begin{equation}
x_j \simeq \exp(n_j \frac{2 \pi i}{4}) \, x_j  \label{ideqnK3}
\end{equation}
with
\begin{equation}
(n_0,n_3,n_4,n_5)=m_0 (1,1,1,1) +m_2(0,1,-1,0) +m_3(0,1,0,-1 )\ .
\label{identK3}
\end{equation}
We stress that $\zeta$, defined in \eqn{x1x2x0z}, is invariant under
the identifications \eqn{equivalentie2} so that the change of variables
\eqn{genident} is one to one.

\subsection{Description and fibrations of $X^*_{24}[1,1,2,8,12]$}
\label{ss:descrCY}
As a second example, we start from the polynomials of degree 24 in the
variables $x_1,
\ldots, x_5$ with weights $1,1,2,8,12$, respectively, which is in the class
(\ref{genCY}) with $K=12$ and $k_3=1$, $k_4=4$, $k_5=6$.
As mentioned before, in section~\ref{ss:K3-generalities},
we impose the global identifications \eqn{identif}.
This leads to  a Calabi--Yau manifold with three complex structure moduli.
The generic  polynomial compatible with the identification group $G$ contains
11 arbitrary coefficients:
\begin{eqnarray}
W^{(2)}&=&\frac{1}{24} (b_1x_1^{24} + b_2 x_2^{24}) + \frac{b_3}{12} x_3^{12}
+\frac{b_4}{3} x_4^3 +\frac{b_5}{2} x_5^2 \nonumber \\
&&- \psi_0 \, x_1 x_2 x_3 x_4 x_5 -
\frac{1}{6} \psi_1 \, (x_1 x_2 x_3)^6 - \frac{1}{12} \psi_s \, (x_1 x_2)^{12}
\nonumber\\
&&-\frac{1}{4} \psi_3 \left ( x_1 x_2 x_3 x_4 \right )^2 - \frac{1}{4}
\psi_4  \left ( x_1 x_2 x_3 \right )^4 x_4 -\frac{1}{3} \psi_5 \left
(x_1 x_2 x_3 \right )^3 x_5 \ .
\label{genpolyn}
\end{eqnarray}
Again we will take $b_1=b_2=B$ and use the form \eqn{genpol}.

\subsubsection{The Calabi--Yau as a $K3$ fibration}\label{ex2:CY2K3}
The general fibration scheme of section~\ref{ss:classK3} then gives
\begin{eqnarray}
W^{(2)}&=&\frac{B'}{12} x_0^{12} + \frac{b_3}{12} x_3^{12}
+\frac{b_4}{3} x_4^3 +\frac{b_5}{2} x_5^2
- \psi_0 \, x_0 x_3 x_4 x_5 -
\frac{1}{6} \psi_1 \, (x_0 x_3)^6 \nonumber\\
&&-\frac{1}{4} \psi_3 \left ( x_0 x_3 x_4 \right )^2 - \frac{1}{4}
\psi_4  \left ( x_0 x_3 \right )^4 x_4 -\frac{1}{3} \psi_5 \left
(x_0 x_3 \right )^3 x_5 \ .
\label{K3genpolyn2}
\end{eqnarray}
In this case the identification group that restricts the available deformations
to those listed above acts on the homogeneous coordinates as
follows:
\begin{equation}
x_j \simeq \exp(n_j \frac{2 \pi i}{12}) \, x_j  \label{neuqnK3}
\end{equation}
with
\begin{equation}
(n_0,n_3,n_4,n_5)=m_0 (1,1,4,6)+m_2(-1,1,0,0)\ .
\label{neuntK3}
\end{equation}
The group $G$ is as in \eqn{GisZKGp} with $K=12$ and $G'=\ZZ_6$
(as $m_0=m_{2}=6$ gives the identity).
The gauge transformations on the moduli space are induced by redefinitions
\begin{equation}
\begin{array}{lcl}
\tilde x_0 = \lambda_0 x_0~, & \null\hskip 0.4cm &
\tilde x_4 = \lambda_4 x_4+ \lambda'_4 (x_0 x_3)^2~, \\
\tilde x_3 = \lambda_3 x_3~, & \null &
\tilde x_5 = \lambda_5 x_5+ \lambda'_5 x_0 x_3 x_4
+ \lambda^{\prime\prime}_5 (x_0 x_3)^3~,
\end{array}
\end{equation}
so that the moduli space of the fibre is 2-complex dimensional.
The polynomial is invariant in the sense that
\begin{equation}
W^{(2)}\left(\tilde x; B', b, \psi\right)  =
W^{(2)}\left( x; \tilde B',\tilde b, \tilde\psi\right)~,
\end{equation}
where
\begin{eqnarray}
\tilde B'&=&B' \lambda_0^{12} \ ;\qquad \tilde b_3=b_3 \lambda_3^{12} \ ;\qquad
\tilde b_4=b_4 \lambda_4^3 \ ;\qquad
\tilde b_5=b_5 \lambda_5^2      \nonumber\\
\tilde \psi_0&=&\psi_0 \lambda_{03} \lambda_4 \lambda_5-
b_5 \lambda'_5 \lambda_5       \nonumber\\
\tilde \psi_1&=&\psi_1 \lambda_{03}^6
    -3 b_5 \lambda^{\prime\prime\,2}_5  -2 b_4 {\lambda'_4}^3 +
       6 \lambda^{\prime\prime}_5  \lambda'_4 \lambda_{03}
         \psi_0  + \ft32 {\lambda'_4}^2 \lambda_0^2
         \lambda_3^2 \psi_3 +
        \ft32 \lambda'_4 \lambda_{03}^4
         \psi_4
         +2 \lambda^{\prime\prime}_5  \lambda_{03}^3 \psi_5
         \nonumber\\
\tilde \psi_3&=&\psi_3 (\lambda_{03} \lambda_4)^2-
2 b_5 {\lambda'_5}^2 - 4 b_4 \lambda'_4 \lambda_4^2 +
         4 \lambda'_5 \lambda_{03} \lambda_4
          \psi_0              \nonumber\\
\tilde \psi_4&=&\psi_4 \lambda_{03}^4 \lambda_4
- 4 b_4 {\lambda'_4}^2 \lambda_4
         \psi_0 +4 \lambda^{\prime\prime}_5  \lambda_{03}
         \lambda_4 \psi_0 +
        2 \lambda'_4 \lambda_{03}^2
         \lambda_4 \psi_3
         +4\lambda'_5\left(
 \lambda'_4  \lambda_{03}
         - b_5 \lambda^{\prime\prime}_5
         +\ft13 \lambda_{03}^3 \psi_5 \right)     \nonumber\\
\tilde \psi_5&=&(\psi_5\lambda_{03}^3 -
3 b_5 \lambda^{\prime\prime}_5  +3 \lambda'_4 \lambda_{03} \psi_0) \lambda_5\ ,
\end{eqnarray}
where $\lambda_{03}=\lambda_0\lambda_3$.
Introducing the auxiliary combinations
\begin{eqnarray}
I_0&=&\psi_3+2\frac{\psi_0^2}{b_5}\ ;\qquad
I_1= \psi_4+ \frac{I_0^2}{ 4 b_4} +\frac{4}{3}\frac{\psi_0\psi_5}{b_5}\ ,
\nonumber\\
I_2&=& \psi_1 +\frac{3}{8}
\frac{I_0I_1}{ b_4}-\frac{1}{32} \frac{I_0^3}{b_4^2}
+\frac{1}{3}\frac{\psi_5^2}{b_5}\ ,
\end{eqnarray}
we parametrize the moduli space with the invariants
\begin{equation}
\nu_1=I_1 ^3 \left(B' b_3
b_4\right)^{-1}\ ;\qquad
\nu_2= I_2^2 \left( B'b_3\right) ^{-1} \ .   \label{invnu12}
\end{equation}

For most of the paper we will use the gauge
\begin{equation}
b_3=b_4=b_5=1\ ;\qquad \psi_3=\psi_4=\psi_5=0\ ,   \label{usualgauge2}
\end{equation}
which is also the one used in earlier work.
For later reference we give the polynomial in this gauge,
\begin{equation}
W^{(2)}(x;B', \psi_0, \psi_1) = \frac{1}{12}
(B' x_0^{12} + x_3^{12})
+\frac{1}{3} x_4^3 +\frac{1}{2} x_5^2
-\psi_0 \, x_0 x_3 x_4 x_5 - \frac{1}{6} \psi_1 \,x_0^6 x_3^6\ ,
\label{WW121146}
\end{equation}
and the expressions for the invariants,
\begin{eqnarray}
&&I_0=2\psi_0^2\ ;\qquad I_1=\psi_0^4\ ,\qquad I_2=
\psi_1+\ft12\psi_0^6~; \nonumber\\
&&\nu_1= \psi_0^{12}/B'\ ,\qquad \nu_2=
\left(\psi_1+\ft12\psi_0^6 \right)^2 /B'\ .
\label{invnu12usgauge}
\end{eqnarray}
The remaining projective transformations on the moduli
space coordinates are  as in \eqn{projmoduli} with $K=12$, $a_0=1$ and $a_1=6$.
\par
With this gauge choice there is, however,
a remaining discrete transformation identifying
\begin{equation}
(\psi_0^6, \psi_1) \sim (-\psi_0^6, \psi_1+\psi_0^6) .
\label{discreteGT}
\end{equation}

For some purposes, the alternative gauge choice
\begin{equation}
b_3=b_4=b_5=1\ ;\qquad \psi_0=\psi_3=\psi_5=0\ ,   \label{gaugefibr2}
\end{equation}
is more convenient: that choice fixes the gauge completely.
Note that in this gauge $\psi_4$ and $\psi_1$ are $I_1$ and $I_2$ and thus
according to \eqn{invnu12usgauge}
correspond to $\psi_0^4 $ and $(\ft12\psi_0^6 + \psi_1)$
in the gauge \eqn{usualgauge2}.
\par
The CY has a $\ZZ_2$ symmetry, which is most clearly exhibited in
the alternative gauge as $x_5\rightarrow -x_5$.
When one of the invariants $\nu_i$ vanishes, this symmetry is
enhanced. If $\nu_1$ vanishes, which in the alternative gauge
corresponds to $\psi_4=0$, then the symmetry becomes $\ZZ_6$,
represented as $x_3\rightarrow e^{2\pi\ii/6}x_3$. This includes the
above-mentioned $\ZZ_2$ as $x_3\rightarrow -x_3$ is by \eqn{neuntK3}
with $m_0=m_2=3$ the same as $x_5\rightarrow -x_5$.
If $\nu_2=0$ (or $\psi_1=0$ in this gauge), the symmetry is enhanced
to $\ZZ_4$, represented by $x_3\rightarrow e^{2\pi\ii/4}x_3$.
\par
This fibration has been used in \cite{KLMVW} to develop the reduction
of the CY manifold, in the neighbourhood of a singular point in its moduli
space. In that neighbourhood the $K3$ fibre is seen to degenerate into an
ALE space. We will continue to work without taking this singular limit at
this stage, and therefore will not approximate the $K3$ fibre as an ALE
space. Instead, we develop additional techniques to work with the $K3$
exactly. One of these techniques is a further fibration of the $K3$ itself.

\subsubsection{The $K3$ as a torus fibration}\label{ss:ex2torusfibr}
\begin{figure}
\begin{center}
\null\hskip -1pt
\epsfysize=5cm
\epsffile{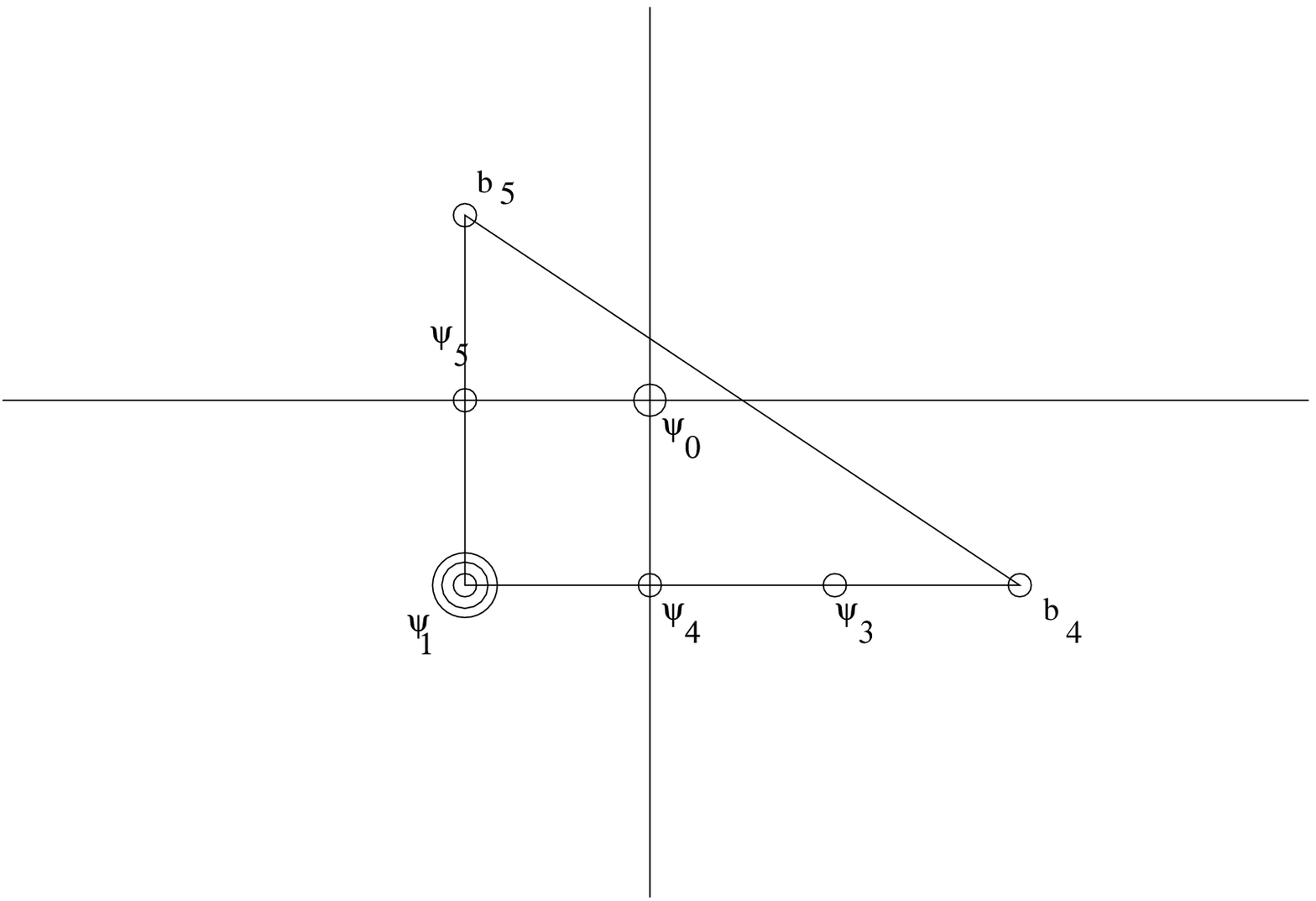}
\end{center}
\vskip -1.1cm
\mycaptionl{The polyhedron of $K3$. Each point corresponds to a monomial in
\eqn{K3Lpol2}. The axes correspond to the powers of $x$ and $y$
in the monomials, while the $\xi$-axis is orthogonal to the plane of the
drawing.
The vertices corresponding to the $B'$ and $b_3$
term are above and below the $\psi_1$ point.}
\label{fig:fibrk3}
\end{figure}

In this  example the $K3$ manifold itself is an elliptic fibration.
Actually, it can be fibred further in two different ways.
The possibility of these fibrations can be seen from toric geometry
considerations (see, for instance, \cite{hosono}).
To this end, we construct the (reflexive) polyhedron that
describes this manifold. The following variables are invariant under
projective and global identifications:
\begin{eqnarray}
\xi=\left( \frac{x_3}{x_0}\right) ^6\ ;\qquad
x= \frac{x_4}{y_0^2}\ ;\qquad y=\frac{ x_5}{y_0^3}\ ,
\end{eqnarray}
where the introduction of
\begin{equation}
y_0= x_0 x_3\
\end{equation}
and $\xi$ again follows the general pattern of section~\ref{ss:classK3},
and the other variables are introduced to write the Laurent polynomial
defining the manifold in a standard form:
\begin{eqnarray}
\frac{W^{(2)}}{x_0 x_3 x_4 x_5}&=&-\psi_0-\ft13\psi_5 x^{-1}+\ft13 b_4x^2y^{-1}
 -\ft14\psi_4 y^{-1}-\ft14\psi_3xy^{-1}+\ft12b_5x^{-1}y
\nonumber\\
&&-\ft16\psi_1x^{-1}y^{-1}+\ft1{12}b_3\xi x^{-1}y^{-1}
+\ft1{12}B'\xi^{-1}x^{-1}y^{-1}\ .
\label{K3Lpol2}
\end{eqnarray}
{}From this expression, we read off the polyhedron in figure~\ref{fig:fibrk3}.
Fibrations of the $K3$ correspond to any 2-plane in this 3-dimensional picture
on which the restriction of the polyhedron
is itself reflexive, which implies that it should include a central point.
This is clearly the case for the plane of the drawing
(orthogonal to the $\xi$-axis), which corresponds to
a first elliptic fibration, with $\xi$ parametrizing the base.
The second possibility is  the plane
perpendicular to the picture through the line of $b_5$ and $\psi_1$.
The base manifold in this case can be parametrized by $x$.
\par
The first fibration takes $\xi$ as the coordinate on the base,
and the fibre is a torus presented as
$X_6[1,2,3]$, in the projective space of
$\{ y_0,x_4,x_5\}$. Using the gauge \eqn{usualgauge2}
the polynomial takes the form
\begin{equation}
W^{(2)}=\frac{1}{6} B^{^{\prime\prime}} y_0^6 + \frac{1}{3} x_4^3
+ \frac{1}{2} x_5^2 - \psi_0 y_0 x_4 x_5 \ ,
\label{WX6123}
\end{equation}
where
\begin{equation}
B^{^{\prime\prime}} = \frac{1}{2} \left(\frac{B^\prime}{\xi} + \xi
- 2\psi_1\right)\ .
\end{equation}
The second fibration takes $x$ as the coordinate of the base.
In the alternative gauge \eqn{gaugefibr2}
 the $K3$-polynomial turns into
\begin{equation}
W^{(2)}y_0^{-6} = \frac{1}{2} y^2 + \frac{1}{12} \left(\xi
 + \frac{B^\prime}{\xi}\right)
+ \frac{1}{6} P(x)\ ,
\label{torusfib-mod2}
\end{equation}
where
\begin{equation}
P(x) = 2 x^3 - \ft{3}{2} \psi_4 x - \psi_1\ .
\label{K3pol3}
\end{equation}

\subsection{Singularities \label{Singuls}}

The rigid limit of the supergravity theory based on a Calabi--Yau manifold
will correspond to the description of the behaviour of the theory in
the vicinity of a certain singularity in the moduli space.
At singular loci in the moduli space the CY manifold develops
a singularity%
\footnote{Recall that singularities occur when at the same time $W=0$
and $dW=0$.
Since we are using homogeneous coordinates the above conditions
are satisfied if $\frac{\partial W}{\partial x_i}=0$.}.
Hence we will first describe the singularity structure of the two
CY 3-folds introduced above, relying on their $K3$-fibred structure.
Therefore we will first study the singularity structure
on the $K3$-fibre  and then examine the total CY space.
In the second example, we will make use of the fact that the $K3$ fibre
itself is an elliptic fibration.
\par
The rigid-limit singularity corresponds \cite{KLM,KKLMV}
to the point at infinity on a line in moduli space where the CY develops a
conifold singularity.
At this point, the CY manifold becomes singular on the whole
$\IP^1$ base space of the fibration. This is the situation we are interested in.
\par
In addition to the singularity structure itself,
for both examples we will include a first look
at the neighbourhood of such a singularity.
This will be given by not only making an expansion around the rigid-limit
singular point in moduli space, but also by approximating the $K3$ fibre
by an ALE space.
Although we stress that this approximation will not be used in the remainder of
this paper, we include it to make contact with \cite{KLMVW}, and also
because it provides a useful picture to direct the ideas.
\subsubsection{Singularities of $X_{8}^*[1,1,2,2,2]$}\label{ss:singex1}
\paragraph{Singularities in the $K3$.}
We start from the defining polynomial of \eqn{K31111},
with \eqn{defz} giving the relevant modulus.
Imposing the conditions for the simultaneous vanishing of
derivatives of  $W^{(1)}$, \eqn{K3-I}, we find,
the following special points in $K3$ moduli space:
\begin{itemize}
\item[a)] $\psi_0^4=B'$, i.e. $z = -1$. This is the conifold singularity.
The point on the $K3$ where the singularity develops is
$(x_0,x_3,x_4,x_5)\sim(1,\psi_0,\psi_0,\psi_0)$ up to identifications
\eqn{equivalentie2}. The monodromy of the periods around $z=-1$ is
diagonalizable (see after \eqn{M-1+}).
\item[b)] $B'=0$, i.e.%
\footnote{There is some danger in this identification, since
the proper treatment requires desingularization of the
CY ambient space. This has been taken into account in the analysis of the
CY singularity structure of the next paragraph.}
$z=0$. Around  this point
the monodromy matrix is not
diagonalizable and the singularity is referred to as
`the large complex structure limit'. The point  on the $K3$
at which the singularity develops is $\sim(x_0,0,0,0)$
where $x_0$ can be scaled to~1.
\end{itemize}

\paragraph{Singularities in the Calabi--Yau.}
Using the results on singularities of the $K3$ fibre, we now
examine the Calabi--Yau 3-fold.
{}From \eqn{B'} we find that each of the above $K3$ fibre singularities
occurs at two points in the $\zeta$-plane:
$B'=0$ (l.c.s.) and $B'=\psi_0^4$ (conifold) occur, respectively, at
\begin{equation}
e_0^\pm =
\tilde\psi_s \pm \sqrt{\tilde\psi_s^2 - 1}
\ ;\qquad
e_1^\pm =
\tilde u \pm \sqrt{\tilde u^2 - 1}~,  \label{e01pm}
\end{equation}
where we have introduced the gauge-invariant
combinations
\begin{equation}
\tilde\psi_s = {\psi_s\over B}  \ ;\qquad \tilde u
\equiv {\psi_s + \psi_0^4\over B}\ .
\label{deftildepsiu}
\end{equation}
Note that, in terms of these, the $K3$ modulus is given by
\begin{equation}
\label{fredag2}
z =  {\tilde\psi_s - \ft12(\zeta + {1\over \zeta})\over
\tilde u - \tilde\psi_s}\ .
\end{equation}
In the CY, $\zeta$ is an additional coordinate,
and singularities occur if, beside the $K3$ fibre being singular,
$\partial_\zeta W^{(1)}=0$. This gives $\partial_\zeta B'=0$,
which imposes that $\zeta^2=1$ or $B=0$. Thus the discriminant becomes
\begin{equation}
\Delta_{CY}\propto B^2 \left(B^2-\psi_s^2 \right)
\left(B^2-(\psi_s+\psi_0^4)^2 \right)=B^6 (e_0^+-e_0^-)^2
(e_1^+-e_1^-)^2\ .
\end{equation}
We now take into account the desingularization mentioned in the previous
paragraph, and also (temporarily) reinstate the $b_i$ parameters,
see~\eqn{w11222g}, to retain the
homogeneous description of the moduli space.
The singular loci, the three components solving $\Delta_{CY}=0$,
are then given by
\begin{eqnarray}
S_0:&& b_1 b_2 b_3 b_4 b_5=0\ ,\nonumber\\
S_1:&& (b_3 b_4 b_5  \psi_s + \psi_0^4 )^2=
            b_1 b_2 ( b_3 b_4 b_5 )^2\ ,\nonumber\\
S_b:&& \psi_s^2=b_1 b_2 \ .
\label{singCpm}
\end{eqnarray}
The first locus has many components, but naturally splits into
$b_1 b_2=0$ and $b_3 b_4 b_5 =0$. We will only be interested in the former,
and therefore limit the analysis to $b_3 b_4 b_5 \ne 0$: we will then
choose the gauge $b_3= b_4= b_5=1$, as before.
Also, we will largely ignore $S_b$ in the following since the rigid limit
in which we are interested in is the intersection of $S_0$ and $S_1$.
\par
We now indicate the location of the singularity on the CY
for $S_1$ and $S_0$. Taking into account the
identifications~(\ref{equivalentie2}), we find that $x_3=x_4=x_5$,
and up to rescalings according to the weights in \eqn{genCY}we have:
\begin{eqnarray}
S_1-S_1 \cap S_0:&& (x_1^8,x_2^8,x_3)\sim(1,1,\psi_0)\nonumber\\
S_0-S_1 \cap S_0:&& (x_1^8,x_2^8,x_3)\sim(1,0,0) \mbox{ if }b_1=0\nonumber\\
                 && (x_1^8,x_2^8,x_3)\sim(0,1,0) \mbox{ if }b_2=0\nonumber\\
S_1 \cap S_0:&& \mbox{ as  $S_0-S_1 \cap S_0$, with in addition } \nonumber\\
        && (x_1^8,x_2^8,x_3)\sim(\zeta,\zeta ^{-1},\psi_0)
        \mbox{ if }b_1=0=b_2\ .
\label{sing_pts_ex1}
\end{eqnarray}
\paragraph{Rigid limit.}
For $b_1=0= b_2$ the CY becomes singular along a whole $\IP_1$,
(with inhomogeneous coordinate $\zeta $), that
is the base space used for the fibration of the CY.
All singular points in
\eqn{sing_pts_ex1} can be parametrized as
$(\zeta,\zeta ^{-1},\psi_0)$, where $\zeta =\pm 1$ correspond
to the generic point on $S_1$, and $\zeta =0$ and $\zeta =\infty$
similarly correspond to $S_0$.
The rigid-limit point of interest is most easily described by using
$b_1=0=b_2$. The vicinity of this point can be explored by putting
$b_1=b_2=B$, and we will use this simplification in all of what follows%
\footnote{\label{b1=0}The rigid limit can also be explored by starting from
the values $b_1=1, b_2=0$, in which case
the singularity expressed in the $x$-coordinates is given by
$(x_1,x_2,x_3,x_4,x_5)=(0,1,0,0,0)$.
In this case one may fix $x_2=1$ as a gauge choice.
The region around the singular point on the CY is parametrized to first order
by a (common) value  of $x_i, i=3,4,5 $ proportional to $x_1$, which
reinstates
the coordinate $\zeta$. To capture all the structure of the rigid limit,
this is not sufficient however: the expansion has to be continued to higher
order.
This makes it less transparent, though equivalent
to the one we give explicitly in the next paragraph, as can be
seen by making a transformation of coordinates to map $b_1=\epsilon$,
$b_2=\epsilon$ to $b_1=1$, $b_2=\epsilon^2$.}.
\remark{ALE expansion.}
To see how the ALE approximation of the $K3$ fibre arises in the vicinity
of the
rigid-limit point in moduli space, we first fix coordinates on the CY manifold.
At $B=0$, the position of the singularity in the $K3$ fibre
is at $(x_0,x_3,x_4,x_5)\sim(1,\psi_0,\psi_0,\psi_0)$.
We choose the gauge $x_0=1$, and will approach the limit by keeping $\psi_0$
fixed while letting $B$ become small.
The following expansion parametrizes the neighbourhood of the singular point
in moduli space:
\begin{equation}
B = 2 \epsilon~, \hskip 1cm
\psi_s + \psi_0^4  = 2 \epsilon\tilde u~,
\label{psp02u}
\end{equation}
where $\tilde u$, defined in \eqn{deftildepsiu},
is kept fixed\footnote{The coordinate $\tilde u$ may be
introduced on the grounds that the ratio $(\psi_s + \psi_0^4)/B$
is a gauge-invariant parameter of the moduli space
(in the sense discussed in section~\ref{ss:descrCY2}) and left unspecified
for the intersection $S_0 \cap S_1$.
Alternatively, one may study the resolution of the singular point
in moduli space by blowup as in \cite{KKLMV,Klemmreview}.}.
Then the powers of $\epsilon$ in the expansion of the other variables
around their critical values can be
determined by examining explicitly the limiting form of the defining polynomial
\eqn{K31111}, and adjust\footnote{Some
care is needed in this procedure.
In principle, one should show that the omitted terms disappear {by
changes of coordinates} which are locally well defined.
For example, for $W=x^2 +x^3 +y^2$: the $x^3$ term can be omitted because
of the coordinate change $x'=x(1+x)^{1/2}$, which is well defined around $x=0$.
Consider, however, $W=x^3+x^2 +2 x y + y^2$.
Deleting the $x^3$ on the grounds that the `more relevant'
$x^2$ is present would also be too naive: after the change of
coordinates $y'=y+x$, the polynomial is $W=x^3 +y'^2$,
which shows that the $x^3$ term {\it is\/} relevant.}
the powers of $\epsilon$ so that as many terms
as possible are contributing in leading order in $\epsilon$.
The expansion that turns out to preserve the
relevant structure of the fibre in the neighbourhood of its singular
point is
\begin{eqnarray}
x_3&=&\psi_0+\sqrt{\epsilon}\left( \sqrt{\frac{1}{3}} y_1 -\sqrt{\frac{1}{6}} y_2
\right) \nonumber\\
x_4&=& \psi_0+\sqrt{\epsilon}\left( \sqrt{\frac{1}{3}} y_1 +\frac{1}{2\sqrt{6}} y_2
-\frac{1}{2\sqrt{2}} y_3\right) \nonumber\\
x_5&=& \psi_0+\sqrt{\epsilon}\left( \sqrt{\frac{1}{3}} y_1
           +\frac{1}{2\sqrt{6}}y_2
+\frac{1}{2\sqrt{2}} y_3\right)\ ,
\label{x345y123}
\end{eqnarray}
in the gauge $x_0=1$,
and the ALE limit of the fibre is given by
\begin{equation}
W^{(1)}_{ALE}=\frac{1}{2}\epsilon\left[\frac{1}{2}(\zeta+\frac{1}{\zeta}) +
y_1^2+y_2^2+y_3^2-\tilde u \right]\ ,
\label{A_1-blowup2}
\end{equation}
which shows the $A_1$ singularity structure.
\subsubsection{Singularities of  $X^*_{24}[1,1,2,8,12]$}
The double fibration can be used to derive the
singularity structure of the 2-modulus Calabi--Yau
manifold $X^*_{24}[1,1,2,8,12]$ in a simple way.

\paragraph{Torus singularities.}
Working out the conditions for
the simultaneous vanishing of derivatives of \eqn{WX6123} we find:
\begin{itemize}
\item[a)] $B^{\prime\prime} = \psi_0^6$
\item[b)] $B^{\prime\prime} = 0$
\end{itemize}

\paragraph{$K3$ singularities.}
Now we look at the $K3$ as an elliptic fibration with the above torus as fibre.
The fibre is singular on the $\xi$ base space at the following
points in the $K3$ base:
\begin{itemize}
\item[a)] $\xi_{\alpha\pm}= (\psi_1 + \psi_0^6) \pm \sqrt{(\psi_1 + \psi_0^6)^2
-B'}$.
The subscript $+$ is for the solution outside $|\xi|=\sqrt{|B^\prime|}$,
$-$ for the one inside this circle.
\item[b)] $\xi_{\beta\pm}=\psi_1 \pm \sqrt{\psi_1^2 -B'}$.
\end{itemize}

Now the total $K3$ space is singular if in addition one has
$\frac{\partial B^{\prime \prime}}{\partial\xi}=0$.
This implies $\xi^2=B'$ and
colliding singularities, i.e.\
\begin{itemize}      \label{it:singK3}
\item[a1)] $B'=(\psi_1+\psi_0^6)^2$ and thus
$\xi_{\alpha+} = \xi_{\alpha-}= \psi_1+\psi_0^6$.
\item[a2)] $B'=\psi_1^2$ and thus
$\xi_{\beta+} = \xi_{\beta-} = \psi_1$.
\item[b)] $B'=0$ and thus $\xi_{\alpha -}=\xi_{\beta -}$.
\end{itemize}
The cases a1) and a2) are identified by the gauge transformation
\eqn{discreteGT}.
It will turn out that the monodromy matrices around this singularity are
diagonalizable, and therefore we will refer to these as `conifold
points'. Those around b) are not diagonalizable;
they are referred to as as the `large complex structure limit'.

Case a) generically exhibits an  $A_1$ singularity.
At the points where the manifold (both the $K3$ and the CY, in fact)
has an enhanced symmetry however (see section~\ref{ex2:CY2K3}),
the character of the singularity changes. In the gauge we use,
these points correspond to the intersections of a1) and a2).
The change in the two cases is as follows:
\begin{itemize}
\item[a,i)]: $\nu_1=0$: $\psi_0=0$ and $B'=\psi_1^2$ or $\xi_{\alpha+}=
\xi_{\alpha -}= \xi_{\beta +}=\xi_{\beta -}$. This is an
$A_2$ singularity. The singularity on the $K3$ is a single point.
\item[a,ii)]:$\nu_2=0$: $\psi_0^6=-2\psi_1$ and $B'=\psi_1^2$ or
$\xi_{\alpha+}=
\xi_{\alpha -}= -\xi_{\beta +}=-\xi_{\beta -}$. This is an
$A_1\times A_1$ singularity. There are two singular points on the $K3$
manifold.
\end{itemize}

\paragraph{CY singularities.}

Let us now consider the Calabi--Yau 3-fold as a $K3$-fibration with the
above
$K3$ as typical fibre. The above analysis says that the fibre has a
singularity on $\zeta$-base space when $B' = \ft12(\zeta + 1/\zeta)-\psi_s$
satisfies a1) or a2) or b).
The total space (i.e.\ the CY 3-fold) is singular when
moreover $\frac{\partial B^{\prime}}{\partial\zeta}=0$,
which implies $\zeta=\pm 1$ or $B=0$. In the latter case,
it is not difficult to see that the CY is singular without any further
conditions. Summarizing, we obtain%
\footnote{If one does not fix $b_1=b_2=B$ from the start,
the right-hand side in these is, in fact, $\sqrt{b_1 b_2}$,
hence the signs. See also, {\it mutatis mutandis},
footnote~\ref{b1=0}. }
\begin{eqnarray}
S_{a1}^{\pm}:  (\psi_0^6+\psi_1)^2 + \psi_s & = & \pm  B \nonumber\\
S_{a2}^{\pm}:  \psi_1^2 + \psi_s & = & \pm  B  \nonumber\\
S_b^{\pm}:  \psi_s & = & \pm  B  \nonumber\\
S_0 : B & =& 0\ .
\label{sing}
\end{eqnarray}
\paragraph{Rigid limit(s).}
Generically,  the singularities on the CY are isolated,
i.e.\ they occur for a
discrete set of values (usually a single one) of $\zeta$ and $x_i$.
If $B = 0$, however, $W^{(2)}$ becomes
independent of $\zeta$ and the singularity on the CY
becomes a $\IP^1$, parametrized by $\zeta$
(the point at infinity corresponding to $x_2=0$):
if the fibre is singular, it is singular
over the {\it whole\/} base space of the fibration.
The intersection points of $S_0$ with $S_{a1}$ and $S_{a2}$ correspond
to rigid limits. We will treat in detail the point with the
$A_2$ singularity, but will also comment on the $A_1 \times A_1$ point.
\remark{ALE expansion.}
In the neighbourhood of the singular point one may \cite{KLMVW} approximate
the $K3$ fibre by an ALE space. We first fix coordinates on the CY manifold.
At the $A_2$ point in moduli space
\begin{equation}
A_2\mbox{ point: }B=0 \ ,\ \psi_0=0 \ ,\ \psi_s=-\psi_1^2\ ,
\label{A2point}
\end{equation}
the position of the singularity in the $K3$ fibre
is at $(x_0,x_3,x_4,x_5)\sim(1,(-\psi_s)^{1/12},0,0)$.
We choose the gauge $x_0=1$, and approach the limit by keeping $\psi_s$ fixed.
The deviations from the singular point in the moduli space can be
parametrized as
\begin{equation}
B= 2\epsilon\ ;\qquad \psi_0^6  \psi_1= \epsilon\tilde u_0^6  \ ;
\qquad \psi_1^2 + \psi_s = \epsilon\tilde u_1\ .\label{epsilonex2}
\end{equation}
where $\tilde u_i$ are kept fixed.
Then the powers of $\epsilon$ in the expansion of the other variables
around their critical values can be
determined by examining explicitly the limiting form of the defining polynomial
$W^{(2)}$, and adjusting the powers of $\epsilon$ so that as many terms
as possible contribute in leading order in $\epsilon$.
The relevant structure in the fibre is preserved by setting
\begin{eqnarray}
 x_3&=&(\psi_1)^{1/6}
 \left(1+\left( \frac{\epsilon}{6} \right)^{1/2}\frac{y_3}{\psi_1}\right)
 \nonumber\\
 x_4&=& (\epsilon)^{1/3} (y_4+\ft12\tilde u_0^2)\nonumber\\
x_5&=&\sqrt{\epsilon} \left(y_5 + \tilde u_0 (y_4+\ft12\tilde u_0^2)\right)\ .
\label{expand}
\end{eqnarray}
The relevant terms in the expansion of $W^{(2)}$ are
\begin{equation}
W^{(2)}_{ALE}=\epsilon
\left[
 \frac{1}{12}(\zeta+\frac{1}{\zeta}) +
 \frac{ y_3^2}{2}+\frac{y_5^2}{2}+
 \frac{y_4^3}{3}-\frac{\tilde u_0^4}{4} y_4 -\frac{\tilde u_1+\tilde u_0^6}{12}
\right]\ ,
\label{A_2-blowup}
\end{equation}
which shows the $A_2$ singularity structure.
\section{Periods, monodromies and intersection matrix \\ in the first example}
\label{ss:pmiex1}

In this section we consider $X^*_8[1,1,2,2,2]$.
We compute periods and intersection matrices
in an integral basis, constructing the results on the
CY manifold in terms of the corresponding ones on the $K3$ fibre.
We will thus have first to determine an integral basis of (transcendental)
$K3$ periods. We will, as done in \cite{candmirror1}
for the CY (see appendix~\ref{ss:z8c}), compute the {\em fundamental\/}
period by explicit integration over an integer homology cycle and
obtain the other periods by analytic continuation.
We also use explicit solutions of the
Picard--Fuchs equations, which by itself, while providing
exact expressions, does not lead to an integer basis.
The information about the fundamental cycle and on the
PF solutions  together allow us
to determine a basis of periods for which the monodromies and
the intersection matrix are integer-valued.

We will then construct
a basis of CY 3-cycles by fibring the $K3$ 2-cycles (associated to the above
periods) as indicated in section~\ref{ss:K3fperiods}.
We want to determine the structure of the
monodromies, which are the essential ingredients for the rigid limit, and
in addition to obtain exact expressions of the periods that could be used
in the future to analyse the expansion of the full supergravity
action.

The $K3$ fibration of this CY manifold was described in detail in
section~\ref{ss:ALEfromCY} and we refer the reader to that section for the
relevant definitions and formulae.
Here we just recall that the potential
of this manifold can be written as the potential for a $K3$ manifold
$X^*_4[1,1,1,1]$:
\begin{equation}
\label{fredag1}
W^{(1)} = {1\over 4} \left(B' x_0^4 + x_3^4 + x_4^4 + x_5^4\right) -
\psi_0 x_0 x_3 x_4 x_5~.
\end{equation}
where the complex structure modulus $z= -B'/ \psi_0^4$, see \eqn{defz},
depends on the coordinate $\zeta$ of the $\IP^1$ base, see
\eqn{B'}.

Let us now study cycles and periods of the $K3$ fibre
itself.
\subsection{Cycles and periods of the $K3$ fibre}
As discussed in \cite{Aspinwall} (and see \eqn{algcomstruc}), the
one-dimensional moduli space of complex structures is
$\Gamma_D\backslash O(2,1)/O(2)$. The $O(2,1)$ vectors
(generalized `Calabi--Visentini coordinates') are the periods
$\vartheta_I(z)$ ($I=0,1,2$) of the (2,0) form computed on an integer basis of
transcendental cycles.  They are subject to the quadratic constraint
(\ref{interfono}),
involving the inverse of the intersection matrix ${\cal I}_{IJ}$ of the
transcendental cycles $c_I$, which has signature $(2,1)$.
The discrete group $\Gamma_D$, the invariance group of the lattice of
transcendental cycles, is generated by the monodromies of the periods around
the singular points in the $z$-plane.

Now we explicitly construct a basis of periods
$\hat\vartheta_I(z)$ with integer-valued monodromies and intersection matrix
which therefore admit the above interpretation as periods
over an integer basis of the transcendental lattice.
\paragraph{The fundamental period.}
We present the construction of the fundamental period of the $K3$ fibre
following closely the steps in  \cite{candmirror1} where the corresponding
period of the quintic CY manifold was defined and determined.
One starts considering an integer homology cycle
for large complex structure $\psi_0$
\begin{eqnarray}
\label{fredag6}
b_0&=& \left\{ (x_0,x_3,x_4,x_5) | x_5 =
\mbox{const.}, |x_0|=|x_3|=\delta, \right. \nonumber\\
&& \left. \mbox{$x_4$ the solution to $W^{(1)}(x)=0$ that tends
to $0$ as $\psi_0 \to \infty$} \right\}/|G'| ~.
\end{eqnarray}
We do not restrict the range of $x_i$, and therefore factor out
$|G'|=16$, the number of elements in the
discrete identification group $G'$, as they act non-trivially
on $b_0$. We thus obtain an elementary cycle.

The holomorphic 2-form $\Omega^{(2,0)}$ of $X^*_4[1,1,1,1]$ can
be expressed via the Griffith map (\ref{griffiths30}). We use here a
rescaled 2-form
\begin{equation}
\label{20form}
\hat\Omega^{(2,0)} =\psi_0 \,\Omega^{(2,0)} =
\frac{|G'|\,\psi_0}{(2\pi\ii)^3} \int {\omega_{K3}\over W^{(1)}} \ ,
\end{equation}
where $\omega_{K3}$ is given in \eqn{omegaCYK3}. This choice
ensures that $\lim_{z \to 0} \hat \vartheta_0(z) = 1$.
Then the fundamental period
$\hat\vartheta_0$ is (see \cite{candmirror1}, section~6)
\begin{eqnarray}
\hat\vartheta_0(z)&\equiv &   \int_{b_0} \hat\Omega^{(2,0)}  \label{tirsdag11}
\\
&=&- \frac{\psi_0}{(2 \pi i)^3}
\int_{\gamma_0 \times \gamma_3 \times \gamma_4}
\frac{x_5 \dd x_0 \, \dd x_3 \, \dd x_4}{W^{(1)}}
=- \frac{\psi_0}{(2 \pi i)^4}
\int_{\gamma_0 \times \gamma_3 \times \gamma_4 \times \gamma_5}
\frac{\dd x_0 \,\dd x_3 \,\dd x_4\,\dd x_5}{W^{(1)}}~,
  \nonumber
\end{eqnarray}
where $\gamma_i$ is the loop $|x_i|=\delta$.
The second equality has been obtained by inserting unity in the form
$(1/2\pi i) \int_{\gamma_5}(dx_5/x_5)$ \cite{candmirror1}.
Finally, writing $W^{(1)}= \hat W^{(1)} -\psi_0 x_0 x_3 x_4 x_5$,
expanding in powers
of $\hat W^{(1)}/(\psi_0 x_0 x_3 x_4 x_5)$ and performing the Cauchy
integrations, one obtains for large $\psi_0$, i.e.\ for small\footnote{That
the convergence radius of the
series (\ref{g0-def}) is 1 can be seen using Stirling's formula for
$n \to \infty$.} $z$,
\begin{equation}
\label{g0-def}
\hat\vartheta_0 =
\sum_{n=0}^{\infty} \frac{(4n)!}{n!^4} \left( \frac{1}{4} \right)^{4n}
(-z)^n~, \hskip 2cm \left( |z|<1\right)~.
\end{equation}

Our aim is now to find the desired integral basis of periods
$\hat\vartheta_I(z)$
via analytic continuation to $|z|>1$ of the fundamental one
$\hat\vartheta_0(z)$. In order to achieve this we have found it
convenient to rewrite first the fundamental period in terms of the
solutions of the Picard--Fuchs equation, discussed in detail
in Appendix~\ref{ss:permonK3}. At that stage the analytic properties of
$\hat\vartheta_0(z)$ can be easily
determined since the PF solutions are given by quadratic combinations of
hypergeometric functions whose asymptotic forms are well known.
\paragraph{An integral basis via the Picard--Fuchs equation
and analytic continuation.}
Making reference to the results presented in Appendix~\ref{ss:permonK3},
we start with the generic expression of the periods
$\hat\vartheta$, thought of as solutions of the appropriate PF equation,
\begin{equation}
\label{2add2}
\hat \vartheta(z) = C_{\alpha\beta} U_\alpha(z)\, U_\beta(z)~,\hskip 0.5cm
(\alpha,\beta=1,2)~.
\end{equation}
where $U_{1,2}$ are two linearly independent solutions of the
hypergeometric equation of parameters $\{{1\over 8},{3\over 8},1\}$,
and $C_{\alpha\beta}$ are arbitrary constants.

We now want to express the integer period $\hat\vartheta_0(z)$, defined for
$|z|<1$ in  (\ref{g0-def}), in terms of $U_1$ and $U_2$.
These solutions in the neighbourhood of $z=\infty$ are given by
\begin{eqnarray}
\label{ig0}
U_{1}(z)&=&
         \frac{\Gamma(\frac{1}{8})\Gamma(\frac{5}{8})}{\Gamma(\frac{3}{4})}
        \left(\frac{1}{z}\right)^{\frac{1}{8}}
        F(\frac{1}{8},\frac{1}{8},\frac{3}{4};-\frac{1}{z})~,
\nonumber\\
U_{2}(z)&=&
       \frac{\Gamma(\frac{3}{8})\Gamma(\frac{7}{8})}{\Gamma(\frac{5}{4})}
       \left(\frac{1}{z}\right)^{\frac{3}{8}}
       F(\frac{3}{8},\frac{3}{8},\frac{5}{4};-\frac{1}{z})\ .
\end{eqnarray}
We consider them
for $|z|>1$ and $-\pi-\epsilon<\arg z<\pi-\epsilon$.
Having defined $z^\alpha={\rm e}^{\alpha~\log z}$  we
have chosen the cuts of $\log z$ in
$\arg(z)=\pm\pi-\epsilon$.
Thus  in the fundamental
domain $z\to {\rm e}^{\ii\pi}$ corresponds to $z\to -1^+$
`above the cut', and $z\to {\rm e}^{-\ii\pi}$ to $z\to -1^-$ `below the
cut'.

In order to make contact with $\hat\vartheta_0(z)$ we
analytically continue these solutions
near $z=0$, i.e.\ in the large complex structure limit of the $K3$.
Using standard formulae for $|z|<1$ and $ -\pi-\epsilon<\arg z<\pi-\epsilon$
one has
\begin{eqnarray}
\label{ig1}
U_{1}(z)&=&
\log(\frac{1}{z})~F(\frac{1}{8},\frac{3}{8},1;{-z}) +
      \sum_{n=0}^\infty
      \frac{\left(\frac{1}{8}\right)_n ~ \left(\frac{3}{8}\right)_n }
           {(n!)^2}h_{1,n} (-z)^n~,
\nonumber\\
U_{2}(z)&=&
\log(\frac{1}{z})~F(\frac{1}{8},\frac{3}{8},1;{-z}) +
      \sum_{n=0}^\infty
      \frac{\left(\frac{1}{8}\right)_n ~ \left(\frac{3}{8}\right)_n }
           {(n!)^2}h_{2,n} (-z)^n~,
\end{eqnarray}
where
\begin{eqnarray}
\label{dom4}
h_{1n}&=&2\psi(n+1)-\psi\left(n+\frac{1}{8}\right)
          -\psi\left(n+\frac{3}{8}\right)
          -\pi~\cot\left(\frac{3}{8}\pi\right)~,
\nonumber\\
h_{2n}&=&2\psi(n+1)-\psi\left(n+\frac{1}{8}\right)
         -\psi\left(n+\frac{3}{8}\right)
         -\pi~ \cot\left(\frac{1}{8}\pi\right)~.
\end{eqnarray}
Since the fundamental period has to be quadratic in $U_{1,2}$ and,
as shown in (\ref{g0-def}), it has a regular
behaviour for $z\to 0$, from the expressions in
(\ref{ig1}) we easily conclude that it
must be proportional to $(U_1 - U_2)^2$. More precisely, by
comparing explicitly the
coefficients of the series expansions, we find
\begin{equation}
  \label{tu3}
  \hat\vartheta_0 = {1\over 4\pi^2} (U_1 - U_2)^2 =
  F^2({1\over 8},{3\over 8},1;-z)~.
\end{equation}
At this point the knowledge of the asymptotic forms of the functions $U_i$
for large values of $z$ given in (\ref{ig0})
allows us to obtain
the desired continuation of $\hat\vartheta_0(z)$ in the region $|z|>1$.

Once the expressions in (\ref{ig0}) are introduced in (\ref{tu3})
we find the analytic continuation of the fundamental period in
the form $\sum_{k\in \ZZ_4} z^\frac{k}{4}
R_k(z)$ with $R_k(z)$ regular functions. Due to the presence of the
fourth roots $\hat\vartheta_0(z)$ is no longer single valued.
We choose to have the cut running from $z = -1$ to
$z =\infty$ and define for $k=1,2,3$,
\begin{equation}
\label{gk-def}
\hat\vartheta_k(z) \equiv \hat\vartheta_0(e^{-2 \pi \ii k}z)~,\hskip 0.5cm
|z|>1~, \pi(2k - 1) < {\rm arg} z < \pi (2k+1)~.
\end{equation}
We immediately obtain
\begin{equation}
  \label{tu4}
  \hat\vartheta_1 = {\ii\over 4\pi^2} (U_1 -\ii U_2)^2~,\hskip 0.5cm
  \hat\vartheta_2 = -{1\over 4\pi^2} (U_1 + U_2)^2~,\hskip 0.5cm
  \hat\vartheta_3 = -{\ii\over 4\pi^2} (U_1 +\ii U_2)^2~.
\end{equation}
Being connected by a $\ZZ_4$ transformation, the $\hat\vartheta_k$
satisfy $\sum_{k=0}^3\hat\vartheta_k = 0$, and only three of them are
independent, for instance the first three $\hat\vartheta_I$, $I=0,1,2$.
In the $\hat\vartheta_I$ basis, the $\ZZ_4$ monodromy,
$z\to{\rm e}^{2\pi\ii} z$ for $z$ large, acts thus by definition as
$\hat\vartheta \to {\cal M}_\infty \hat\vartheta$, with
\begin{equation}
\label{Minf}
{\cal M}_\infty= \left(
\begin{array}{ccc}
0 & 1 & 0 \\
0 & 0 & 1 \\
-1 & -1 & -1
\end{array}
\right)\ .
\end{equation}

The analytic continuation of $\hat\vartheta_I(z)$ back to the region $|z|<1$,
$-\pi < \arg z < \pi$, obtained via
 (\ref{ig1}), exhibits $\log$ and $\log^2$ cuts in the $\hat\vartheta_k$
periods with $k\neq 0$ ($\hat\vartheta_0$,
is given by the regular expansion in (\ref{g0-def})).
Under $z\to{\rm e}^{2\pi\ii}z$ with $|z|<1$ we see
from (\ref{ig1}) that $U_\alpha\to U_\alpha - \ii (U_1 - U_2)$.
Via (\ref{tu3},
\ref{tu4}) this induces on the basis $\{\hat\vartheta_I\}$ the monodromy
$\hat\vartheta\to {\cal M}_0\hat\vartheta$ with
\begin{equation}
\label{M0}
{\cal M}_0 = \left(
\begin{array}{ccc}
1 & 0 & 0 \\
-1 & -1 & -1 \\
6 & 4 & 3
\end{array}
\right).
\end{equation}
\begin{figure}
\begin{center}
\null\hskip -1pt
\epsfxsize=5cm
\epsfysize=4.7cm
\epsffile{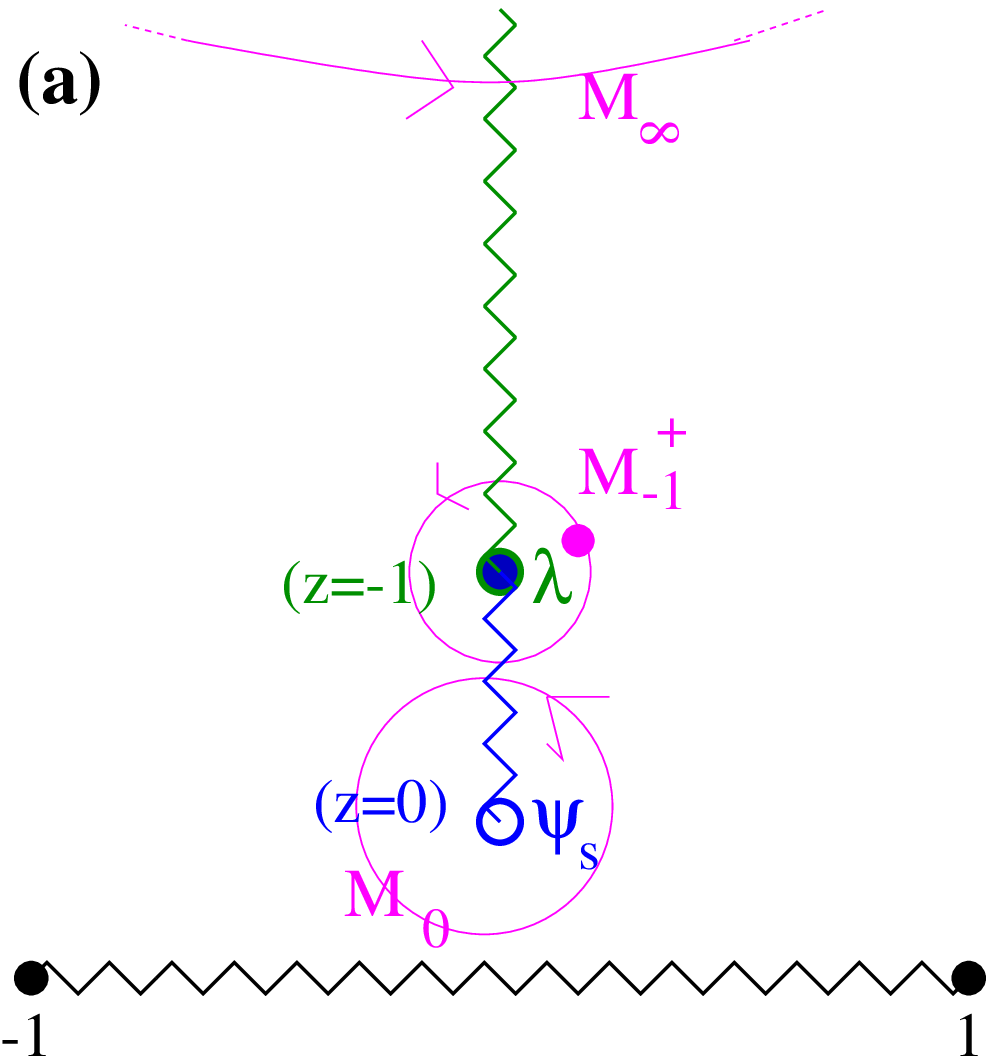}
\end{center}
\vskip -1.1cm
\mycaptionl{ Cuts in the $w=\ft12(\zeta+\zeta^{-1})$ base space;
the cut from $-1$ to 1 is
associated to the `measure' factor $\protect\sqrt{1 - w^2}$; the periods of
the $K3$ fibre, of modulus $z(w)$ have a fourth-root cut from
$z(\tilde\psi_s)\equiv 0$ to $z(\tilde u)\equiv -1$ and a logarithmic one
to $z(\infty)\equiv\infty$; they have non-trivial monodromies
along the indicated paths.}
\label{bigfiga}
\end{figure}
Since two different cuts depart from the conifold point $z=-1$ (see
figure~\ref{bigfiga}), in defining the monodromy at the conifold
we have to specify the position of the basepoint of the small
loop around $z=-1$. In both cases, i.e.\ if the starting point is `above'
or `below' the
cut, such a monodromy is obtained by combining
the monodromies about $z=0$ and $z=\infty$, but the order is
different. If it is above (lower sign below) the cut, we have
$\hat\vartheta\to {\cal M}^\pm_{-1} \hat\vartheta$, with
\begin{equation}
\label{M-1+}
{\cal M}_{-1}^+ = {\cal M}_\infty^{-1} {\cal M}_{0}^{-1} =
\left(
\begin{array}{rrr}
0 & 1 & 0 \\
1 & 0 & 0 \\
-3 & 3 & 1
\end{array}
\right) \ ;\qquad
{\cal M}^-_{-1} = {\cal M}_{0}^{-1}{\cal M}_\infty^{-1}=
\left(
  \begin{array}{rrr}
  -1 & -1 & -1 \\
  6 & 4 & 3 \\
  -6 & -3 & -2
  \end{array}
  \right)\ .
  \end{equation}
The matrices ${\cal M}_{-1}^\pm$ are diagonalisable with eigenvalues $(-1,1,1)$.
\par
The condition (\ref{interfono})  with \eqn{tu3},
\eqn{tu4} determines the inverse intersection matrix ${\cal I}^{IJ}$ up to a
constant. In this basis of integer cycles, the intersection matrix
${\cal I}_{IJ}$ has to be an integer, which implies that it is a multiple of
\begin{equation}
\label{intersectionmatrix1}
{\cal I} =
\left(\matrix{0 & 1 & -2\cr
              1 & 0 & 1\cr
              -2 & 1 & 0}\right)~.
\end{equation}
Around $z=-1$ there is the vanishing cycle (above the cut)
$\nu=\hat \vartheta _0- \hat \vartheta _1$, which should have self-intersection
$-2$, which is the case with \eqn{intersectionmatrix1}. One can then check that
the Picard--Lefshetz formula \eqn{Picard-LefshetzK3CY} leads also to \eqn{M-1+}.
\paragraph{Expansion near the conifold.}
Now we want to obtain the expansion of the $K3$ periods $\hat\vartheta_I$
near the conifold singularity $z \to -1$. In order to do this we need
consider the asymptotic expressions of
the functions $U_{1,2}$ in the vicinity of%
\footnote{Some care is required in using
the analytic continuation formulae because  the range of ${\rm
arg}\,z$  has to be appropriate to $z\to -1^+$ (or to $z\to -1^-$).}
$z =-1^+$ (or $z = -1^-$). This can be obtained in two steps, by first
re-expressing $U_1$ and $U_2$ in terms of the basis
\begin{equation}
\label{extra51}
u_1(z)= 2\pi~F(\frac{1}{8},\frac{3}{8},1;{-z})~,\hskip 0.5cm
u_2(z)= \frac{\Gamma(\frac{1}{8})\Gamma(\frac{3}{8})}{\Gamma(\frac{1}{2})}
F(\frac{1}{8},\frac{3}{8},\frac{1}{2};{1+z})
\end{equation}
by comparing the $U$'s and the $u$'s in the
neighbourhood\footnote{The analytic continuation of $u_2$ reads
$u_{2(0)}(z)=-\log(-z)~F(\frac{1}{8},\frac{3}{8},1;{-z}) +
      \sum_{n=0}^\infty
      \frac{\left(\frac{1}{8}\right)_n ~ \left(\frac{3}{8}\right)_n }
           {(n!)^2}h_{n} (-z)^n $,
for $|z|<1$ and  $-\pi+\epsilon<\arg(-z)<\pi+\epsilon$, i.e.
$\epsilon<\arg(z)<2\pi+\epsilon$,
with $h_{n}=2\psi(n+1)-\psi(n+\frac{1}{8})-\psi(n+\frac{3}{8})$.}
of $z=0$. The result is
\begin{equation}
\label{extra52}
U_1(z)=\sqrt{\frac{2-\sqrt{2}}{2}} e^{\frac{11\pi}{8}i} u_1 +u_2~,
\hskip 0.5cm
U_2(z)=-\sqrt{\frac{2+\sqrt{2}}{2}} e^{\frac{\pi}{8}i} u_1 +u_2 ~.
\end{equation}
The desired continuation to the neighbourhood of $z=-1^+$ is now obtained
by the analytic continuation of $u_1$, valid for $|\!\arg(1+z)|<\pi$,
to this region:
\begin{equation}
\label{extra53}
u_{1(-1)}(z) =
   \frac{1}{\sqrt{2}}~u_{2(-1)}(z)+
   2\pi\frac{\Gamma(-\frac{1}{2})} {\Gamma(\frac{1}{8})\Gamma(\frac{3}{8})}
         ~(1+z)^{\frac{1}{2}} ~
         F(\frac{7}{8},\frac{5}{8},\frac{3}{2};{1+z})~.
\end{equation}
Substituting (\ref{extra53}) into (\ref{extra52})
one may rewrite the resulting expression as
\begin{eqnarray}
\label{ig3}
U_{1}(z)&=&
K_1 \left[ F(\frac{1}{8},\frac{3}{8},\frac{1}{2};{1+z})
    -\ii (\sqrt{2} -1)
   (1+z)^{\frac{1}{2}} ~  F(\frac{7}{8},\frac{5}{8},\frac{3}{2};{1+z})
   \right]~,
\nonumber\\
U_{2}(z)&=&
K_1 \left[ {\rm e}^{-\ii\frac{\pi}{4}} (\sqrt{2} - 1) ~
    F(\frac{1}{8},\frac{3}{8},\frac{1}{2};{1+z}) -
    {\rm e}^{\ii\frac{\pi}{4}}
   (1+z)^{\frac{1}{2}} ~  F(\frac{7}{8},\frac{5}{8},\frac{3}{2};{1+z})
   \right]~,
\end{eqnarray}
with $K_1 = {\rm e}^{-\ii{\pi\over 8}} \sqrt{2 + \sqrt{2}\over 4}
{\Gamma({1\over 8})\Gamma({3\over 8})\over \Gamma({1\over 2})}$.

We have the ingredients to expand the periods $\hat\vartheta_I$
near $z=-1$, but before
doing so we find it useful to define another integral basis,
different from the one in \eqn{tu3} and
\eqn{tu4}, a basis that will enable us to compute  the rigid
limit efficiently. In particular, one of the periods in the new basis will be the
`vanishing period' at the conifold point, which will be fibred on
the base space to give a CY cycle with $S^3$ topology and one with
$S^2\times S^1$, as described in general in section~\ref{ss:stratK3per}.
Ultimately, the periods along these two CY cycles are the ones that
reduce in the rigid limit to the periods of the meromorphic 1-form
of the $SU(2)$ Seiberg--Witten theory, as we will show.

Thus we introduce the periods $\hat\vartheta'_I$ ($I=0,1,2$) related to the
previous ones by the integer change of basis
\begin{equation}
  \label{bas2}
  \hat\vartheta'=F\,\hat\vartheta~,\hskip 0.5cm
  F=\left(\matrix{-1 & 1 & 0 \cr 1 &-2 & -1\cr 1 & 0 & 0}\right)~.
\end{equation}
These periods are distinguished by their behaviour near the
singularities. Indeed, $\hat\vartheta'_0=\hat\vartheta_1-\hat\vartheta_0$
is the vanishing period `above the cut' at the conifold point $z\to -1^+$;
$\hat\vartheta'_2=\hat\vartheta_0$ is the fundamental cycle;
$\hat\vartheta'_1=\hat\vartheta_0 - 2\hat\vartheta_1 - \hat\vartheta_2$
is regular at the conifold and  has no $\log^2$ term in its small-$z$
expansion. This can be seen both by the form of the monodromies
and by explicit expansions.

In the $\hat\vartheta'$ basis the monodromy matrix at the conifold,
${\cal M}_{-1}^+$ and the monodromy at large complex structure,
${\cal M}_0$ become
\begin{equation}
  \label{bas3}
  {\cal M}_{-1}^{+\prime} =
  \left(\matrix{-1 & 0 & 0\cr 0 & 1 & 0\cr 1 & 0 & 1}\right)~,
  \hskip 0.5cm
  {\cal M}_0' =
  \left(\matrix{1 & 1 & -2\cr 0 & 1 & -4\cr 0 & 0 & 1}\right)~.
\end{equation}
This is in agreement with the behaviours described above.
The ${\bf \ZZ}_4$ transformation and the intersection matrix ${\cal I}'$,
(such that $\vartheta_I' {{\cal I}'}^{-1\,IJ} \vartheta_J' = 0$)
become
\begin{equation}
  \label{bas4}
  {\cal M_{\infty}}'=\left(\matrix{-3 & -1 & -2\cr 4 & 1 & 4\cr 1 & 0 & 1}\right)
  ~,\hskip 0.5cm
  {\cal I}' = \left(\matrix{-2 & 0 & 1\cr 0 & 4 & 0\cr 1 & 0 & 0}\right)~.
\end{equation}

\subsection{Cycles and periods of the CY}
\label{cy1cp}
Now we describe the 3-cycles of the CY manifold $X^*_8[1,1,2,2,2]$
by fibring $K3$ 2-cycles over
paths in $\zeta$ in the two ways explained at the end of
section~\ref{ss:K3fperiods}. Correspondingly,
the periods are integrals over such paths of the $K3$ periods. We will
consider periods of the rescaled $(3, 0)$-form
\begin{equation}
\hat\Omega^{(3,0)} \equiv  \psi_0\,\Omega^{(3,0)}=\int_{\gamma_4}\frac{\psi_0|G|}{(2\pi\ii)^4}
\frac{\omega_{CY}}{W^{(1)}}=\hat\Omega^{(2,0)}\frac{d\zeta}{2\pi\ii\,\zeta}\ .
\end{equation}
and will sometimes make use of the variable
\begin{equation}
w=\frac12 \left(\zeta+ \frac{1}{\zeta}\right) \ ,\label{defw}
\end{equation}

\iffigs
\begin{figure}
\begin{center}
\null\hskip -1pt
\epsfxsize=11.5cm
\epsffile{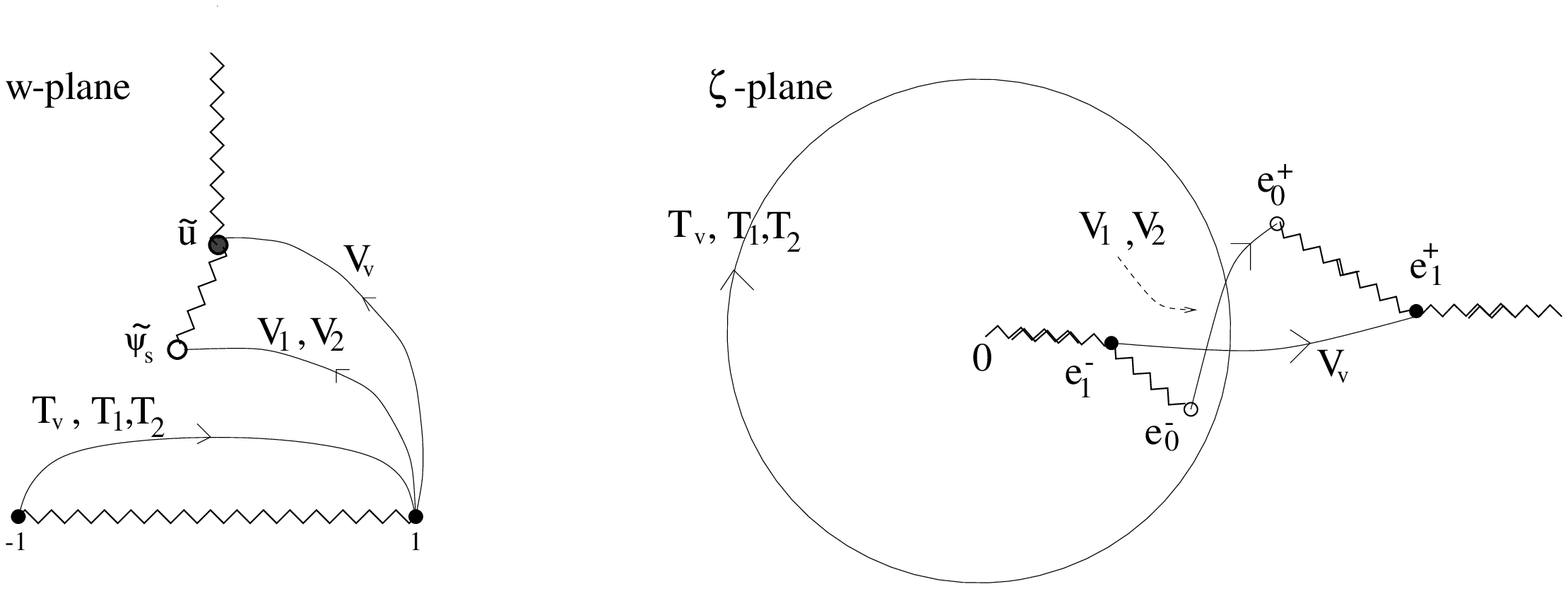}
\vskip -0.1cm
\end{center}
\mycaptionl{An integer basis of cycles, suitable for taking the rigid
limit, is depicted in the base space, parametrized by $\zeta$ or
$w={1\over 2}(\zeta + {1\over\zeta})$.
The cycles $T_v,V_v$ are obtained by fibring over the indicated paths the
$K3$ `vanishing cycle' $c_0'$; the associated periods ${\cal T}_v,{\cal
V}_v$ will reduce, in the rigid limit, to the section of the rigid SU(2)
special geometry, i.e.\ the periods of the meromorphic form $\lambda$. The
cycles $T_1,T_2$ and $V_{t_a},V_{t_b}$ arise by fibring as indicated,
respectively, the cycles $c_1',c_2'$ and $2c_2' - c_1', 4 c_2'$.}
\label{fig:rlbas}
\end{figure}
\fi
First we can transport the three $K3$ cycles over the circle
$|\zeta|=1$. We will use the basis $\hat \vartheta'_I$ and will thus
define
\begin{eqnarray}
{\cal T}_v &\equiv & \frac{1}{2\pi\ii}\int_C\frac{\dd\zeta}{\zeta}\hat
\vartheta'_0=
  {1\over\pi}\int_{-1}^1 {\dd w\over\sqrt{1 - w^2}}\, \hat\vartheta'_0 \nonumber\\
{\cal T}_1 &\equiv & \frac{1}{2\pi\ii}\int_C\frac{\dd\zeta}{\zeta}\hat
\vartheta'_1=
  {1\over\pi}\int_{-1}^1 {\dd w\over\sqrt{1 - w^2}}\, \hat\vartheta'_1 \nonumber\\
{\cal T}_2 &\equiv & \frac{1}{2\pi\ii}\int_C\frac{\dd\zeta}{\zeta}\hat
\vartheta'_2  =
  {1\over\pi}\int_{-1}^1 {\dd w\over\sqrt{1 - w^2}}\, \hat\vartheta'_2 \ .
  \label{nbas2}
\end{eqnarray}
 Then we may transport $K3$ cycles
between points where they vanish (see figure~\ref{fig:rlbas}).
First we consider the $K3$ period
$\hat\vartheta'_0 = \hat\vartheta_1 - \hat\vartheta_0$,
introduced in  (\ref{bas2}) and let $c_0'$ be its corresponding
2-cycle (a 2-sphere). It vanishes at the conifold point $z(w)=-1$, i.e.\
$\zeta=e_1^\pm$ (see \eqn{e01pm} and figure~\ref{bigfiga}), and we can thus consider
its transport between these two points in $\zeta$ or, equivalently,
on a path starting at $w=\tilde u$ and ending again
at the same point (on the other sheet) after having crossed the
square-root cut that runs  from $-1$ to 1.
Clearly this 3-sphere collapses when $\tilde u = 1$.
It is easy to see that the corresponding period is given by
\begin{equation}
  \label{nbas1}
  {\cal V}_v \equiv \frac{1}{2\pi\ii}\int_{e_1^-}^{e_1^+}
\frac{\dd\zeta}{\zeta}\,\hat\vartheta '_0=
  {1\over \pi}\int_{1}^{\tilde u} {\dd w\over\sqrt{1 - w^2}}\, \hat\vartheta'_0~.
\end{equation}
Similarly, we can consider which cycles
vanish at the points $e_0^\pm$. This can be read from the
monodromy matrix ${\cal M}_0'$ in \eqn{bas3}. Indeed, comparing with the
Picard--Lefshetz formula \eqn{Picard-LefshetzK3CY} shows that vanishing cycles are
proportional to $\hat\vartheta '_1$ and $\vartheta '_2$. So we define
\begin{eqnarray}
{\cal V}_{1}&\equiv &\frac{1}{2\pi\ii}\int_{e_0^-}^{e_0^+}
\frac{\dd\zeta}{\zeta}\,\hat\vartheta '_1=
{1\over\pi}\int_{1}^{\tilde\psi_s} {\dd w\over\sqrt{1 - w^2}}\,
  \, \hat\vartheta'_1\nonumber\\
{\cal V}_{2}&\equiv &\frac{1}{2\pi\ii}\int_{e_0^-}^{e_0^+}
\frac{d\zeta}{\zeta}\,  \hat \vartheta '_2 =
{1\over\pi}\int_{1}^{\tilde\psi_s} {\dd w\over\sqrt{1 - w^2}}\,
  \, \hat\vartheta'_2 \ .
\end{eqnarray}
We have thus introduced the basis
\begin{equation}
v=\left\{ {\cal V}_v,{\cal V}_1,{\cal V}_2,{\cal T}_v,{\cal T}_1,{\cal T}_v\right\}\ .
\label{basisvex1}
\end{equation}

\paragraph{Intersection matrix.}

The intersection matrix can easily be obtained from considering the
figure~\ref{fig:rlbas}. We obtain in the basis \eqn{basisvex1}:
\begin{equation}
q=\pmatrix{ &&&&&\cr &
q_{vv} &&&-{\cal I}'&\cr&&&&&\cr &{\cal I}'&&&0&\cr&&&&&}=
\pmatrix{0&0&1&2&0&-1\cr 0&0&-2&0&-4&0\cr -1&2&0&-1&0&0\cr
-2 & 0 & 1&0&0&0\cr 0 & 4 & 0&0&0&0\cr 1 & 0 & 0&0&0&0} \ .
\label{explicitICYex1}
\end{equation}
Only the derivation of the intersection between ${\cal V}_1$ and ${\cal V}_2$ is not
trivial. We will explain in more detail the similar computation in
the second example in section~\ref{CYcycles}. In appendix~\ref{ss:z8c}
we derive cycles in another way, and obtain also an expression for the part
$q_{vv}$ in terms of a monodromy matrix and the intersection matrix of the $K3$
fibre.
\section{Periods, monodromies and intersection matrix \\
in the second example}
\label{ss:pmiex2}

We will work for the second example starting from its formulation as
a double fibration: after the $K3$ fibration, we take the second
torus fibration explained in section~\ref{ss:ex2torusfibr}, and thus
use the polynomial in the form \eqn{torusfib-mod2}.
The holomorphic $3$-form on the Calabi--Yau manifold is then \eqn{griffiths30}
(we take $y_0$ constant)
\begin{eqnarray}
\Omega^{(3,0)} &=& \frac{|G|}{(2\pi\ii)^{4}} \int  \frac{\omega_{CY}}{W^{(2)}}=
   \frac{|G'|}{(2\pi\ii)^{4}} \int  \frac{\omega_{K3}}{W^{(2)}}\frac{\dop\zeta}{\zeta}
= \frac{12}{(2\pi\ii)^{4}} \frac{\dop\zeta}{\zeta}\int \frac{1}{12}
\frac{\dop\xi}{\xi} \dop x^4 \dop x^5 \frac{1}{W^{(2)}y_0^{-6}}
\nonumber\\ &=&
\frac{1}{(2 \pi i)^4} \frac{\dop \zeta}{\zeta} \wedge \frac{\dop\xi}{\xi} \wedge \dop x
\int \dop y  \frac{1}{W^{(2)}y_0^{-6}} =
\frac{1}{(2 \pi i)^3} \frac{\dop \zeta}{\zeta} \wedge \dop x
\wedge \frac{1}{y(\zeta,\xi,x)} \frac{\dop \xi}{\xi}\ , \label{Omega30ex2}
\end{eqnarray}
where we used the form \eqn{torusfib-mod2} for $W^{(2)}$, and as path around the
surface for Griffiths' residue theorem we used a loop in $y$, such
that at the end $y(\zeta,\xi,x)$ is the solution to $W^{(2)}=0$.
Note that $\Omega^{(1,0)} \equiv \frac{1}{2 \pi i} \frac{1}{y}
\frac{\dop \xi}{\xi}$
is the (up to a constant factor unique) holomorphic 1-form
on the torus fibre, and $\Omega^{(2,0)} \equiv \frac{1}{2 \pi i} \dop x
\wedge \Omega^{(1,0)}$
the holomorphic 2-form on the $K3$.

We will explicitly construct a basis of 3-cycles and their corresponding
periods. To achieve this, we exploit the fibration structure of
the model; first we study the cycles and periods of the torus fibre,
then those of the $K3$, and finally those of the Calabi--Yau manifold itself.
The explicit construction of the cycles allows us to compute monodromies
and intersection forms in a straightforward way.
For the torus and $K3$ fibres, we furthermore obtain closed expressions of
the periods in terms of hypergeometric functions.

\subsection{Torus cycles and periods}
\label{torus-section}

\begin{figure}
\begin{center}
\null\hskip 1pt
\epsfxsize 7.8cm
\epsffile{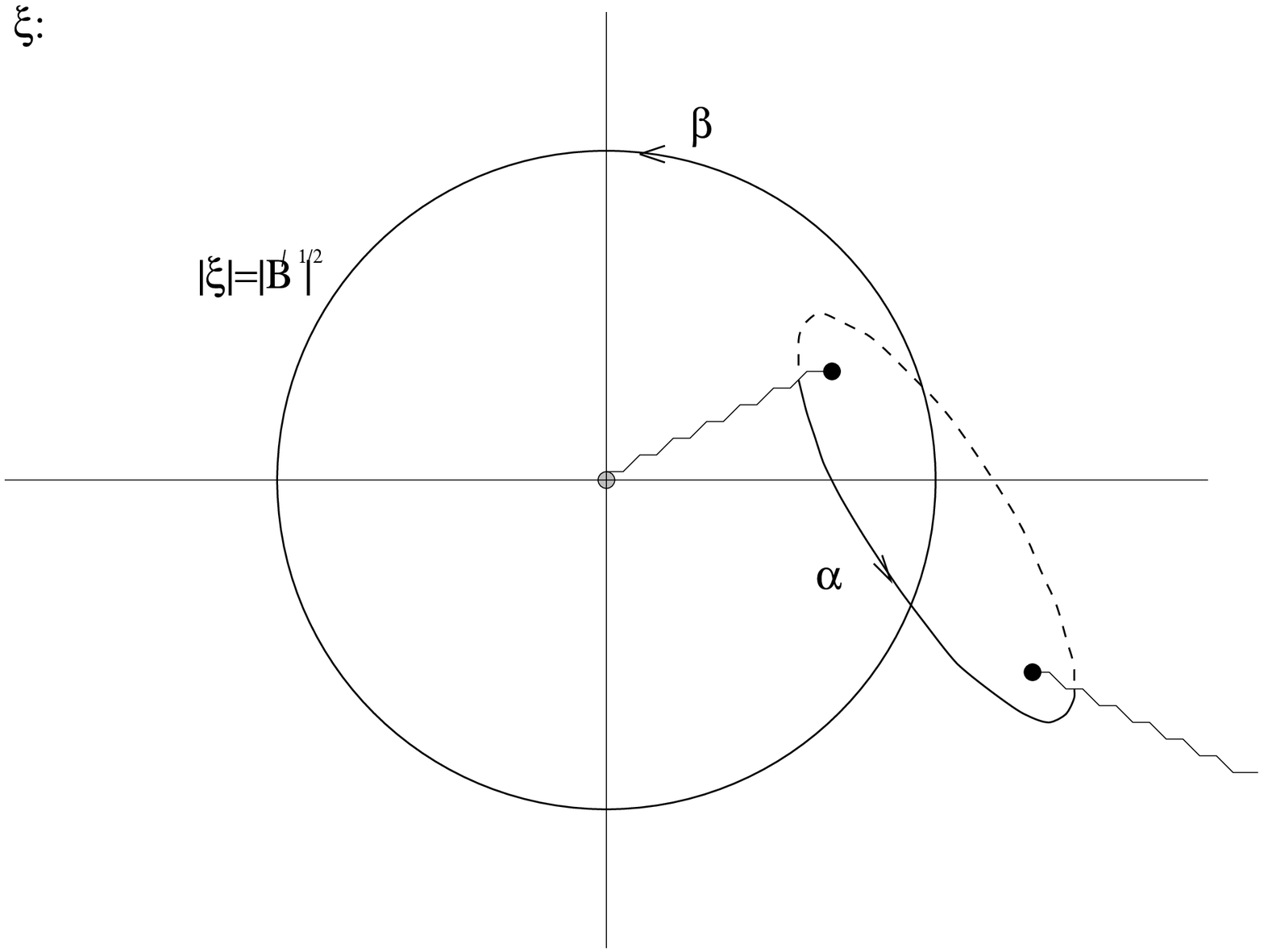}
\epsfxsize 7.8cm
\epsffile{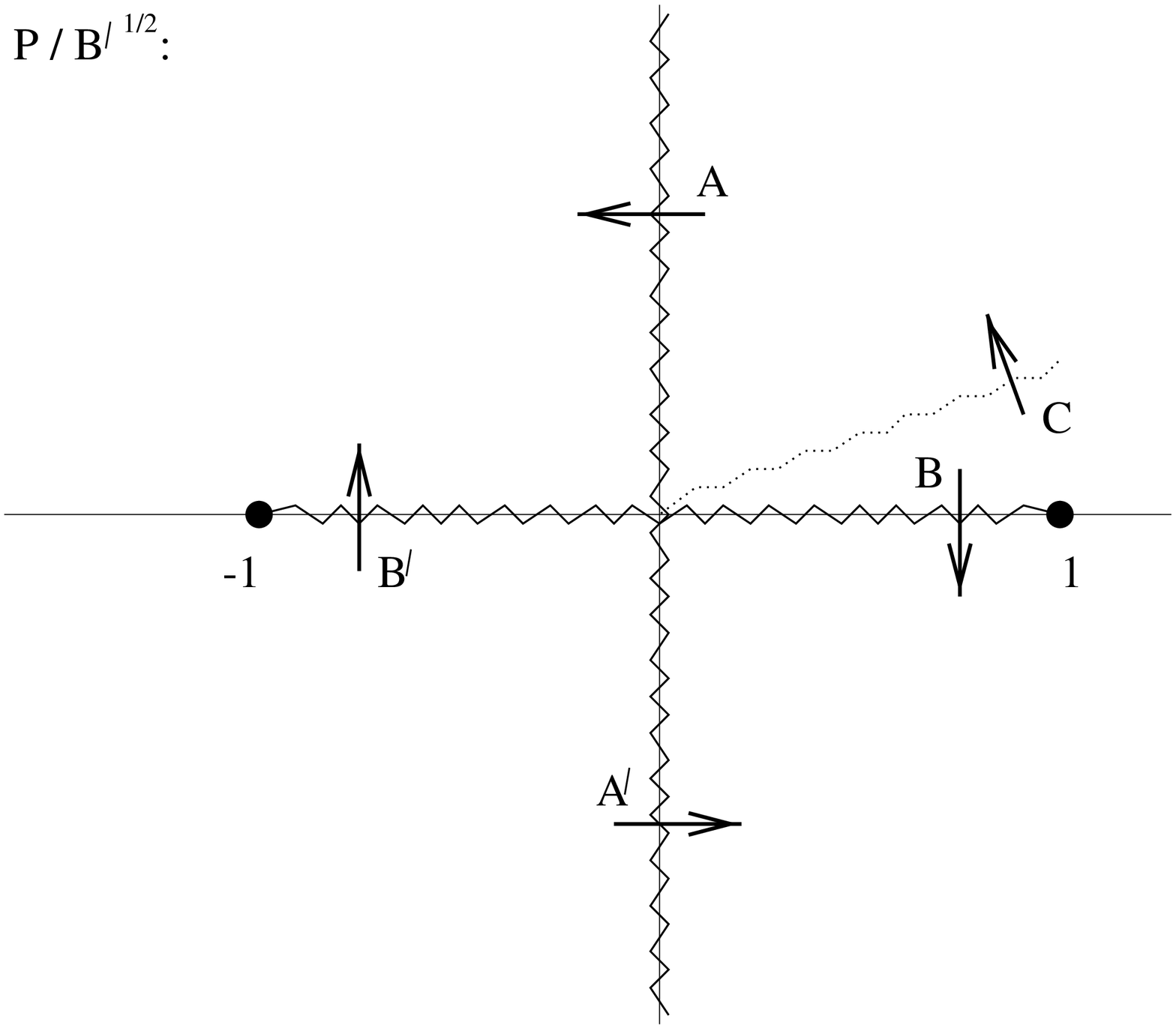}
\end{center}
\mycaptionl{{\bf (a)} Branch points, cuts and cycles of the torus
fibre in the $\xi$-plane. {\bf (b)} Singularities and cuts
in the $P / (B^\prime)^{1/2}$-plane.}
\label{toruscuts}
\end{figure}

\paragraph{Branch points:}

If we consider the defining equation~(\ref{torusfib-mod2})
to be an equation for $y$ as a function of
$\xi$, we get the torus as a 2-sheeted cover of the $\xi$ plane.
The sheets coincide at the branch points, which come in pairs symmetric under
$\xi \leftrightarrow \frac{B^\prime}{\xi}$, and are located at
\begin{equation}
\xi = -P \pm \sqrt{P^2 - B^\prime} \label{bp-pos}
\end{equation}
and at
\begin{equation}
\xi = 0, \infty.
\end{equation}

\paragraph{Cycles.}

Define $\beta$ to be the 1-cycle $|\xi| = |B^\prime|^{1/2}$ passing
counterclockwise through the point%
\footnote{In the following, $\sqrt{z}$ denotes the  square root
of $z$ with positive real part.}
$(\xi=i\sqrt{B^\prime},y=\frac{1}{\sqrt{3}} \sqrt{-P})$,
and $\alpha$ the shortest cycle encircling the pair of branch points
(\ref{bp-pos}), with orientation such that $\alpha \cdot \beta = +1$.
This is shown in figure \ref{toruscuts}{\bf (a)}.

\paragraph{Singularities and vanishing cycles.}

The torus degenerates when two branch points coincide. There are two
possibilities:
\begin{itemize}
\item ${\bf P^2 - B^\prime = 0}$: the branch points (\ref{bp-pos})
coincide and $\alpha$ vanishes. The locus in the $(\zeta,x)$-plane
where this occurs is a {\it genus-5\/} Riemann surface, as can be seen by
substituting the expressions for $P(x)$ and $B^\prime(\zeta)$. We denote
this surface by $\Sigma$. Since $\alpha$ collapses to a point on $\Sigma$,
this surface can be lifted trivially to the full Calabi--Yau. By slight
abuse of notation, we denote the lifted surface by $\Sigma$ as well.
Thus $\Sigma$ is the locus of elliptic fibre singularities
of the Calabi--Yau. We could also view the full 10D spacetime
$M_4 \times {\rm CY}$ as an elliptic fibration. Then the locus of elliptic
fibre singularities gets promoted to a $(5+1)$-dimensional manifold
$M_4 \times \Sigma$.
\item ${\bf B^\prime = 0}$:  one of the branch points (\ref{bp-pos})
coincides with the branch point $\xi = 0$, and $\beta$ vanishes.
The surface where this occurs consists of two copies of the $x$-plane
at fixed $\zeta$ positions, and is denoted by $\Sigma^\prime$.
Again, this surface can be lifted to the full Calabi--Yau or promoted to a
$(5+1)$-dimensional submanifold of spacetime.
\end{itemize}

\paragraph{Cuts.}

There are jumps (or cuts) in the definition of $\alpha$ at values of
$P/\sqrt{B^\prime}$
where there are two homologically different cycles encircling the two
branch points with equal minimal length. This occurs when the branch points
(\ref{bp-pos}) are collinear, i.e.\ when
$\frac{P}{\sqrt{B^\prime}}$ is imaginary (type~A cut).
Jumps in the definition of $\beta$ occur when the branch points lie on
the circle $|\xi| = |B^\prime|^{1/2}$, i.e.\ when
$\frac{P}{\sqrt{B^\prime}} = 0$ lies on the real interval $[-1,1]$ (type~B cut).
There is yet another type of cut, for both $\alpha$ and $\beta$,
namely where our prescription for the location of $\beta$ (and hence
$\alpha$) is ambiguous. For fixed $B^\prime$, this is the case when $P$ is
real and positive (type~C cut).

The cut structure in the $P/\sqrt{B^\prime}$-plane is shown in
figure~\ref{toruscuts}({\it b\/}). For $B^\prime \approx 1$ (which will be of
special interest later), the C cut runs approximately over the positive
real axis.
The transformation rules for continuous transport of torus cycles across the
cuts (yielding the monodromies) are
\begin{eqnarray}
A, A^\prime:&& \alpha \to \alpha + 2 \beta \\
B, B^\prime:&& \beta \to \beta + \alpha \\
C:&& \alpha, \beta \to -\alpha, -\beta
\end{eqnarray}
Notice that the origin of the $P/\sqrt{B^\prime}$-plane is not really singular,
as the monodromy about this point is, in fact, trivial.

\paragraph{Periods.}

The following expressions for the periods can be obtained by direct
integration. Denoting $k_\mp^2=\frac{1}{2} (1 \mp \frac{P}{\sqrt{B^\prime}})$,
we find
\begin{eqnarray}
\int_\alpha \Omega^{(1,0)} &=& \frac{2 \sqrt{6} i }{\pi}
(\sqrt{B^\prime})^{-1/2}
\, \, {\bf K} (k_+^2) \mbox{ for } |k_+^2|<1
\label{alpha-period}
\\
&=&  \pm \frac{2 \sqrt{6}  }{\pi}
(\sqrt{B^\prime})^{-1/2}
\, \, {\bf K} (k_-^2) \mbox{ for } |k_-^2|<1 \mbox{ and } \pm \Im P >0
\end{eqnarray}
where the imaginary part of $P$ denotes a position above or below the C-cut,
respectively,
\begin{equation}
\int_\beta \Omega^{(1,0)} = \frac{\sqrt{6}}{\pi} (\pm \sqrt{B^\prime})^{-1/2}
\, (\pm k_\mp^2)^{-1/2}
\, \, {\bf K} ( k_\mp^{-2} )
\label{beta-period}
\end{equation}
where the expressions are valid when the argument of the $K$-function
is in the unit circle, and for $P$ near the C-cut the square root
in the prefactor is equal to $\pm i (\mp k_\mp^2)^{-1/2}$.
Here ${\bf K}(u)=\frac{\pi}{2}  F(\frac{1}{2},\frac{1}{2},1;u)$ is the
complete elliptic integral of the first kind.
These expressions can be extended to other values of $P/\sqrt{B^\prime}$ by
standard analytic continuation. To obtain the values for the $\alpha $
and $\beta$ periods one has to take into account the changes of
definition of the cycles across the A-, B- and C-cuts.

\subsection{K3 cycles and periods}\label{sec6.2}

As explained earlier, the $K3$ fibre (at a fixed generic value of $\zeta$,
and hence of $\sqrt{B^\prime}$) is itself an elliptic
fibration, with base parametrized by $x$. Accordingly,
the relevant 2-cycles of the $K3$ fibre, that is, those which are in the
transcendental lattice, can be constructed as circle fibrations over
certain paths in the $x$-plane, where the circle is a 1-cycle in the torus
fibre.

\begin{figure}
\begin{center}
\null\hskip 1pt
\epsfxsize 14cm
\epsffile{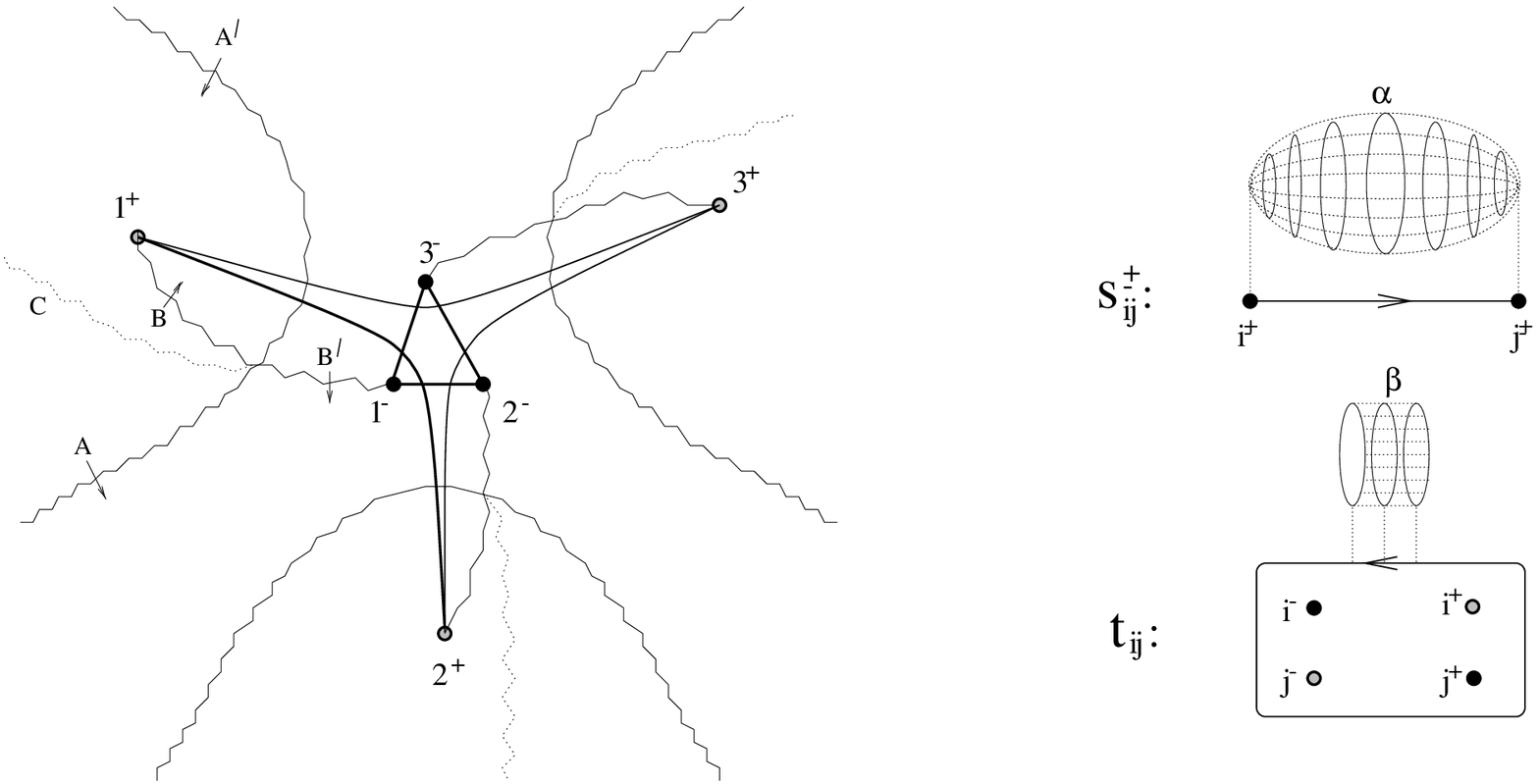}
\end{center}
\mycaptions{(Elliptic) fibre singularities, cuts and cycles of the $K3$ manifold
in the $x$-plane.}
\label{cuts-in-x}
\end{figure}

\paragraph{Points with degenerating elliptic fibre.}

The $x$-plane can be viewed as a 3-fold covering of the
$P/\sqrt{B^\prime}$-plane considered in section~\ref{torus-section}, as $\zeta$
is fixed and $P$ is of degree 3 in $x$. Therefore, in the $x$-plane,
there are three copies of every ingredient (cuts, singularities, etc)
of figure~\ref{toruscuts}({\it b\/}). This is shown in figure~\ref{cuts-in-x}. In
particular,
the $g=5$ Riemann
surface $\Sigma$, on which $\alpha$ vanishes, intersects
the $x$-plane
(having a fixed value of $\zeta$) in $6$ points. As $\Sigma$ splits in
two branches $\Sigma_{\pm}$ corresponding to the solutions of
$P(x)/\sqrt{B^\prime}=\pm 1$, we can divide those six points accordingly
in two groups of three, which we label
by $1^+,2^+,3^+$ and $1^-,2^-,3^-$. Choose numbering such that the B-type
cuts connect $i^+$ with $i^-$ (see the figure).

\paragraph{Cycles.}

The idea is to construct $K3$ 2-cycles as circle fibrations by transporting
a torus 1-cycle $c$ along a path $\gamma$ in the $x$-plane. The path
$\gamma$ can either be a closed loop without monodromy for $c$,
or an open path terminating on
the points $i^{\pm}$, with $c$ vanishing at both endpoints.
The first possibility will produce a torus and the second a sphere.

This gives us the following 2-cycles (see figure~\ref{cuts-in-x}):
\begin{itemize}
\setlength{\itemsep}{-5pt}
\item ${\bf s^+_{ij}}:$ $c=\alpha$ (at $i^+$) and $\gamma$ running between
$i^+$ and $j^+$. This is a sphere.
\item ${\bf s^-_{ij}}:$ $c=\alpha$ (at $i^-$) and $\gamma$ running between
$i^-$ and
$j^-$; again a sphere.
\item ${\bf t_{ij}}:$ $c=\beta$ (at $i^+$) and $\gamma$ a closed path
encircling $i^+, i^-, j^+, j^-$. This is a torus.
\end{itemize}
Note that it is not possible to construct a 2-cycle by taking $\gamma$ to run from
$i^+$ to $j^-$ because such a path necessarily passes through an A-type cut:
an $\alpha $-cycle at one endpoint, when continued to the other
endpoint, becomes $\alpha +2 \beta$ and does not vanish there.

Actually we have not yet given a precise description of the
above cycles in general, only for the specific case of
figure~\ref{cuts-in-x}.
To define this set in the general case, we can
proceed as follows: first,
we require the cycles to be compatible with the above description
including the homological relations and intersections. Any set of cycles
obtained by continuation from the specific set of figure~\ref{cuts-in-x}
will satisfy this. Because continuation is, in general, not uniquely
defined (due to monodromies), there are still many possibilities, corresponding
to the choice of cuts.
We will not try to fix the remaining ambiguity here in general, but assume
that in each case a prescription is adopted such
that $s_{ij}^\pm$ vanishes whenever $i^\pm$ approaches $j^\pm$.

At the level of homology, we have the following relations:
\begin{eqnarray}
t_{ij}=-t_{ji} = s^+_{ij} - s^-_{ij}  & \ ;\qquad &
s^\pm_{ij} = -s^\pm_{ji} \nonumber\\
s^\pm_{12} + s^\pm_{23} + s^\pm_{31} = 0  &  \ ;\qquad &
t_{12} + t_{23} + t_{31} = 0  \ .      \label{t-in-s}
\end{eqnarray}
This implies that of the 2-cycles constructed above, only four are independent.

\paragraph{Intersections.}

Combining the intersections of the paths $\gamma$ in the base
(as shown in figure~\ref{cuts-in-x})
with the known intersections of the torus 1-cycles (taking into account
their transformation when passing a cut), all 2-cycle intersections
can be calculated in a straightforward way:
\begin{equation}
\begin{array}{rclrclrcl}
s^\pm_{12} \cdot s^\pm_{12} &=& -2 &
s^-_{12} \cdot s^+_{12} &=& -2 &
s^\pm_{12} \cdot t_{12} &=& 0 \\
s^\pm_{12} \cdot s^\pm_{23} &=& 1 &
s^-_{12} \cdot s^+_{23} &=& 2 &
s^\pm_{12} \cdot t_{23} &=& 1 \\
s^-_{12} \cdot s^+_{31} &=& 0 &
s^\pm_{12} \cdot t_{31} &=& -1 &
t_{ij} \cdot t_{kl} &=& 0
\end{array}
\label{intrsct}
\end{equation}
and all cyclic permutations hereof. The first two equations can be summarized as
\begin{equation}
s^\pm_{ij} \cdot s^\pm_{kl} = \delta_{il} + \delta_{jk} - \delta_{ik} - \delta_{jl}.
\label{int-corr}
\end{equation}
Note that this is precisely (minus) the intersection of the 0-cycles
$j^\pm - i^\pm$ and
$l^\pm - k^\pm$. This fact will prove to be important for the reduction
of the geometrical
data from the Calabi--Yau to a Riemann surface in the rigid limit.

\paragraph{Singularities and vanishing cycles.}

The $K3$ degenerates when two (or more) of the points $1^\pm,2^\pm,3^\pm$
coincide. There are several possibilities:
\begin{itemize}
\item ${\bf B^\prime = 0}$. Here
$1^+=1^-$,
$2^+=2^-$, $3^+=3^-$, and the tori $t_{ij}$ degenerate to lines (recall that
$\beta$ vanishes at $B^\prime=0$).
\item ${\bf B^\prime = (\psi_0^6 + \psi_1)^2}$.
Keeping in mind that we have defined $\sqrt{B^\prime}$ to have positive real
part, there are two possibilities:
\begin{itemize}
\item if $\real (\psi_0^6 + \psi_1) < 0$, two of
the zeros $1^+,2^+,3^+$ of $P=+\sqrt{B^\prime}$ coincide, and the
corresponding sphere $s_{ij}^+$ vanishes.
\item if $\real (\psi_0^6 + \psi_1) > 0$, two of
the zeros $1^-,2^-,3^-$ of $P=-\sqrt{B^\prime}$
coincide, and the corresponding sphere $s_{ij}^-$ vanishes.
\end{itemize}
In each case, we call the vanishing sphere at this point $v_a$.
\item ${\bf B^\prime = \psi_1^2}$:
Again, there are two cases: if $\real \psi_1$ is positive (negative),
there is a vanishing $s_{ij}^-$ ($s_{ij}^+$). Call this vanishing sphere
$v_b$.
\end{itemize}

Define $t_a \equiv t_{ji}$ if $v_a=s^\pm_{ij}$, and similarly
for $t_b$. The tori $t_a$ and $t_b$ degenerate at $B^\prime = 0$.
The set
\begin{equation}
   c '=\pmatrix{v_a \cr v_b \cr t_a \cr t_b} \label{basisK3cex2}
\end{equation}
forms a basis of the transcendental lattice, which has rank 4 here.

We have constructed different bases depending on the signs of
$\real (\psi_0^6 + \psi_1)$ and $\real \psi_1$. We therefore also expect
the corresponding intersection matrix to depend on these signs.
To calculate this matrix when, say, $\real (\psi_0^6 + \psi_1) > 0$
and $\real \psi_1 >0$, we consider the case where $\psi_1$
is very close to $\psi_1 + \psi_0^6$ (which is the case shown
in figure~\ref{cuts-in-x}). Then it is easy to see, by direct inspection of
the roots of $P(x)^2 - B^\prime$, that, choosing a suitable numbering of the
roots
\begin{equation}
v_a=s^-_{12} \ ;\qquad v_b=s^-_{23} \ ;\qquad  t_a= t_{21} \ ;\qquad
t_b=t_{32}\ . \label{vt-in-st}
\end{equation}
This identification, together with (\ref{intrsct}), provides the complete
intersection matrix. An analogous procedure can be followed for the other cases.

The results are (in the basis $c'$ \eqn{basisK3cex2}):
\begin{itemize}
\item If $\real \psi_1$ and $\real (\psi_1 + \psi_0^6)$ have the same sign:
\begin{equation}
\label{SU3int}
{\cal I} = \left(
\begin{array}{rrrr}
-2 & 1 & 0 & -1 \\
1 & -2 & 1 & 0 \\
0 & 1 & 0 & 0 \\
-1 & 0 & 0 & 0
\end{array} \right)\ .
\end{equation}
We recognize the $SU(3)$ Cartan matrix in the upper block and
therefore call this part of moduli space the `$SU(3)$ sector'.
In terms of the invariants of \eqn{invnu12}, this sector includes
the region where $\nu_1$ is small with respect to  $\nu_2$.

\item If the real parts of $\psi_1 + \psi_0^6$ and $\psi_1$ have opposite sign:
\begin{equation}
{\cal I}^\prime = \left( \begin{array}{rrrr}
-2 & 0 & 0 & -1 \\
0 & -2 & 1 & 0 \\
0 & 1 & 0 & 0 \\
-1 & 0 & 0 & 0
\end{array} \right)\ . \label{SU22int}
\end{equation}
Here we recognize the $SU(2) \times SU(2)$ Cartan matrix; accordingly we
call this part of moduli space the `$SU(2) \times SU(2)$ sector'.
This sector includes the region where $\nu_2$ is small with respect to $\nu_1$.
\end{itemize}

Notice that our division of the moduli space in an $SU(3)$ and
an $SU(2) \times SU(2)$ sector is dependent on the sign convention
for $\sqrt{B^\prime}$ (except on the subspace
$\psi_1^2 = (\psi_1 + \psi_0^6)^2$).
Therefore, though the convention we have taken is quite natural, especially
when we are close to a rigid limit, one can not
expect the boundary between these sectors to have
any physical significance\footnote{A physically significant definition of,
for example, the $SU(3)$ sector would be the region of moduli space where BPS
states exist which can be identified as $SU(3)$ gauge bosons. Unfortunately,
for a generic point of moduli space, the existence of these states is
very difficult to check analytically, if not impossible.}.
However, we shall see that there exists a certain
region inside the $SU(3)$ sector (close to the $SU(3)$ rigid limit) where
a four-dimensional low-energy observer
indeed sees $SU(3)$ Yang--Mills physics (weakly) coupled to gravity, and
similarly for $SU(2) \times SU(2)$. Outside these regions, four-dimensional
low-energy
physics might not look at all like a particular non-Abelian gauge
theory.

\paragraph{Monodromies.}

We consider four monodromies:
\begin{enumerate}
\setlength{\itemsep}{-3pt}
\item The monodromy around $B'=\infty$, $\quad$
$ M_\infty :\  \  B' \rightarrow e^{- 2 \pi i} B'$.
\item The  monodromy around the
conifold point a1 on page~\pageref{it:singK3}, with generator
\newline
$ \quad\quad T_a \quad : \quad \left((\psi_0^6+\psi_1)^2-B' \right) \, \longrightarrow \,
 \exp \left[2 \, \pi \, {\rm i}\right] \left((\psi_0^6+\psi_1)^2-B' \right) $.
\item The  monodromy around the  conifold point a2, with generator
\newline
$ \quad\quad T_b \quad : \quad \left(\psi_1^2 -B' \right) \, \longrightarrow \,
 \exp \left[2 \, \pi \, {\rm i}\right] \left(\psi_1^2 -B' \right)$.
\item The monodromy around the large complex structure
point $B'=0$, with generator $B_0$.
\end{enumerate}
The matrices realizing the monodromies depend on the choice of
basis, which we have taken to depend on the sector:
we will indicate the monodromy matrices corresponding
to the choices in the $SU(2)\times SU(2)$ sector with a prime.

It is clear that these four monodromies are related:
The exact relation depends on the choice of base point. We find
\begin{equation}
M_\infty^{-1} = T_b T_a B_0\ ,\label{Afromothers}
\end{equation}
which can be taken to define  $M_\infty$ in terms of the others.
\par
The results for $T_a$ and $T_b$ can be computed from the Picard--Lefshetz
formula \eqn{Picard-LefshetzK3CY}: the vanishing cycles
are, in both sectors, $v_a$ and $v_b$, respectively.
For the $B_0$ monodromy both $t_a$ and $t_b$
vanish, and the matrix is obtained by drawing some pictures.
Finally the $M_\infty$ matrix is obtained from \eqn{Afromothers}.
Working in the basis \eqn{basisK3cex2}, the results are
\begin{eqnarray}
 T_a = \left(
\begin{array}{rrrr}
-1 & 0 & 0 & 0 \\
1 & 1 & 0 & 0 \\
0 & 0 & 1 & 0 \\
-1 & 0 & 0 & 1
\end{array}
\right)  &\ ;\qquad &
T_b = \left(
\begin{array}{rrrr}
1 & 1 & 0 & 0 \\
0 & -1 & 0 & 0 \\
0 & 1 & 1 & 0 \\
0 & 0 & 0 & 1
\end{array}
\right)  \nonumber\\
T_a^\prime = \left(
\begin{array}{rrrr}
-1 & 0 & 0 & 0 \\
0 & 1 & 0 & 0 \\
0 & 0 & 1 & 0 \\
-1 & 0 & 0 & 1
\end{array}
\right)   &\ ;\qquad &
T_b^\prime = \left(
\begin{array}{rrrr}
1 & 0 & 0 & 0 \\
0 & -1 & 0 & 0 \\
0 & 1 & 1 & 0 \\
0 & 0 & 0 & 1
\end{array}
\right)\nonumber\\
B_0=B_0'= \left(
\begin{array}{rrrr}
1 & 0 & -1 & 0 \\
0 & 1 & 0 & -1 \\
0 & 0 & 1 & 0 \\
0 & 0 & 0 & 1
\end{array}
\right) &\ ;\qquad &
 M_\infty =\pmatrix{-1& 0& 1& 0\cr 0& -1& 0& 1\cr 0& 1& 1& 0\cr -1& -1& 0& 1}
\ . \label{monodromiesK3ex2}
\end{eqnarray}
The matrices $T_a$, $T_a^\prime$, $T_b$ and $T_b^\prime$ have Jordan form
$\mbox{diag}(-1,1,1,1)$, hence an expansion of the periods in a variable
$z$ around the corresponding singularity has terms of the form
$z^n$ and $z^{1/2+n}$. On the other hand, $B_0$ has Jordan form
\begin{equation}
B_{0,Jordan} = \left(
\begin{array}{rrrr}
1 & 1 & 0 & 0 \\
0 & 1 & 0 & 0 \\
0 & 0 & 1 & 1 \\
0 & 0 & 0 & 1
\end{array}  \ ,
\right)
\end{equation}
so period expansions have terms $z^n$ and $z^n \ln z$.

The matrices $T_a$ and $T_b$ generate the group ${\cal S}_3$, the Weyl group of $SU(3)$,
while $T_a^\prime$ and $T_b^\prime$ generate ${\cal S}_2 \times {\cal S}_2$, the
Weyl group of $SU(2) \times SU(2)$.

\paragraph{Periods.}

An integral representation for the $K3$ periods can be given by making
use of the elliptic fibration structure and (\ref{alpha-period})-(\ref{beta-period}).
Note, in particular, that when $i^\pm$ and $j^\pm$ come close to each other,
we have
\begin{equation}
\int_{s_{ij}^\pm} \Omega^{(2,0)} \approx \frac{\sqrt{6}}{2 \pi i}
 (\pm \sqrt{B^\prime})^{-1/2}
 \int_{i^\pm}^{j^\pm} \dop x
= \frac{\sqrt{6}}{2 \pi i} (\pm \sqrt{B^\prime})^{-1/2} (x_{j^\pm} - x_{i^\pm}),
\label{approx-K3per}
\end{equation}
where the $x_{i^\pm}$ are simply found by solving $P(x) = \pm \sqrt{B^\prime}$.

\subsection{Closed expressions}\label{ss:exprex2}
For the discussion of the main features of the rigid limit it is not
really necessary to find closed expressions for the periods. But as
explained in the introduction we want to also provide the explicit
expressions of the basic quantities in the supergravity action for
later use in explicitly constructing the expansion of that action
around its rigid limit. These closed expressions can be found using
first the Picard--Fuchs techniques (see appendix~\ref{ss:permonK3}
for a large part based on \cite{lianyau}) for the periods of
$K3$. These as usual do not provide an integral basis, and thus
we have to find the transformation to the basis constructed above.
This will be done by comparing the monodromy matrices. This will lead
to a basis transformation up to an overall factor. The latter is then
determined by comparing asymptotic expansions in a limit where these
can be calculated easily on both sides. We will do this in a rigid limit.

\paragraph{The Picard--Fuchs result.}
The analysis of the Picard--Fuchs equations for the $K3$-periods
in appendix~\ref{ss:PFex2} leads to the solution \eqn{dirproduc}:
the periods, which are functions of two variables, are
linear combinations of products of two functions of one variable (each).
Choosing a convenient normalization (which removes the scaling introduced
in \eqn{hatvarthetaex2}) we write
\begin{equation}
{ \vartheta} \equiv \int \Omega^{(2,0)}\equiv
\left( \matrix{ \vartheta_{12} \cr \vartheta_{21} \cr
\vartheta_{11} \cr \vartheta_{22} \cr} \right) =\frac{\sqrt{3}}{\psi_0}
\left( \matrix{ \xi_1(r) \, \xi_2(s) \cr
                 \xi_2(r) \, \xi_1(s) \cr
                 \xi_1(r) \, \xi_1(s) \cr
                 \xi_2(r) \, \xi_2(s) \cr
                 } \right)\ ,  \label{vecvartheta}
\end{equation}
where the variables $r,s$ are given in the `usual gauge' as
\begin{eqnarray}
r&=& \frac{1}{2}
+\frac{\sqrt{(\psi_0^6+\psi_1)^2-B'} - \sqrt{\psi_1^2-B'}}{2\psi_0^6} \nonumber\\
s&=& \frac{1}{2}
+\frac{\sqrt{(\psi_0^6+\psi_1)^2-B'} + \sqrt{\psi_1^2-B'}}{2\psi_0^6} \ . \label{finbran}
\end{eqnarray}
and the functions $\xi_{1,2}$ are given for large values of $r$ and
$s$ in \eqn{hygeobase}, where we choose
\begin{equation}
B_1 = \frac{\Gamma(\frac{2}{3})}{\Gamma^2(\frac{5}{6})}\ ;\qquad
B_2 = \frac{\Gamma(-\frac{2}{3})}{\Gamma^2(\frac{1}{6})}\ ;\qquad
B_1B_2=-\frac{\sqrt{3}}{4\pi}\ . \label{B12}
\end{equation}
We will refer to the basis in \eqn{vecvartheta} as our Picard--Fuchs
basis.

\paragraph{Monodromies in the Picard--Fuchs basis.}
We will denote these monodromies with a superscript $PF$.
Using \eqn{finbran} we immediately obtain for the action on
the variables $r$~and~$s$:
\begin{eqnarray}
M_\infty^{PF} & : & \cases{r \, \longrightarrow \, 1-r \cr
s \, \longrightarrow \, 1-s \cr}\nonumber\\
T_a^{PF} & : & \cases{r \, \longrightarrow \, 1-s \cr
s \, \longrightarrow \, 1-r \cr}\nonumber\\
T_b^{PF} & : & \cases{r \, \longrightarrow \, s \cr
s \, \longrightarrow \, r \cr}\nonumber\\
B_0^{PF} & : &  \cases{r \, \longrightarrow \, r \cr
s \, \longrightarrow \, s \ .\cr}
\label{rsaction}
\end{eqnarray}
To determine the action on the functions $\xi_i$ and the periods, we have
to take into account also the corresponding paths in the complex plane.
For example, for $B_0^{PF}$, $r$ turns around the point $r=1$ (and $s$
around $s=1+\frac{\psi_1}{\psi_0^6}$,
which is however immaterial since this is not a singularity).
Note that encircling the singularities
for the monodromies $T_a$ and $T_b$ we may stay in the region where
both $r$ and $s$ are large provided $|\psi_1|>>|\psi_0^6|$,
so that  $\real \psi_1 \cdot \real (\psi_1 + \psi_0^6)>0$.
This is in the $SU(3)$ sector, which is exactly what interests us here.
The expressions  in \eqn{hygeobase} for $\xi_1(u)$ and
$\xi_2(u)$ in terms of hypergeometric functions are
unambiguous for $|u|>1$ provided we specify
the sixth root. We place the cut from 0 to $\infty$
along the negative imaginary axis, so that
$u^{-1/6}$ will mean $|u|^{-1/6} \exp (-i\phi /6)$, where the phase
of $u$ is taken to obey $|\phi|<\pi$.
The monodromy matrices for the $T_a^{PF}$ and $T_b^{PF}$ monodromies
depend to a large extent on this factor.
{}From expression (2.10.6) of \cite{Bateman} one finds that
$\xi_1(u)= u^{-1/6}{}_2F_1(\frac16,\frac16,\frac13;\frac1u)=
 (u-1)^{-1/6}{}_2F_1(\frac16,\frac16,\frac13;\frac1{1-u})$.
Hence, for large values of $u$ with $\Im u>0$ we have
that $\xi_1(1-u) = \exp\left(\frac{i \pi}{6}\right) \xi_1(u)$
and the opposite phase for $\Im u<0$. Similarly,
$\xi_2(1-u) = \exp\left(\frac{\pm i 5\pi}{6}\right) \xi_2(u)$
when $\pm \Im u >0$.
\par
By use of \eqn{rsaction} the monodromy matrices in the basis
\eqn{vecvartheta} are:
\begin{equation}
T_a^{PF}=\left(
\matrix{ 0 & \omega & 0 & 0 \cr
       \omega^2 & 0 & 0 & 0 \cr
              0 & 0 & 1 & 0 \cr
              0 & 0 & 0 & 1 \cr  } \right) \ ;\qquad
T_b^{PF}=\left(
 \matrix{     0 & 1 & 0 & 0 \cr
              1 & 0 & 0 & 0 \cr
              0 & 0 & 1 & 0 \cr
              0 & 0 & 0 & 1 \cr  }
   \right)\ ,
   \label{natbasis}
\end{equation}
where $\omega=\exp \left( \pm 2\pi i/3\right) $ for starting with
$\pm \Im r >0$ (and thus $\mp \Im s>0$).
These matrices have eigenvalues $(1,1,1,-1)$, as we expected for
$A_1$ monodromies.
\par
The  monodromy $M_\infty^{PF}$ is not so easily computed, but requires
the continuation of the hypergeometric functions from large values of
$u$. Indeed, when $B'$ is large, either
$r$ is large and $s$ is close to $\frac{1}{2}$ or vice versa.
It is easier instead to obtain the matrix $B_0^{PF}$:
the monodromy encircles the hypergeometric function singularity at $r=1$
but leaves $\xi_i(s)$ invariant.
In the rest of this paragraph we only write down the formulae for
$\omega=e^{+2\pi\ii/3}$ in \eqn{natbasis}.
Expanding $\xi_1(r)$ and $\xi_2(r)$ near $r=1$
via eqs. 2.10.6 and 2.10.7 of \cite{Bateman}
\begin{eqnarray}
\label{sr4}
\xi_1(r) & =&\frac{1}{2\pi\sqrt{3}} \left[
- \log(r -1)
F({1\over 6},{5\over 6},1;1 -r) + \sum_{n=0}^\infty H_n
(1 -r)^n\right] \nonumber\\
\xi_2(r) & = & -\frac{1}{2\pi\sqrt{3}} \left[- \log(r -1)
F({1\over 6},{5\over 6},1;1 -r) + \sum_{n=0}^\infty G_n
(1 -r)^n\right]~.
\end{eqnarray}
The monodromy matrix follows from this continuation.
\par
The combination that is regular at $r=1$ is the sum. Using the explicit
expressions for the coefficients one finds that
$ H_n-G_n=2\pi\sqrt{3}  {\left({1\over 6}\right)_n \left({5\over 6}\right)_n\over
  n!^2}$,
and therefore
\begin{equation}
\label{sr18}
\xi_1(r) +\xi_2(r) = F({1\over 6},{5\over 6},1;1 - r) \hskip 1cm
\mbox{for $r\sim 1$}\ .
\end{equation}
This leads to the result
\begin{equation}
B_0^{PF}=\pmatrix{
1-\frac{\ii}{\sqrt{3}}&0&0& -\frac{\ii }{\sqrt{3}}\cr
0&1+\frac{\ii}{\sqrt{3}}&\frac{\ii}{\sqrt{3}}&0\cr
0& -\frac{\ii}{\sqrt{3}}& 1-\frac{\ii}{\sqrt{3}}&0\cr
\frac{\ii}{\sqrt{3}}&0&0&1+\frac{\ii}{\sqrt{3}}  }\ .
\end{equation}

\paragraph{Comparing.}
By comparing the monodromy matrices in the Picard--Fuchs basis, and
those in \eqn{monodromiesK3ex2} one obtains the matrix describing the
basis transformation up to an overall
factor. We wrote the expressions in the previous paragraph in the
$SU(3)$ basis, so we have to solve
$\{T_a,T_b,B_0\} S_v =S_v \{T_a^{PF},T_b^{PF},B_0^{PF}\}$. This leads
to
\begin{equation}
\vartheta '=\int_{c'}\Omega^{(2,0)} =  S_v   \vartheta
\ ;\qquad S_v=  a
 \pmatrix{
-\ii\omega & \ii\omega^2 & 0 & 0\cr
-\ii      &\ii          & 0 & 0\cr
\frac{\omega}{\sqrt{3}}    & \frac{\omega^2}{\sqrt{3}}
& \frac{\omega^2}{\sqrt{3}}& \frac{\omega}{\sqrt{3}}\cr
\frac{1}{\sqrt{3}}           &  \frac{1}{\sqrt{3}}
& \frac{1}{\sqrt{3}}    & \frac{1}{\sqrt{3}} } \ ,   \label{experm}
\end{equation}
where $a$ is a number that will be shown to be  1  by comparing the leading
term in the $SU(3)$ rigid limit in section~\ref{ss:compPLb}.

\subsection{CY cycles and periods}
\label{CYcycles}

\begin{figure}
\epsfxsize 7.7cm
\epsffile{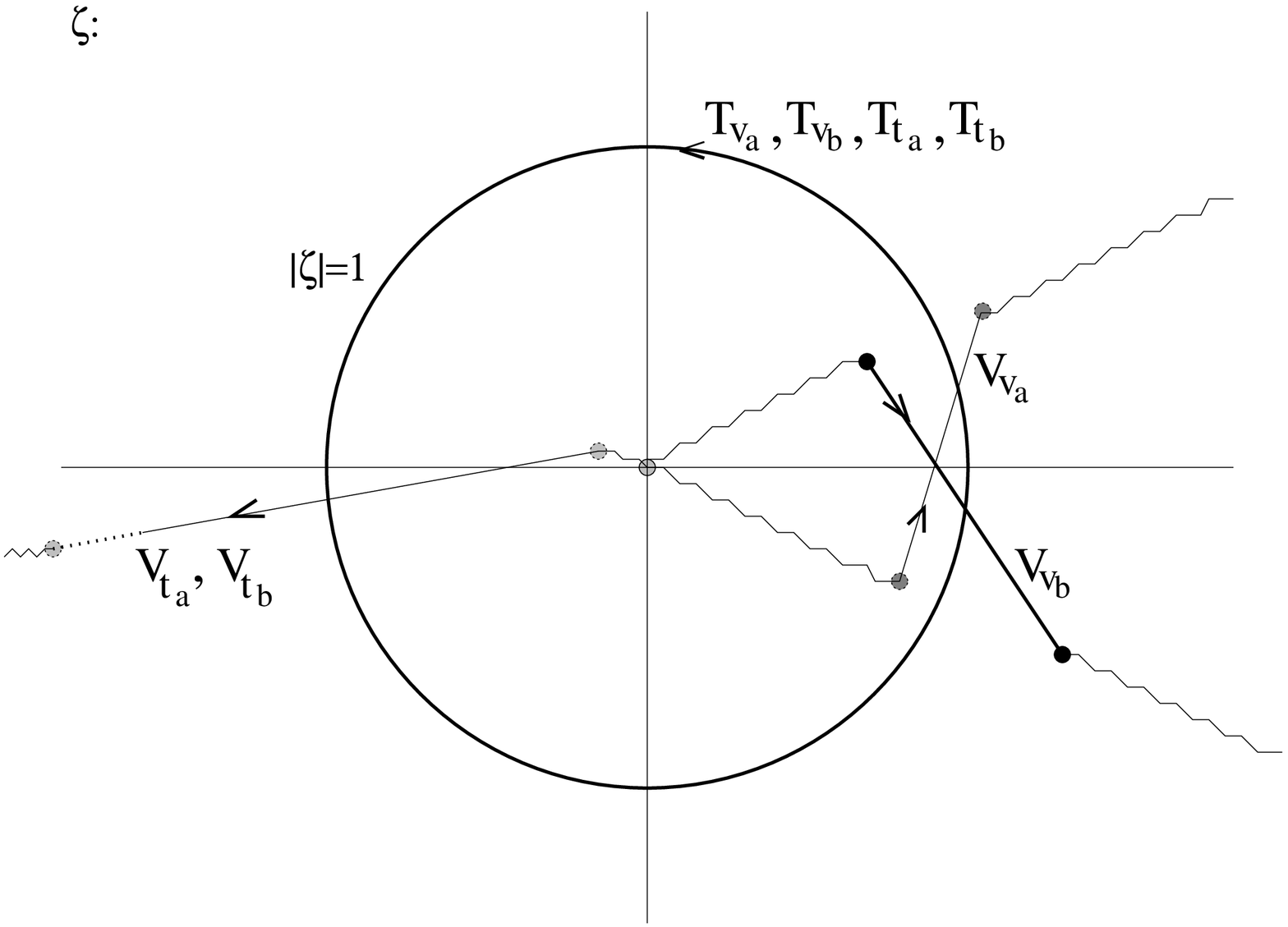}
\null\vskip -5.3cm
\hskip 8cm
\epsfxsize 7.8cm
\epsffile{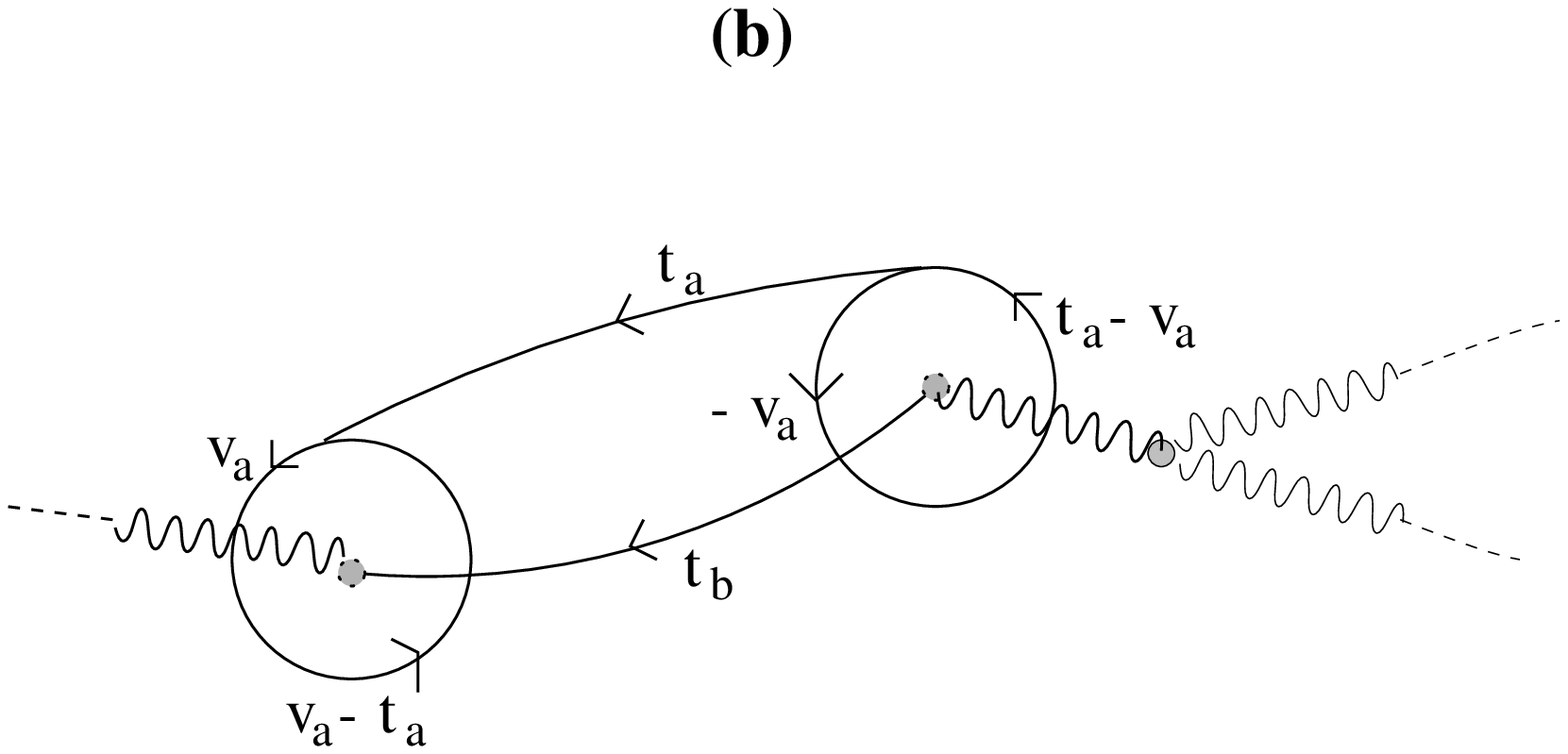}
\vskip 1.3cm
\mycaptionl{{\bf (a)} ($K3$) fibre singularities, cuts and cycles
of the $CY$ manifold in the $\zeta$-plane. {\bf (b)} The intersection
$V_{t_a}\cdot  V_{t_b}$, a detail of figure {\bf (a)}.}
\label{cuts-in-zeta}
\end{figure}

Since the Calabi--Yau 3-fold under consideration is a $K3$ fibration,
one can construct the CY 3-cycles as $K3$ cycle fibrations
over paths in the base manifold (the $\zeta$-plane). Denote the
path in the base by $\gamma$ and the $K3$ cycle which is transported
along $\gamma$ by $c$; $\gamma$ can be open, with $c$
vanishing at both endpoints, or closed, with trivial monodromy for $c$.
The corresponding CY period is given by
\begin{equation}
\frac{1}{2 \pi i} \int_\gamma \frac{\dop \zeta}{\zeta} \theta(\zeta),
\end{equation}
where $\theta(\zeta)$ denotes the period of $c$ in the $K3$ fibre above $\zeta$.

A basis of 3-cycles can be constructed as follows (see fig. \ref{cuts-in-zeta}{\bf (a)}):
\begin{itemize}
\item ${\bf V_{v_a}}$: $c=v_a$ and $\gamma$ running between the two
solutions of $\frac{B}{2} (\zeta + \frac{1}{\zeta}) + 1 = (\psi_0^6 + \psi_1)^2$.
This is an $S^3$.
\item ${\bf V_{v_b}}$: $c=v_b$ and $\gamma$ running between the two
solutions of $\frac{B}{2} (\zeta + \frac{1}{\zeta}) + 1 = \psi_1^2$.
\\ This is also an $S^3$.
\item ${\bf T_{v_a, v_b}}$: $c=v_a, v_b$ and $\gamma$ the unit circle.
This has topology $S^1 \times S^2$.
\item ${\bf V_{t_a, t_b}}$: $c=t_a, t_b$ and $\gamma$ running between the two
solutions of $\frac{B}{2} (\zeta + \frac{1}{\zeta}) + 1 = 0$.
\\ The topology is  $S^2 \times S^1$.
\item ${\bf T_{t_a, t_b}}$: $c=t_a, t_b$ and $\gamma$ the unit circle.
The topology is $T^3$.
\end{itemize}

The intersection matrix in the basis
\begin{equation}
C=\left( V_{v_a}, V_{v_b}, V_{t_a}, V_{t_b} , T_{v_a}, T_{v_b},
T_{t_a}, T_{t_b} \right) \label{basisVT}
\end{equation}
and in the $SU(3)$ region is
\begin{equation}
q=\left(\begin{array}{cc}
\begin{array}{cccc} 0 & -1 & 0 & 0\\ 1 & 0 & 0 & 0 \\ 0 & 0& 0 & -2
\\ 0 & 0 & 2 & 0
\end{array}
&\hspace{4mm} {\cal I}\hspace{4mm}\vspace{5mm}\\[4mm] -{\cal I}& 0
\end{array}\right) \ , \label{intersCYex2}
\end{equation}
where ${\cal I}$ is given in \eqn{SU3int}. The appearance of this
matrix is clear from figure~\ref{cuts-in-zeta}{\bf (a)},
and the $T$ paths in $\zeta$ can be taken to be non-intersecting.
The intersection $V_{t_a}\cdot  V_{t_b}$ can be computed as follows.
Instead of $V_{t_a}$, consider an  8-shaped loop in the $\zeta$-plane
around its end points, and take along the cycle $t_a-v_a$ of the fibre.
After crossing the cut, this combination turns into $-v_a$, as is read
off from the  monodromy matrix $B_0$ in \eqn{monodromiesK3ex2}.
The result is sketched in figure~\ref{cuts-in-zeta}{\bf (b)},
together with $V_{t_b}$.
Shrinking the loops around the branch points to zero one
recovers $V_{t_a}$. The intersection with $V_{t_b}$ can be read off
from the figure before this shrinking.
\section{The rigid limit }   \label{ss:rigidlimitend}
\subsection{General facts and methods}
The singularities of the two CY manifolds we consider,
$X_8^*[1,1,2,2,2]$ and $X_{24}^*[1,1,2,8,12]$, were discussed in
section~\ref{Singuls}, by regarding the CYs as $K3$ fibrations.
The l.c.s. singularity ($B'=0$) of the $K3$ fibre occurs, in both examples, at
two points (symmetric with respect to the exchange $\zeta\to 1/\zeta$) in the base space,
such that $w \equiv 1/2(\zeta + 1/\zeta)= \psi_s/B$. For the first example,
these are the points $e_0^\pm$ in figure~\ref{fig:rlbas};
for the second one, they are the two points to the
left of figure~\ref{cuts-in-zeta}.
In the first example, the fibre $X_4^*[1,1,1,1]$ has, moreover, a conifold
singularity at the points $\zeta=e_1^\pm$, see figure~\ref{fig:rlbas}; in the
second, the fibre $X_{12}^*[1,1,4,6]$ of the second develops two types of
conifolds at two pairs of points, appearing to the right of
figure~\ref{cuts-in-zeta}.

We are now interested in the neighbourhood of those CY singularities,
named `rigid limits' in section~\ref{Singuls}, that are parametrized by
\eqn{psp02u} and \eqn{epsilonex2}, respectively, for the two examples.
In these limits, the position of the l.c.s. in the $K3$ fibre is given by
\begin{equation}
w\equiv \frac12\left(\zeta+\frac{1}{\zeta}\right)=
\tilde\psi_s=\frac{\psi_s}{2\,\epsilon}=-\frac{1}{2\,\tilde\epsilon}\ ,
\label{poslcsp}
\end{equation}
where we have introduced
an expansion parameter $\tilde\epsilon$ invariant under \eqn{projmoduli}
(also for the moduli we have indicated by a tilde the invariant moduli,
i.e.\ $\tilde u$)
\begin{equation}
\epsilon=-\psi_s\tilde\epsilon\ . \label{deftileps}
\end{equation}
Thus in the rigid limit the positions of the l.c.s singularity move to
$\zeta=0$ and $\zeta=\infty$ in the base.

The position of the conifold singularities is instead independent of
$\tilde\epsilon$:
\begin{equation}
\mbox{example 1:}~ w = \tilde u~;
\null\hskip 0.8cm
\mbox{example 2:}~
\begin{array}[c]{l}
 w=\ft12\left(\tilde u_1 + 2\tilde u_0^6\right) +{\cal O}(\epsilon)~,\\
 w =\ft12 \tilde u_1~.
\end{array}
\label{posconsp}
\end{equation}
The $\tilde u$ moduli will be the moduli of the rigid special geometry
that emerges in the limit, that in the examples will be the one associated to
the Seiberg--Witten low-energy effective action for the $SU(2)$, respectively
$SU(3)$, $N=2$, $d=4$ SYM theory.

To deduce the behaviour of the periods in the limit
$\epsilon\to 0$ one can look to the direct evaluation of the period integrals
or one can analyse their monodromies under
$\epsilon\to\rm{e}^{2\pi\ii}\epsilon$.
For example, suppose we have two periods
$v_1$ and $v_2$, which under $\epsilon \to e^{2\ii\pi} \epsilon$ transform
as
\begin{equation}
\left(
\begin{array}{c}
v_1 \\
v_2
\end{array}
\right)
\to
\left(
\begin{array}{rr}
e^{2\ii\pi q} & 0 \\
a e^{2\ii\pi q} & e^{2\ii\pi q}
\end{array}
\right)
\left(
\begin{array}{c}
v_1 \\
v_2
\end{array}
\right)
\end{equation}
with $q>0$. Then \cite{Arnold}
\begin{eqnarray}
v_1(\epsilon) &=& \epsilon^q (a_{-n} \epsilon^{-n} + \ldots + a_0 + a_1 \epsilon + \ldots) \\
v_2(\epsilon) &=& \frac{a}{2 \pi i} \log \epsilon \, \, v_1(\epsilon) +
\epsilon^q (b_{-m} \epsilon^{-m} + \ldots + b_0 + b_1 \epsilon + \ldots)
\end{eqnarray}
where the coefficients $a_i$ and $b_i$ are independent of $\epsilon$. The value of
the integers
$n, m$ can usually be deduced rather easily by inspection of the behaviour of the cycles and
the integrated holomorphic form. For example, when both the cycle and the holomorphic
form (at a generic point of the manifold) stay bounded when $\epsilon \to 0$,
the period has a finite limit \cite{Arnold}, and therefore cannot have poles.

As we will see, such expressions for the periods contain just enough
information to obtain the rigid limit of the \Ka\ potential. With
the exact solutions we only obtain some constants, be able to compare
different bases, and show the compatibility of the two methods.
The extra information of the exact solutions is of course necessary
to calculate gravitational corrections.

\subsection{Rigid limit of the first example}
\subsubsection{Expansion of $K3$ periods}
The CY manifold of the first example has been analysed in section~\ref{ss:pmiex1}
as a $K3$ fibration. The modulus of the $K3$ was $z$, see \eqn{fredag2}, whose
expansion in the rigid limit is
\begin{equation}
  \label{dom2}
  z = -1 + 2\tilde\epsilon(\tilde u - w) + {\cal O}(\tilde\epsilon^2)~.
\end{equation}
The periods of the $(3,0)$-form $\hat \Omega^{(2,0)}$, \eqn{20form},
were determined as integrals over integer cycles in the form $\hat \vartheta _I$
(see \eqn{tu3} and \eqn{tu4} with \eqn{ig3} for their explicit values)
or as $\hat \vartheta' _I$ after the basis transformation \eqn{bas2}.
For this $K3$ manifold
the $\epsilon$-monodromy thus corresponds to the monodromy around
$z=-1$, i.e.\ ${\cal M}_{-1}$ in \eqn{M-1+} or \eqn{bas3}. We can diagonalize
this monodromy, obtaining that the $\epsilon$ dependence of the periods is of the form
\begin{eqnarray}
\label{add4}
\hat \vartheta '_0 & = &
\eta ~ \sqrt{2\tilde\epsilon(\tilde u - w)} + {\cal O}\left(\tilde\epsilon^{3\over 2}\right)~,
\nonumber\\
\hat \vartheta '_1 & = &
k_1 +\ell _1\,\tilde\epsilon(\tilde u - w) + {\cal O}\left(\tilde\epsilon^2\right)~,
\nonumber\\
2\hat \vartheta '_2+\hat \vartheta '_0 =\vartheta_0+
\vartheta _1 & = &
 k_2  + \ell _2\,\tilde\epsilon(\tilde u - w)
+ {\cal O}\left(\tilde\epsilon^{2}\right)~.
\end{eqnarray}
The last two periods have a regular behaviour, i.e.\ starting with a
constant and a $\tilde\epsilon$ term. In the second example the one but leading terms
will be of order $\tilde\epsilon^{2/3}$ such that the $\epsilon$-monodromy
diagonalization will be sufficient to distinguish the constant term from the
one but leading term. Here, however, we need the explicit expansions. From the
explicit expressions, referred to above, we can find that
\begin{eqnarray}
\hat \vartheta '_0 & = &
\eta (1+z)^{\frac{1}{2}}F(\frac{1}{8},\frac{3}{8},\frac{1}{2};{1+z})
~  F(\frac{7}{8},\frac{5}{8},\frac{3}{2};{1+z})
\nonumber\\
\hat \vartheta '_1 & = &
k_1 \left(F^2(\frac{1}{8},\frac{3}{8},\frac{1}{2};{1+z})- (1+z)\,
F^2(\frac{7}{8},\frac{5}{8},\frac{3}{2};{1+z})\right)
\nonumber\\
2\hat \vartheta '_2+\hat \vartheta '_0 & = &
k_2\left(  F^2(\frac{1}{8},\frac{3}{8},\frac{1}{2};{1+z}) +(1+z)\,
F^2(\frac{7}{8},\frac{5}{8},\frac{3}{2};{1+z})\right) \ ,
\end{eqnarray}
 and
\begin{eqnarray}
\eta&=&\frac{1}{\pi^2} (1+\ii)(1-\sqrt{2}) K_1^2= -{1\over 2\pi^2}
{\Gamma^2({1\over 8})\Gamma^2({3\over 8})\over\Gamma^2({1\over
2})}\nonumber\\
k_1&=&{\ii \eta\over\sqrt{2}} \ ;\qquad \ell _1=-\frac{13}{8} k_1  \ ;\qquad
k_2=-{\eta\over 2} \ ;\qquad \ell _2=\frac{19}{8} k_2   \ .
\end{eqnarray}
\subsubsection{Rigid limit of CY periods.}
The CY periods are obtained in section~\ref{cy1cp}
by integrating $K3$ cycles over paths
indicated in figure~\ref{fig:rlbas}. Let us first consider
their $\epsilon$-monodromy. From \eqn{posconsp} and
\eqn{poslcsp} we see that the the conifold point does not change its
position under $\epsilon\rightarrow e^{2\pi\ii}\epsilon$, but the
large complex structure point makes a complete tour. This means that
for the periods ${\cal V}_v$ and the three ${\cal T}_I$ (integrated
over the circle), we just have to consider the $\epsilon$-monodromy
of the $K3$ fibre. On the other hand, the path for ${\cal
V}_{t_a,t_b}$ is deformed in its beginning and end point in the
$\zeta$-plane. In this deformation, the line crosses cuts, and
thus on the new path one has to consider a different cycle of the fibre.
To make this concrete, let us denote the path between $e_0^\pm$ (or between 1
and $\tilde\psi_s$ in $w$-plane) by $L_0$ (with $e_0^+\sim\ft1\epsilon$
now very large in the
same direction, and $e_0^-$ very small), and the line between $e_1^\pm$ (or
between 1 and $\tilde u $ in $w$-plane) by $L_1$. So for a cycle $c$ on the
path $L_0$ we first have to use the $\epsilon$-monodromy of the fibre, i.e.\
${\cal M}^{+\prime}_{-1} c$. Following then the path starting from the point 1,
it now first makes a circle in a clockwise direction for large $\zeta$. This can
be deformed to first turning around the $e_1^+$ point, following the outer half
of the $L_1$ contour first with $K3$ cycle ${\cal M}^{+\prime}_{-1} c$, and
coming back after having crossed the cuts, thus with cycle $c$. Then this cycle
$c$ is making its tour on the $|\zeta|=1$ circle, before finally being
transported on the outer part of $L_0$ to $e_0^+$. On the inner part of the
contour the same deformation takes place. Adding this we obtain that the
$K3$ cycle $c$ on path $L_0$, denoted as $L_0(c)$ has $\epsilon$ monodromy
\begin{equation}
L_0(c) \rightarrow L_1\left({\cal M}^{+\prime}_{-1} c\right) -L_1(c)+2\, T(c)
+L_0(c)\ . \label{ex1L0cmono}
\end{equation}
The difference ${\cal M}^{+\prime}_{-1} c-L_1(c)$ is proportional to the
vanishing cycle in $z=-1$, and thus that part of the CY cycle can be written in
terms of $V_v$. The result in the basis \eqn{basisvex1} is
\begin{equation}
v\rightarrow {\cal M}^\epsilon_{CY}v=\pmatrix{&&&0&0&0 \cr &{\cal M}^{+\prime}_{-1} &
&0&2&0\cr &&&0&0& 2\cr &&&&& \cr &0& && {\cal M}^{+\prime}_{-1} & \cr
&&&&&}v\ .
\end{equation}
This matrix can be brought in Jordan form by defining
\begin{equation}
{\cal T}_2'=2\, {\cal T}_2 + {\cal T}_v\ ;\qquad {\cal V}_2'= 2\,
{\cal V}_2 +{\cal V}_v-{\cal T}_v\ .
\end{equation}
Indeed, reordering to the basis
\begin{equation}
{\cal C}'=\{ {\cal V}_v, {\cal T}_v, {\cal T}_1, {\cal V}_1,
{\cal T}'_2, {\cal V}'_2 \}\ ,
\end{equation}
the $\epsilon$-monodromy matrix is
\begin{equation}
{\cal M}^\epsilon_{CY,J}=\pmatrix{
-1& 0& 0& 0& 0& 0\cr 0& -1& 0& 0& 0& 0\cr  0& 0& 1& 0& 0& 0\cr
0& 0& 2& 1& 0& 0\cr 0& 0& 0& 0& 1& 0\cr 0& 0& 0& 0& 2& 1} \ .
\end{equation}

This allows us to write the following expressions for the periods:
\begin{equation}
\pmatrix{{\cal V}_v\cr {\cal T}_v\cr {\cal T}_1\cr {\cal V}_1\cr
{\cal T}'_2\cr {\cal V}'_2}=\pmatrix
{\tilde\epsilon^{1/2}{\cal A}_1(\tilde\epsilon)  \cr
\tilde\epsilon^{1/2}{\cal A}_2(\tilde\epsilon)  \cr
{\cal A}_3(\tilde\epsilon) \cr
\frac{\log \tilde\epsilon}{\pi\ii}{\cal A}_3(\tilde\epsilon)+{\cal A}_4(\tilde\epsilon) \cr
{\cal A}_5(\tilde\epsilon) \cr
\frac{\log \tilde\epsilon}{\pi\ii}{\cal A}_5(\tilde\epsilon)+{\cal A}_6(\tilde\epsilon)  }
=\pmatrix{
\tilde\epsilon^{1/2}\left( V_1+{\cal O}(\tilde\epsilon)\right) \cr
\tilde\epsilon^{1/2}\left( V_2+{\cal O}(\tilde\epsilon)\right) \cr
-k_1 - \tilde\epsilon\ell _1(\tilde u)+{\cal O}(\tilde\epsilon^2)\cr
-\frac{1}{ \pi \ii}  \log \tilde\epsilon \left( k_1+\tilde\epsilon\ell _1(\tilde u)
\right) + k'_1 +\tilde\epsilon\ell' _1(\tilde u)
+{\cal O}(\tilde\epsilon^2)\cr
-k_2 - \tilde\epsilon\ell _2(\tilde u)+{\cal O}(\tilde\epsilon^2)\cr
-\frac{1}{ \pi \ii}  \log \tilde\epsilon\left(  k_2+\tilde\epsilon\ell _2(\tilde u)
\right) + k'_2 +\tilde\epsilon\ell' _2(\tilde u)
+{\cal O}(\tilde\epsilon^2)}\ ,
\label{limperex1}
\end{equation}
where ${\cal A}_\Lambda(\tilde\epsilon)$ stand for analytic functions of
$\tilde\epsilon$.  It is not immediately clear which of the
coefficients $k,k',\ell ,\ell '$ are functions of $\tilde u$.
However, using the expressions \eqn{add4}, most integrals \eqn{nbas1} and
\eqn{nbas2} over the base space paths
are easy to perform. For example, it is trivial to see that $k_1$ and
$k_2$ in the above expressions are the constants present in \eqn{add4}, and that
$\ell _{1,2}(\tilde u)=\tilde u\ell _{1,2}$.
When gravity is decoupled%
\footnote{This
scaling would also have been obtained if from the start we were to leave the
ratio
between $b_1$ and $b_2$ in \eqn{w11222g} arbitrary, as that would lead to an
arbitrary scale in the $\zeta$-plane, which can be identified
with $\Lambda$ in the Seiberg--Witten curves. },
\begin{equation}
{\rm Tr} \phi^2 \sim u= \Lambda^{2}\tilde u\ ,
\end{equation}
where  $\phi$ is the adjoint scalar and $u$
the modulus of the SW theory.
One immediately obtains
\begin{eqnarray}
  \label{bas17}
 V_1 & = &\eta
  {\sqrt{2}\over \pi} \int_{1}^{\tilde u}
  \dd w\,\sqrt{\tilde u -  w\over 1 - w^2} =  \eta
  {\sqrt{2}\over \pi\Lambda} \int_{\Lambda^2}^u
  \dd t\,\sqrt{u -  t\over \Lambda^4 - t^2} = \eta \, {a_D(u;\Lambda)\over\Lambda}~,\nonumber\\
V_2 & = &\eta
  {\sqrt{2}\over \pi} \int_{-1}^1
  \dd w\,\sqrt{\tilde u -  w\over 1 - w^2} = \eta
  {\sqrt{2}\over \pi\Lambda}\int_{-\Lambda^2}^{\Lambda^2}
  \dd t\,\sqrt{u -  t\over \Lambda^4 - t^2} =\eta \, {a(u;\Lambda)\over \Lambda}~,
\end{eqnarray}
where $a_D(u;\Lambda)$ and $a(u;\Lambda)$ are the $SU(2)$ Seiberg--Witten
periods, in the
form given in the original paper \cite{SeiWit}.
One may check that in the basis
$(a_D,a)$, the monodromy $M_u$ around the
conifold, (\ref{monodrex1CY}), restricted to the periods ${\cal V}_v,{\cal T}_v$,
becomes the SW monodromy around the
massless monopole point $u=\Lambda^2$,
\begin{equation}
  \label{mb2}
  \left(\matrix{1 & 0\cr -2 & 1}\right)~.
\end{equation}

The only integrals which need clarification are ${\cal V}_1$
and ${\cal V}'_2$. We have to evaluate integrals of the form
\begin{equation}
I= \int_{e_0^-}^{e_0^+} \frac{\dd\zeta}{\zeta} f(x)\qquad\mbox{with}\qquad
x=2\tilde\epsilon(\tilde u-w) \ .
\end{equation}
The function $f$ is regular, i.e.\ can be expanded as $f(x)=\sum_{n=0}^{\infty}
f_n x^n$, and we get
\begin{eqnarray}
I&=& -2\log\tilde\epsilon \left(f(0)+2\tilde\epsilon f'(0)\tilde u+{\cal
O}(\tilde\epsilon^2)\right)\nonumber\\ &&
+2 \int_0^1\frac{dx}{x}\left( f(x)-f(0)\right)
+2\tilde\epsilon\tilde u   \int_0^1\frac{dx}{x}\left( f'(x)-f'(0)\right)+{\cal
O}(\tilde\epsilon^2)\ .
\end{eqnarray}
This allows one to obtain the expression of ${\cal V}_1$, and one finds
that $k'_1$ is indeed independent of $\tilde\epsilon$. Finally one
has\footnote{Note that in this example the path of $T$ was the unit circle
in the {\it clockwise\/} direction.}
\begin{equation}
  {\cal V}'_2  =    {1\over2\pi\ii}\int_{e_0^-}^{e_0^+}
  {\dd \zeta\over \zeta}  (2\hat \vartheta '_2+ \hat \vartheta '_0)
+{1\over2\pi\ii}\left[-\int_{e_0^-}^{e_0^+}
  +\int_{e_1^-}^{e_1^+}+ \oint_{|\zeta|=1}\right]
  {\dd \zeta\over \zeta} \hat \vartheta '_0~.   \label{evalcV'2}
\end{equation}
In the second term the first two integrals add up to lines between
$e_1^-$ and $e_0^-$ and between $e_0^+$ and $e_1^+$. Because of the
symmetry $\zeta\rightarrow \zeta^{-1}$ (which changes the sign of $\zeta^{-1}
\dd \zeta$), the integral from $e_0^+$ to $e_1^+$ is equal to the
first one. Because for the function which we are considering,
there is only a square root cut between $e_0^-$ and $e_1^-$ (and no
cut to $\zeta=0$), these first two parts add up to an integral over a
cycle surrounding the point $e_0^-$ and $e_1^-$. That can be deformed
to the circle $|\zeta|=1$, and thus remains in \eqn{evalcV'2} only
the integral over $(2\hat \vartheta '_2+ \hat \vartheta '_0)$,
evaluated in the same way as for ${\cal V}_1$. One checks then again
that the constants in \eqn{limperex1} agree with their earlier
definition, and that $k'_2$ is independent of $\tilde u$.

We thus have for ${\cal C}'$
the decomposition as in \eqn{rigidexpv} with $a=1/2$ and
\begin{equation}
v_0=\pmatrix{0\cr 0\cr -k_1 \cr -\frac{1}{ \pi \ii} k_1 \log \tilde\epsilon + k'_1\cr
-k_2 \cr -\frac{1}{ \pi \ii} k_2 \log \tilde\epsilon + k'_2\cr} \ ;\qquad
v_1=\pmatrix{V_1\cr V_2\cr
0\cr 0\cr 0\cr 0}\ ;\qquad
v_2=\pmatrix{ {\cal O}(\tilde\epsilon^{3/2})\cr {\cal O}(\tilde\epsilon^{3/2})\cr
 - \tilde\epsilon\ell _1\tilde u+{\cal O}(\tilde\epsilon^2)\cr
 -\frac{1}{ \pi \ii}  \ell _1 \tilde u\tilde\epsilon \log \tilde\epsilon
 +\tilde\epsilon\ell' _1\tilde u+{\cal O}(\tilde\epsilon^2)\cr
 - \tilde\epsilon\ell _2\tilde u+{\cal O}(\tilde\epsilon^2)\cr
-\frac{1}{ \pi \ii} \ell _2 \tilde u\tilde\epsilon\log \tilde\epsilon
 +\tilde\epsilon\ell' _2 \tilde u +{\cal O}(\tilde\epsilon^2)}
\ . \label{v012ex2}
\end{equation}

The intersection matrix in the basis ${\cal C}'$ is
\begin{equation}
q'=\pmatrix
{0& 2& 0& 0& 0& 0\cr
 -2& 0& 0& 0& 0& 0\cr
 0& 0& 0& 4& 0& 0\cr
 0& 0& -4& 0& 0& -4\cr
 0& 0& 0& 0& 0& 2\cr
 0& 0& 0& 4& -2& 0} \ ;\qquad q'^{-1}=\frac14 \pmatrix
 {0& -2& 0& 0& 0& 0\cr  2& 0& 0& 0& 0& 0\cr
 0& 0& 0& -1& -2& 0\cr 0& 0& 1& 0& 0& 0\cr
 0& 0& 2& 0& 0& -2\cr  0& 0& 0& 0& 2& 0}\ .
\end{equation}
Observe that this is (up to a reordering and rescaling) nearly a canonical
intersection matrix (see \eqn{canQ}). To obtain the canonical form,
we only have to replace ${\cal V}_1$ by ${\cal V}'_1\equiv {\cal
V}_1+2{\cal T}'_2$.

The calculation of $\langle  v_0+\tilde\epsilon^{1/2} v_1+v_2,\bar v_0+\bar{\tilde\epsilon}^{1/2}
\bar v_1+\bar v_2 \rangle $ does not immediately lead to the structure
of \eqn{expvbarv}. Indeed there is an extra term due to the $\ell
$-terms in $v_2$:
\begin{eqnarray}
\langle v,\bar v\rangle &=&\ii M^2 (\tilde\epsilon)+\ft12 |\tilde\epsilon| \left(V_2\bar
V_1- V_1\bar V_2\right)  +R(\tilde\epsilon,\tilde u ,\bar{\tilde u} )
+ F(\tilde\epsilon,\tilde u)-\bar F(\bar {\tilde\epsilon},\bar {\tilde u})
   \nonumber\\
 M^2 (\tilde\epsilon)&=& -\ii v_0^T \, q'^{-1}  \, \bar v_0 = -\frac{1}{2\pi}\left(
 |k_1|^2 + 2|k_2|^2\right)\log |\tilde\epsilon |
 +\Im \left( \ft12 k_1\bar k_1'+ k_2\bar k_2'-k_1\bar k_2\right)
 \nonumber\\
R(\tilde\epsilon,\tilde u ,\bar{\tilde u} ) &=& {\cal O}(\tilde\epsilon^2)  \nonumber\\
F(\tilde\epsilon,\tilde u) &=& \tilde\epsilon\tilde u (f_1+f_2\log\tilde\epsilon) \ .
\label{expvbarvex1}
\end{eqnarray}
where $f_1$ and $f_2$ are constants.
If we proceed to expand the K\"ahler potential
${\cal K} = -\log(-\ii <v,\bar v>)$ as in (\ref{Kahler_expansion})
we find that
\begin{eqnarray}
\label{rigagain}
{\cal K}(\tilde\epsilon,\tilde u) &=&-\log M^2(\tilde\epsilon)
+\frac{|\tilde\epsilon|}{M^2(\tilde\epsilon)}
K (\tilde u,\bar {\tilde u}) +   {\cal O}(\tilde\epsilon^2)
+\frac{\ii}{M^2} F(\tilde u) - \frac{\ii}{M^2}\bar F(\bar {\tilde u})\nonumber\\
K&=& \frac{\ii|\eta|^2}{2\Lambda^2} \left(a(u,\Lambda)\bar a_D(\bar  u,\Lambda) -
a_D(u,\Lambda)\bar a(\bar  u,\Lambda)\right) \ .
\end{eqnarray}
The terms with $F$ and $\tilde F$ amount to an irrelevant \Ka\ transformation, and the
supergravity \Ka\ potential reduces to the \Ka\ potential of the rigid
$SU(2)$ theory, with a multiplicative `renormalization'.

\subsection{The $SU(3)$ rigid limit}
\subsubsection{Definition}

In section~\ref{ss:pmiex2} we have considered the Calabi--Yau manifold
as a torus fibration. The embedding space was parametrized by $(y,\xi,\zeta,x)$, and
for each value of $(\zeta,x)$ the equation $y(\xi)$ defines a torus.
This torus degenerates on two (one complex dimensional) surfaces in
the $(\zeta,x)$-space: on $\Sigma'$ the $\beta$-cycles of the torus
vanish, while on $\Sigma$ the $\alpha$ cycles vanish. The latter is a
genus-5 surface defined by the equation
\begin{equation}
\Sigma\ :\ P^2(x)-B'(\zeta)=0\ ,
\end{equation}
where $B'(\zeta)$ is defined in \eqn{B'}, while $P(x)$ is given in
\eqn{K3pol3} in the alternative gauge where $\psi_4$ and $\psi_1$ take the place of $I_1$ and
$I_2$. In in the usual gauge they are expressed in \eqn{invnu12usgauge}, and we thus get
\begin{equation}
P(x)=2x^3-\ft32 \psi_0^4 x - (\psi_1-\ft12 \psi_0^6)\ .
\end{equation}
At the $A_2$ point \eqn{A2point}, the equation for $\Sigma$ thus
degenerates to
\begin{equation}
x^6 - \psi_1 x^3 =0 \ .
\end{equation}
The $SU(3)$ rigid limit is thus reached when three sheets of the Riemann surface
$\Sigma$ coincide, or equivalently, when the CY considered as an elliptic fibration
acquires an $I_3$ singularity according to the Kodaira classification (i.e.\ a curve
of
$A_2$ singularities). However, as some periods are ill defined
in this limit, we have to specify how we approach this point. This
has been defined in \eqn{epsilonex2} with
$\epsilon \to 0$ while keeping  $\tilde u_1$ and $\tilde u_0$ finite.
In the equation for the CY \eqn{torusfib-mod2} we thus have
\begin{eqnarray}
P(x)&=& 2 x^3 - \frac{3}{2} \epsilon^{\frac23} \left(\tilde u_0^4+{\cal O}(\epsilon)\right)
\, x\, (-\psi_s)^{-\frac13} -\sqrt{\epsilon\tilde u_1-\psi_s}
- \frac{\epsilon\tilde u_0^6 }{2\sqrt{-\psi_s}} + {\cal
O}(\epsilon^{2})          \nonumber\\
B'(\zeta)&=&\epsilon  (\zeta + \frac{1}{\zeta}) -\psi_s \ .\label{PB'epsilon}
\end{eqnarray}

The choice of $\epsilon$ dependence keeps the branch points in the
$\zeta$-plane with vanishing $v_b$ or $v_a$ at finite positions, respectively
given by
\eqn{posconsp},
while the branch points with vanishing $t_a$ and $t_b$, given in
\eqn{poslcsp}
 are sent to infinity. This choice gives rise to light BPS states
(namely $D$-3-branes wrapped around the basis cycles $V_{v_a}$, $V_{v_b}$,
$T_{v_a}$, $T_{v_b}$)
which can be identified as the massive gauge bosons and dyons of the pure
$N=2$ $SU(3)$ Yang--Mills theory. Furthermore, we will show that in this
limit, local special geometry indeed reduces to $SU(3)$ rigid special geometry
on the rigid moduli space parametrized by  $\tilde u_1$ and $\tilde u_0$,
justifying the above choice of $\epsilon$ dependence.

In the region of moduli space under consideration, the paths in the $x$-plane
defining the $K3$ 2-cycles $v_a$ and $v_b$ stretch between sheets of
$\Sigma_-$, the branch of the $g=5$ Riemann surface $\Sigma$ given by the
equation $P(x) +\sqrt{B^\prime(\zeta)}=0$.
Rescaling as in \eqn{deftileps} and
\begin{equation}
x = \tilde\epsilon^{1/3} (-\psi_s)^{1/6}\tilde{x} \ ,
\label{tildeepsilonx}
\end{equation}
and expanding the square roots in \eqn{PB'epsilon} for finite $\zeta$:
\begin{equation}
\Sigma_-\ :\ \tilde{x}^3 - \frac{3}{4} \tilde u_0^4 \tilde{x}
- \frac{1}{4} (\tilde u_0^6 + \tilde u_1)
+ \frac{1}{4} (\zeta + \frac{1}{\zeta}) + {\cal O}(\tilde \epsilon) = 0 \ ,
\label{genus2SWeqn}
\end{equation}
which is the equation for the genus-2 $SU(3)$ Seiberg--Witten Riemann surface.
Thus we see that in the $SU(3)$ rigid limit, a genus-2 branch of our general
genus-5 Riemann surface $\Sigma$ degenerates and produces the Seiberg--Witten
surface,
with punctures `at infinity', where the rest of the genus-5 surface is attached.

\subsubsection{$\epsilon$-expansion of periods and K\"{a}hler potential}
\label{ss:epsexpex2}
Let us calculate this `$\epsilon$-monodromy' for the Calabi--Yau periods
corresponding
to the 3-cycles \eqn{basisVT}.
First note that the equation for the Calabi--Yau \eqn{torusfib-mod2} with
\eqn{PB'epsilon} is transformed (to an isomorphic equation) when
$\epsilon \to e^{2\ii\pi} \epsilon$. By the transformation \eqn{tildeepsilonx}
one gets rid of this complication. But we also
have to perform this transformation in the $(3,0)$ form \eqn{Omega30ex2}
\begin{equation}
\Omega^{(3,0)} =
\frac{\tilde\epsilon^{1/3}(-\psi_s)^{1/6}}{(2 \pi i)^3} \, \frac{\dop \zeta}{\zeta} \wedge \dop \tilde{x}
\wedge \frac{1}{y} \frac{\dop \xi}{\xi}\ .
\end{equation}
Therefore under the $\epsilon$-monodromy
\begin{equation}
\Omega^{(3,0)} \to e^{2 \pi\ii/ 3} \Omega^{(3,0)} \ .
\end{equation}
So if the basis of cycles $C$ is transformed as $C \to M C$, the corresponding
periods ${\cal C}$ transform as ${\cal C} \to \omega M {\cal C}$
where $\omega = e^{2 \pi\ii/ 3}$

To calculate the action of the monodromy on the cycles, we make use of their fibred
structure. In the case at hand, the $K3$ 2-cycle fibres (at fixed $\zeta$) as well as
the base paths in the $\zeta$-plane are transformed by the monodromy. The transformation
of the base paths is easily obtained by following the $K3$ degeneration points in the
$\zeta$-plane when $\epsilon \to e^{2 \pi i} \epsilon$. Only the path corresponding
to $V_{t_a}$ and $V_{t_b}$ is affected. The monodromy of our basis of $K3$ 2-cycles (at
fixed $\zeta$) can be found analogously by making use of their fibred structure. Of
the four base paths in the $\tilde{x}$-plane, only the ones corresponding to $t_a$ and
$t_b$ are transformed. Finally, the torus cycles $\alpha$ and $\beta$ at $\tilde{x} = 0$
transform as follows
\begin{equation}
\left(
\begin{array}{c}
\alpha \\
\beta
\end{array}
\right)
\to
\left(
\begin{array}{rr}
1 & 0 \\
1 & 1
\end{array}
\right)
\left(
\begin{array}{c}
\alpha \\
\beta
\end{array}
\right) \ .
\end{equation}
Indeed, this can be seen, because the singular points in figure~\ref{toruscuts}
are from \eqn{bp-pos} at
\begin{equation}
\xi_\pm = -\sqrt{-\psi_s}(1+ {\cal O}(\tilde\epsilon))\pm\sqrt{\tilde\epsilon(
2\tilde u_1+\tilde u_0^6)}\ .
\end{equation}
Thus the two points interchange position under the epsilon monodromy. $\alpha$
is therefore the vanishing cycle, to be used in the Picard--Lefshetz formula
\eqn{Picard-LefshetzK3CY}.

For the $K3$ cycles we note that in figure~\ref{cuts-in-x} the
inner points do not move, as the equation for these is now \eqn{genus2SWeqn}.
The outer points are determined by $\Sigma_+$, i.e.\
\begin{equation}
\Sigma_+\ :\ -1+\tilde\epsilon\left( \tilde{x}^3 - \frac{3}{4} \tilde u_0^4 \tilde{x}
- \frac{1}{4} (\tilde u_0^6 + \tilde u_1)
- \frac{1}{4} (\zeta + \frac{1}{\zeta})\right)  + {\cal O}(\tilde \epsilon^2) = 0 \ ,
\end{equation}
and the solutions rotate for small $\epsilon$ under the
$\epsilon$-monodromy. Therefore the $v_a$ and $v_b$ paths do not
change, while for $t$ paths the rotated cycles are re-expressed as%
\footnote{One should see from the monodromy that
the torus $\beta$ cycle above $t_{12}$ is deformed to a $\beta$ cycle above $t_{31}$
and an $\alpha$ cycle above $s^-_{23}$.}
\begin{equation}
t_{21}\rightarrow t_{31}-s^-_{23}      \ .
\end{equation}
This equation and its cyclic permutations with \eqn{t-in-s} and \eqn{vt-in-st} give
\begin{equation}
\left(
\begin{array}{c}
v_a \\
v_b \\
t_a \\
t_b
\end{array}
\right)
\to M^\epsilon_{K3}
\left(
\begin{array}{c}
v_a \\
v_b \\
t_a \\
t_b
\end{array}
\right)  =
\left(
\begin{array}{rrrr}
1 & 0 & 0 & 0 \\
0 & 1 & 0 & 0 \\
0 &-1 &  -1 & -1 \\
1 & 1 & 1 & 0
\end{array}
\right)
\left(
\begin{array}{c}
v_a \\
v_b \\
t_a \\
t_b
\end{array}
\right)
\end{equation}

For the CY cycles, we consider figure~\ref{cuts-in-zeta}. As
mentioned already, only the beginning and end points of the $V_{t_a}$ and
$V_{t_b}$ cycles are (in lowest order) $\epsilon$ dependent. They
make a complete tour in this plane. Hence, for all but $V_{t_a}$ and
$V_{t_b}$, we just have to take the $K3$ result. The remaining two
are modified by circles around the origin and infinity, and we have to
consider the monodromies \eqn{monodromiesK3ex2}
of the transported $K3$ cycles when crossing cuts.
This is similar to the manipulations in \eqn{ex1L0cmono}. This leads to
\begin{equation}
\pmatrix{V_{v_a}\cr V_{v_b}\cr V_{t_a}\cr V_{t_b} \cr T_{v_a}\cr T_{v_b}\cr
T_{t_a}\cr T_{t_b} }  \rightarrow M^\epsilon_{CY}
\pmatrix{V_{v_a}\cr V_{v_b}\cr V_{t_a}\cr V_{t_b} \cr T_{v_a}\cr T_{v_b}\cr
T_{t_a}\cr T_{t_b} } =
\pmatrix{ M^\epsilon_{K3}&\begin{array}{cccc}
0 & 0 & 0& 0\cr
0 & 0 & 0& 0\cr
0 & 1 & 2& 2\cr
-1 & -1 & -2& 0
\end{array}\cr \begin{array}{cccc}
&&&\cr
&&&\cr
&0&&\cr
&&&
\end{array}& M^\epsilon_{K3}}
\pmatrix{V_{v_a}\cr V_{v_b}\cr V_{t_a}\cr V_{t_b} \cr T_{v_a}\cr T_{v_b}\cr
T_{t_a}\cr T_{t_b} }\ .
\end{equation}
The monodromy matrix of the periods has an extra factor $\omega$ as
explained above.

We can bring this monodromy matrix for the periods in Jordan form
\begin{equation}
{\cal M}^\epsilon_{CY,J} =S\,  \omega M^\epsilon_{CY} \, S^{-1}=
\left(
\begin{array}{rrrrrrrr}
\omega &  &  &  &  &  &  & \\
 & \omega &  &  &  &  &  & \\
 &  & \omega &  &  &  &  & \\
 &  &  & \omega &  &  &  & \\
 &  &  &  & 1 & 0 &  & \\
 &  &  &  &-2 & 1 &  & \\
 &  &  &  &  &  & \omega^2 & 0\\
 &  &  &  &  &  & -2\omega^2 & \omega^2 \\
\end{array}
\right)
\end{equation}
by the transformation
\begin{equation}
{\cal C}'=\pmatrix{ {\cal V}_{v_a}\cr {\cal V}_{v_b}\cr {\cal T}_{v_a}\cr {\cal T}_{v_b}
\cr {\cal T}'_{11}\cr {\cal V}'_{11}\cr {\cal T}'_{22}\cr {\cal V}'_{22}}=
S{\cal C}= \pmatrix
{1&0&0&0&0&0&0&0\cr 0&1&0&0&0&0&0&0\cr 0&0&0&0&1&0&0&0\cr
0&0&0&0&0&1&0&0\cr 0&0&0&0&-\frac{\ii
\omega^2}{\sqrt{3}}&\frac{\ii}{\sqrt{3}}&-\omega^2&1\cr
-\frac{\ii \omega^2}{\sqrt{3}}&\frac\ii{\sqrt{3}}&-\omega^2&1&\frac{\omega^2}3&-1/3&0&0\cr
0&0&0&0&\frac{\ii \omega}{\sqrt{3}}&-\frac\ii{\sqrt{3}}&-\omega&1\cr
\frac{\ii \omega}{\sqrt{3}}&-\frac\ii{\sqrt{3}}&-\omega&1&\frac\omega 3&-\frac13&0&0}
\pmatrix{{\cal V}_{v_a}\cr {\cal V}_{v_b}\cr {\cal V}_{t_a}\cr {\cal V}_{t_b}
\cr {\cal T}_{v_a}\cr {\cal T}_{v_b}\cr
{\cal T}_{t_a}\cr {\cal T}_{t_b} } \ .
\label{basisCprime}
\end{equation}
We will explain the choice of names for the basis vectors of ${\cal C}'$.
The first four basis vectors, eigenvectors of eigenvalues
$\omega$, correspond to integrals over $\{V_{v_a},V_{v_b},T_{v_a},T_{v_b}\}$,
i.e.\ related to
the $v_a$ and $v_b$ cycles in $K3$. The last four are simpler in
terms of the $K3$ cycles in Picard--Fuchs basis. Indeed, defining the
periods on the $|\zeta|=1$ circle (see the inverse of \eqn{experm})
\begin{equation}
\pmatrix{{\cal T}_{12}\cr{\cal T}_{21}\cr{\cal T}_{11}\cr {\cal T}_{22}}=
\int_{|\zeta|=1}\frac{1}{2\pi\ii}\frac{d\zeta}{\zeta}
\pmatrix{\vartheta_{12} \cr \vartheta_{21} \cr
\vartheta_{11} \cr \vartheta_{22} }\ .
\end{equation}
Similarly we define ${\cal V}_{11}$ and ${\cal V}_{22}$
\begin{eqnarray}
{\cal V}_{11} &=&\frac{1}{\sqrt{3}}\left( -{\cal V}_{v_a}+\omega
{\cal V}_{v_b}\right) +\ii \left( {\cal V}_{t_a}-\omega{\cal V}_{t_b}\right)
\nonumber\\
{\cal V}_{22} &=&\frac{1}{\sqrt{3}}\left( -{\cal V}_{v_a}+\omega^2
{\cal V}_{v_b}\right) -\ii \left( {\cal V}_{t_a}-\omega^2{\cal V}_{t_b}\right) \ ,
\end{eqnarray}
where ${\cal V}_I$ are the integrals of the $(3,0)$ form on $V_I$
\footnote{It may be possible to deform paths in a way similar to what we did in the
first example to combine these integrals to one of $\vartheta _{11}$ and
$\vartheta_{22} $.}. The last four elements of
${\cal C}'$ are
\begin{equation}
\pmatrix{
{\cal T}'_{11}\cr {\cal V}'_{11}\cr {\cal T}'_{22}\cr {\cal
V}'_{22}}= a\pmatrix{\ii\omega^2 {\cal T}_{11}\cr \ii\omega^2 {\cal V}_{11}
+\frac{\omega^2}{\sqrt{3}}{\cal T}_{21} \cr -\ii \omega{\cal
T}_{22}\cr -\ii\omega{\cal V}_{22}+\frac{\omega}{\sqrt{3}}{\cal T}_{12}
} \ .    \label{calTV1122}
\end{equation}

The complex transformation of basis between the periods ${\cal C}$
corresponding to the cycles \eqn{basisVT} and those in \eqn{basisCprime},
i.e.\ ${\cal C}'=S{\cal C}$, changes the inner product in
\eqn{Kahlerpotentials}. The K\"{a}hler potential on the CY complex structure moduli
space is thus
\begin{equation}
{\cal K} = - \log \left( -\ii \langle {\cal C},\bar {\cal C}\rangle \right) \ ;\qquad
\langle {\cal C},\bar {\cal C}\rangle  =
{\cal C}^T q^{-1} \bar {\cal C} ={\cal C}^{\prime \, T} q'^{-1} \bar{{\cal C}^\prime} \ ,
\end{equation}
with $q$ in \eqn{intersCYex2}, and the anti-Hermitian matrix $q'$ is given by
\begin{equation}
q'^{-1}= S^{-1\,T} q^{-1} \bar S^{-1}
= \left(
\begin{array}{cc}
  \begin{array}{ccc}
  &&\\&Q^{-1}&\\&&
  \end{array} & \\
 & \begin{array}{cccc}
\frac{7 \ii}{\sqrt{3}} & 1 & 0 & 0 \\
-1 & 0 & 0 & 0 \\
0 & 0 & \frac{-7 \ii}{\sqrt{3}} & 1 \\
0 & 0 & -1 & 0
\end{array}
\end{array}
\right)\ . \label{qprimeex2}
\end{equation}
This anti-Hermitian matrix has a $4+2+2$ block diagonal form with as
upper block
\begin{equation}
Q = \left(
\begin{array}{rrrr}
0&-1&-2&1 \\
1&0&1&-2\\
2&-1&0&0\\
-1&2&0&0
\end{array} \right)
\ ;\qquad
Q^{-1}=\frac{1}{3}\pmatrix{ 0& 0& 2& 1\cr 0& 0& 1& 2\cr -2& -1& 0& -1\cr -1& -2& 1& 0}
\ .    \label{matrixQex2}
\end{equation}

{}From the monodromy, we read off the $\epsilon$ dependence of ${\cal C}^\prime$:
\begin{equation}
\pmatrix{  {\cal V}_{v_a}\cr{\cal V}_{v_b}\cr{\cal T}_{v_a}\cr{\cal T}_{v_b}\cr
{\cal T}'_{11}\cr{\cal V}'_{11}\cr{\cal T}'_{22}\cr{\cal V}'_{22} }
=  \pmatrix{
\tilde\epsilon^{1/3}{\cal A}_1(\tilde\epsilon)\cr
\tilde\epsilon^{1/3}{\cal A}_2(\tilde\epsilon)\cr
\tilde\epsilon^{1/3}{\cal A}_3(\tilde\epsilon)\cr
\tilde\epsilon^{1/3}{\cal A}_4(\tilde\epsilon)\cr
{\cal A}_5(\tilde\epsilon)\cr
\frac{-1}{\pi \ii} \log \tilde\epsilon \, {\cal A}_5 + {\cal A}_6(\tilde\epsilon)\cr
\tilde\epsilon^{2/3}{\cal A}_7(\tilde\epsilon)\cr
\frac{-1}{ \pi \ii} \log \tilde\epsilon \, {\cal A}_7 +\tilde\epsilon^{\frac23}{\cal A}_8(\tilde\epsilon) }
=\pmatrix{
\tilde\epsilon^{1/3}\left( V_1+{\cal O}(\tilde\epsilon)\right) \cr
\tilde\epsilon^{1/3}\left( V_2+{\cal O}(\tilde\epsilon)\right) \cr
\tilde\epsilon^{1/3}\left( V_3+{\cal O}(\tilde\epsilon)\right) \cr
\tilde\epsilon^{1/3}\left( V_4+{\cal O}(\tilde\epsilon)\right) \cr
k + {\cal O}(\tilde\epsilon)\cr
\frac{-1}{ \pi \ii} k \log \tilde\epsilon + k' + {\cal O}(\tilde\epsilon,\tilde\epsilon \log \tilde\epsilon)\cr
\tilde\epsilon^{2/3}\left( \ell  + {\cal O}(\tilde\epsilon)\right)  \cr
\tilde\epsilon^{\frac23}\left( \frac{-1}{\pi \ii} \ell   \log \tilde\epsilon
+\ell '   + {\cal O}(\tilde\epsilon,\log \tilde\epsilon)   \right)
}\ ,    \label{eafh}
\end{equation}
where ${\cal A}_\Lambda(\tilde\epsilon)$ stand for analytic functions of
$\tilde\epsilon$.
The absence of poles follows from the boundedness of the $K3$ periods
(for bounded $B^\prime$) in the limit $\tilde\epsilon \to 0$; therefore, ${\cal T}'_{11}$
and ${\cal T}'_{22}$ have a finite limit, and ${\cal V}'_{11}$ and ${\cal V}'_{22}$
diverge logarithmically due to the factor
\begin{equation}
\int_{-\tilde\epsilon}^{-1/(\tilde\epsilon)} \frac{1}{2\pi\ii}\frac{\dop \zeta}{ \zeta} =
\frac{-1}{\pi\ii}\log \tilde\epsilon\ .  \label{longintegral}
\end{equation}
The `constants' $V_A^0$, $k$, $k'$, $\ell $ and $\ell '$ are independent of $\tilde\epsilon$,
but possibly still dependent on the rigid moduli $\tilde u_i$. However, $k$ and $k'$ must be
independent of the rigid moduli: indeed, as can be seen easily from the explicit
form of $\Omega^{(3,0)}$ (and the cycles), taking derivatives of the period integrals
${\cal T}'_{11}$ and ${\cal V}'_{11}$ with respect to the rigid moduli gives a positive power of $\tilde\epsilon$
times something which is at most logarithmically divergent when $\epsilon \to 0$.
Therefore%
\footnote{Actually, we have to be a bit careful here. If our cycles were
such that $y$ was bounded from below in the limit $\epsilon \to 0$, the
conclusion would have been obvious. However, for the cycles under consideration,
this is not the case since they contain unavoidably a piece on which $y$
is of order $\epsilon^{1/2}$ (this is the piece close to the surface of
singularities which develops when $\epsilon \to 0$). The contribution of this
piece can be estimated quite easily by going to the ALE approximation, which is
allowed precisely in the problematic region. Thus, for the expressions we get
by taking derivatives of the periods with respect to rigid moduli, the net contribution
of this piece (including the extra factor $\epsilon^\gamma$ produced by
taking the derivative) turns out to be proportional to $\epsilon^{1/3}$.
So the conclusion holds.},
these derivatives are zero at $\epsilon = 0$. This is only possible if both $k$
and $k'$ are independent of the rigid moduli. This will also be
clear from the explicit expressions in section~\ref{ss:compPLb}.

The periods of $v_a$ and $v_b$ for finite values of $\zeta$ thus become
small in the rigid limit, and the first
four CY basis cycles to give rise to the light $D$-brane states corresponding to
gauge
bosons and dyons. We have found the structure announced in
section~\ref{ss:rigidlimit}. Indeed, we have for ${\cal C}'$ the
decomposition as in \eqn{rigidexpv} with $a=1/3$ and
\begin{equation}
v_0=\pmatrix{0\cr 0\cr 0\cr 0\cr k\cr \frac{-1}{\pi\ii}
k\log \tilde\epsilon+k'\cr 0\cr 0\cr} \ ;\qquad
v_1=\pmatrix{V_1\cr V_2\cr V_3\cr V_4\cr
0\cr 0\cr 0\cr 0}\ ;\qquad
v_2=\pmatrix{0\cr 0\cr 0\cr 0\cr 0\cr 0\cr \tilde\epsilon^{2/3} \ell     \cr
\tilde\epsilon^{\frac23}\left( \frac{-1}{\pi \ii} \ell   \log \tilde\epsilon
+\ell ' \right)  } +{\cal O}(\tilde\epsilon)\ .
\label{v012}
\end{equation}
Note that the $\tilde u$-independence of $k$ and $k'$ is
essential to fit in this scheme. Also important is the $4+2+2$ block diagonal
structure of $q'$ in \eqn{qprimeex2}, which implies the orthogonality of $v_1$,
$v_2$ and $v_3$ (the latter to order $\tilde\epsilon$), such that \eqn{expvbarv}
is applicable.
We find
\begin{equation}
M^2(\tilde\epsilon)=-\ii v_0^T \, q'^{-1}  \, \bar v_0 = |k|^2 \left(
\frac{7}{\sqrt{3}}-\frac{2}{\pi}\log |\tilde\epsilon|\right)
+2\,\Im (k\bar k')\ ,
\end{equation}
to be used in \eqn{Kahler_expansion} with
\begin{equation}
K=\ii v_1^T(\tilde u) \pmatrix{Q^{-1}&0\cr 0&0}\bar v_1(\bar {\tilde u})\ .
\end{equation}
{}From
the monodromy structure of the periods $V_i$ and the light $D$-brane spectrum,
we can identify this as a rigid limit corresponding to a $N=2$ $SU(3)$
pure Yang--Mills theory in four dimensions. In the next section we show how
the Seiberg--Witten solution of this gauge theory is retrieved.

\subsubsection{The Seiberg--Witten expression of the rigid K\"{a}hler potential}
\label{sss:SW}
For finite $\zeta$ we obtained (\ref{approx-K3per})
\begin{equation}
\int_{s_{ij}^-} \Omega^{(2,0)} = (-\psi_s)^{-1/12}\frac{\sqrt{6}}{2 \pi}
\tilde\epsilon^{1/3}(\tilde{x}_{j^-} - \tilde{x}_{i^-}) + {\cal O}(\tilde\epsilon^{4/3})\ .
\label{sij-Omega2}
\end{equation}
So for a CY cycle $\gamma \wedge s_{ij}^-$ obtained by transporting
$s_{ij}^-$ along a path $\gamma$ in the $\zeta$-plane,
we find to leading order
\begin{equation}
\int_{\gamma \wedge s_{ij}^-} \Omega^{(3,0)}
=(-\psi_s)^{-1/12} \tilde\epsilon^{1/3} \int_{\gamma_{j^-} - \gamma_{i^-}}
\lambda_{SW}\ ,
\label{lambdaex2}
\end{equation}
where
\begin{equation}
\lambda_{SW} = \frac{\sqrt{6}}{2 \pi} \frac{1}{2 \pi i}
\, \tilde{x} \frac{\dop \zeta}{\zeta} \ ,
\end{equation}
and
\begin{equation}
\gamma_{i^-} = \gamma \mbox{ lifted to sheet $i$ of $\Sigma_-$}\ .
\end{equation}
Note that $\gamma_{j^-} - \gamma_{i^-}$ is always a closed cycle on the SW
Riemann surface (single loop if $\gamma$ is open, double loop if closed). Thus,
since $\lambda_{SW}$ is also precisely the Seiberg--Witten meromorphic
1-form, we find that the leading-order part of the first four CY periods
(the $V_i$) are nothing but the Seiberg--Witten periods\footnote{Note
that higher-order $\epsilon$-corrections (gravity effects) to, for example, the SW
BPS mass formula can be calculated rather easily in this set-up.
There are two sources of corrections: the surface itself is corrected
as well as the meromorphic 1-form.}.

It is interesting to retrace how the normalization of the
1-form $\lambda_{SW}$ arises. In fact, at the start, the Calabi--Yau
$(3, 0)$-form can be normalized quite arbitrarily, without losing holomorphy,
by multiplying it with an arbitrary holomorphic function of the moduli,
in particular, the rigid moduli. The
normalization we adopted arises naturally from the Griffiths
representation \eqn{griffiths30}, and after integration over 2-cycles
in the $K3$ gives rise to the meromorphic form $\lambda_{SW}$,
\eqn{lambdaex2}.
The normalization chosen in \cite{candmirror1} is different:
$\hat \Omega^{(3,0)}=\psi_0 \Omega^{(3,0)}$.
With this normalization, since
$\psi_0\sim \tilde\epsilon^{1/6}\,\tilde u_0\,(-\psi_s)^{-1/12}$,
it would seem to give an incorrect
($\tilde u$-dependent) normalization of $\lambda_{SW}$.
However, with this normalization,
the leading term of the periods $v_0$ (\eqn{v012})
would not be independent of the rigid modulus $\tilde u_0$.
To recover the rigid-limit scheme of section~\ref{ss:rigidlimit},
one could make a \Ka\ transformation. This effectively amounts
to changing the normalization of $\Omega^{(3,0)}$ with a
factor that depends on the rigid moduli%
\footnote{In the first example this question did not come up
because $\psi_0$ is independent
of $\tilde u$ in a first approximation.}.

Finally, using (\ref{int-corr}) and the comment below it, it is not difficult to
show that the intersection matrix of the first four Calabi--Yau basis cycles is
precisely equal to minus the intersection matrix of the corresponding
four SW cycles, as we did in the previous sections. Thus we find
\begin{equation}
K(\tilde u,\bar {\tilde u}) =\ii\int_{\gamma_A} \lambda_{SW} \, \, (Q^{-1})^{AB}
\, \, \int_{\gamma_B} \bar\lambda_{SW}\ ,
\end{equation}
where $(\gamma_A)_{A=1 \ldots 4}$ is a basis of cycles of the SW Riemann surface
and $Q$ is its intersection matrix. This completes the identification of $K$
with the Seiberg--Witten solution for the K\"{a}hler potential of $N=2$ $SU(3)$
Yang--Mills theory in 4 dimensions.

\subsubsection{Comparison with Picard--Fuchs basis.}\label{ss:compPLb}
Consider the $\epsilon$ expansion in the Picard--Fuchs basis.
Using the expansion \eqn{epsilonex2} writing everything in terms of
$\psi_s$, $\epsilon$ and $\tilde u_{1,2}$,  we have that $\psi_0^6$ is
of order $\epsilon$ (we are in the `$SU(3)$ sector'), and we have as
leading order
\begin{equation}
r,s\approx \sigma_\mp \tilde{\epsilon}^{-1/2}
\ ;\qquad \sigma_{\mp} =
\frac{\sqrt{2\, \tilde u_0^6+ \tilde u_1 - (\zeta +
\frac{1}{\zeta}) }
\mp \sqrt{ \tilde u_1 - (\zeta + \frac{1}{\zeta}) } }
{2\, \tilde u_0^6}\ .  \label{tildeepsilon}
\end{equation}
Because $r$ and $s$ are large, we can use \eqn{hygeobase}. The leading order is
\begin{eqnarray}
\xi_1(r),\xi_1(s) &\approx& B_1\sigma_\mp^{-1/6} \tilde{\epsilon}^{1/12} \nonumber\\
\xi_2(r),\xi_2(s) &\approx& B_2\sigma_\mp^{-5/6} \tilde{\epsilon}^{5/12}
\end{eqnarray}
and  (using $\tilde u_0=(2\sigma_-\sigma_+)^{-1/6}$)
\begin{equation}
\vartheta \approx -\frac{3}{4\pi} (-\psi_s)^{-1/12} 2^{1/6}
\pmatrix{ \sigma_+^{-2/3} \tilde{\epsilon}^{1/3}\cr
\sigma_-^{-2/3}  \tilde{\epsilon}^{1/3}\cr
\frac{B_1}{B_2} \cr
\frac{B_2}{B_1} (\sigma_-\sigma_+)^{-2/3} \tilde{\epsilon}^{2/3}} \ .
\label{varthetaapprox}
\end{equation}

With \eqn{sij-Omega2} and \eqn{vt-in-st} we can write the first two periods as
\begin{eqnarray}
\int_{v_a}\Omega^{(2,0)}&=& (-\psi_s)^{-1/12}\frac{\sqrt{6}}{2 \pi}
\tilde\epsilon^{1/3}(\tilde{x}_{2^-} - \tilde{x}_{1^-}) \nonumber\\
\int_{v_b}\Omega^{(2,0)}&=& (-\psi_s)^{-1/12}\frac{\sqrt{6}}{2 \pi}
\tilde\epsilon^{1/3}(\tilde{x}_{3^-} - \tilde{x}_{2^-}) \ .
\end{eqnarray}
On the other hand, the left-hand sides can also be written in terms
of $\sigma_\mp$ in \eqn{tildeepsilon} using  \eqn{experm}  and
\eqn{varthetaapprox}. We thus get
\begin{eqnarray}
\pmatrix{\tilde{x}_{2^-} - \tilde{x}_{1^-}\cr \tilde{x}_{3^-} - \tilde{x}_{2^-}}
=- 2^{-4/3}\sqrt{3}\,a \pmatrix{-\ii \omega& \ii\omega^2\cr
-\ii&\ii}\pmatrix{\sigma_+^{-2/3}\cr \sigma_-^{-2/3}}\ .
\end{eqnarray}
On the other hand, we have for the solutions of \eqn{genus2SWeqn}
\begin{eqnarray}
\tilde x_{1^-}+\tilde{x}_{2^-}+\tilde{x}_{3^-}&=& 0\nonumber\\
\tilde x_{1^-}\tilde{x}_{2^-} +\tilde{x}_{2^-} \tilde{x}_{3^-}+\tilde{x}_{3^-}
\tilde{x}_{1^-} &=&-\ft34 \tilde u_0^4=-
2^{-8/3}\,3(\sigma_-\sigma_+)^{-2/3}\nonumber\\
\tilde
x_{1^-}\tilde{x}_{2^-}\tilde{x}_{3^-}&=&\ft14\tilde
u_0^{12}(\sigma_-^2+\sigma_+^2)=
2^{-4}\left( \sigma_+^{-2} +\sigma_-^{-2}\right)  \ .
\end{eqnarray}
Combining these equations gives $a=1$ as announced.

These results give an independent derivation of the
$\epsilon$-monodromy results of  section~\ref{ss:epsexpex2}.
The first two basis vectors are a recombination of $v_a$ and $v_b$,
see \eqn{experm}. Knowing that the paths in the $K3$ fibrations
over which these cycles are integrated are independent
of $\epsilon$ to first order, shows that the corresponding four CY periods behave as
$\epsilon^{1/3}$. Furthermore the last two $K3$ cycles are integrated
over the $\epsilon$-independent circle $|\zeta|=1$, which gives the
behaviours of the fifth and seventh period in \eqn{eafh}. There
integrals over the path which for $\epsilon\rightarrow 0$ stretches
to infinite length leads to extra $\log\epsilon$-terms, see \eqn{longintegral}.
The fact that
the leading term is the constant $B_1/B_2$ implies indeed that $k$
is independent of the rigid moduli $\tilde u$.
In fact, this leading term comes immediately from \eqn{calTV1122}, and
we get
\begin{equation}
k= \ii\omega^2 (-\frac{3}{4\pi})   (-\psi_s)^{-1/12} 2^{1/6}
\frac{B_1}{B_2}\ .
\end{equation}

\subsection{Physical interpretation of the rigid limit}

Physically, taking $\tilde\epsilon\to 0$ means decoupling gravity.
By-now-standard arguments\footnote{Originally relying on the chain of string
dualities \cite{2ndqms}
${\rm Het}(K3\times T_2) \stackrel{S}{\rightarrow} {\rm IIA}(X)
\stackrel{T}{\rightarrow} {\rm IIB}(X^*)$
with $X$ and $X^*$ $K3$-fibrations \cite{KLM,KKLMV} that relates  the IIB modulus
$\tilde\psi_s = -1/(2\tilde\epsilon)$ to the heterotic dilaton $S$, whose
v.e.v.\ gives the bare complexified gauge coupling: $\langle S\rangle =
{8\pi^2\over g_0^2}
+ \ii \theta_0$, by $\tilde\psi_s = \exp(S/2)$. Later, for example, in \cite{GE},
remaining within IIA or IIB framework, only generic requirements for
the gauge theory to emerge in the rigid limit were invoked.}
\cite{KLM,KKLMV,KLMVW,GE} suggest that the running of the gauge coupling of
the $N=2$ field theory
requires $-2\tilde\epsilon\sim
(\Lambda/M)^{n}$ for a pure ${\rm SU}(n)$ theory, where $M$ is the scale,
of the order of the Planck mass $M_{\rm P}$, at which the bare
gauge coupling is defined, and $\Lambda$ is the scale at which
the gauge theory becomes strongly coupled.
The $\tilde u$'s are related, in the semiclassical regime of the field theory,
to the masses of the $W$-bosons, and remain finite while gravity is decoupled.

Our analysis leads, however, to some correction to this identification.
By plugging the leading terms in the $\tilde\epsilon$ expansion of the \Ka\
potential
into the
supergravity action, we find that the kinetic terms of matter fields get a coefficient of
order $\tilde\epsilon^{2/n}/\log\tilde\epsilon$. Thus comparing the
Einstein and the
gauge terms in the action one obtains that
$(\Lambda/M_{\rm P})^{2} \sim \tilde\epsilon^{2/n}/\log\tilde\epsilon$.

\section{Conclusions}     \label{ss:conclusions}

Within the framework of special \Ka\ geometry we have discussed
some aspects of the rigid limit of $N=2$ supergravity and
type~IIB string theory compactified on Calabi--Yau 3-folds
which admit a $K3$ fibration.

The building blocks of the corresponding
supersymmetric actions are the periods of the Calabi--Yau
3-forms. We have studied in detail their structure as
fibrations of $K3$ periods.

We have found it convenient to study the complex structure
moduli spaces, of the CY and of its $K3$ fibre,
by taking advantage of its projective structure, i.e.\
without making specific `gauge choices'. Postponing the
gauge-fixing is particularly useful for the
derivation of the Picard--Fuchs equations and for
the discussion of the singularities.
For the latter, some desingularization is automatic in
the non-gauge-fixed moduli space.
The Picard--Fuchs equations used to determine the $K3$ periods
are differential equations
for the integrals of the $(2,0)$ form over the 2-cycles.
Working in the full moduli space, i.e.\ keeping all moduli non-gauge
fixed, allows one to take
derivatives with respect to these parameters and
then to re-express them as derivatives with respect to
the invariant variables. Borrowing ideas
from toric geometry, this method avoids however the introduction of the full
toric geometry machinery and leads directly
to lower-order differential equations.
Some aspects of the structure of the solutions are explained from
group-theoretical considerations of the full moduli space of the $K3$
fibre in appendix~\ref{app:k3fibremodulispace}.

The rigid limit is a description of the behaviour of the theory
in the vicinity of a singularity in the moduli space, more specifically
of a point in moduli space on a conifold line at infinity, where the
Calabi--Yau manifold becomes singular on the whole fibration base space $\IP^1$;
indeed the  $K3$ fibre has everywhere a conifold singularity.
In previous works, starting with  \cite{KLMVW}, the vicinity of this
point was discussed by replacing the fibre by a non-compact resolution
of its singularity, an ALE space.
We have considered the decoupling of gravity and
the reduction of special \Ka\ geometry
from local to rigid without going first to this ALE approximation.
The general procedure by which the limit is
obtained is given in section~\ref{ss:rigidlimit}, the relevant points
being the splitting of the periods according to \eqn{rigidexpv}, and the
special properties of the intersection matrices.
The periods are split in a leading term $v_0$,
a first correction vanishing at the singularity $v_1$ and a remainder $v_2$.
The term $v_0$ should {\it not\/} depend on the {\it rigid\/} moduli.
The term $v_1$ contains the periods which remain in the rigid limit,
and the intersection matrices which we found imply
orthogonality between $v_0$ and $v_1$.
The term $v_0$ is generated by  cycles which do not occur
in the ALE manifolds. These periods
contain $\log \epsilon$ terms which give in
the rigid limit (infinite Planck mass) a
diverging renormalization of the rigid \Ka\ potential. The
requirement that $v_0$ does not depend on the rigid moduli
fixes the moduli dependence of the normalization
(see section~\ref{sss:SW})
of the meromorphic 1-form $\lambda$ whose periods specify the rigid theory.

Indeed, in the ALE approach
one starts with a particular normalization of the $(3,0)$-form, that correctly
gives in the limit the meromorphic form known on the field theory side;
one may wonder what fixes the freedom of rescaling it by a holomorphic
function of moduli. The above requirement on $v_0$ does.

Now we point out a connection with the $M$-theory branes.
In the second example we can consider the 3-cycles as circle fibrations
over certain `2-branes' in the $(x,\zeta)$-space. The circles
vanish at two (one complex dimensional) surfaces in this space: $\Sigma$
($\alpha$
cycles vanish) and $\Sigma^{\prime}$ ($\beta$ cycles vanish).
The 2-branes can therefore either
be closed with trivial monodromy for the circle fibre, or open and ending on
$\Sigma$ or $\Sigma'$.  The topology of the 2-brane can
be either a disc (gives $S^3$ CY 3-cycle), a cylinder ($S^1 \times S^2$ 3-cycle)
or a torus ($T^3$ 3-cycle). This is very similar to the brane picture
solutions of field theories via $M$ theory. Combining $\Sigma$ with
the four-dimensional spacetime $M_4$, we may think of a
5-brane $M_4 \times \Sigma$ ($\alpha$ fibre) or
$M_4 \times \Sigma^\prime$ ($\beta$ fibre) on which the 2-branes end.
There are several other similarities,
like the conditions for a supersymmetric state, but we will not discuss them here.

This connection with the 2/5 brane picture generalises, at least
formally,  the discussion in
\cite{KLMVW} for the $\epsilon \to 0$ ALE approximation.  In
the full Calabi--Yau geometry at finite
$\epsilon$ the surface $\Sigma$ is a genus-5 Riemann surface.

Seiberg--Witten Riemann surfaces corresponding to
different rigid limits emerge as degenerating branches of the
genus-5 Riemann surface, defined for all values of the CY moduli.
{}From a physical point of view we expect to approach the rigid limit
when we tune the moduli in such a way that the higher-genus Riemann
surface develops a branch of almost coinciding sheets. Indeed, in this case
we obtain a set of small-volume 3-cycles, namely those
constructed from a disc or a cylinder stretched between the sheets that
approach each other. Branes wrapped around these cycles are
light compared to the Planck mass and decouple from gravity.
Mathematically, the required degeneration of a branch of the higher-genus
surface means that the Calabi--Yau develops a complex curve of
singularities.

The results of this work can be used for a full expansion of the
$N=2$ supergravity action with vector multiplets, which starts
with a rigid supersymmetry action.
We have developed several methods useful for such an expansion.
These allow us to go beyond the usual general statements of the
embedding of a Riemann surface in a Calabi--Yau manifold.

\medskip
\section*{Acknowledgments.}
\noindent
We had stimulating discussions of various parts of the paper with
Sergio~Cacciatori,
Ansar~Fayazzudin,
Albrecht~Klemm,
Wolfgang~Lerche,
Koen~Vermeir
and Nick~Warner.

The authors acknowledge the hospitality from several institutes where
they could meet and discuss. In particular, CERN, the Ecole Normale
Superieure in Paris, Nordita in Copenhagen, and the Newton Institute
in Cambridge.

\appendix
\section{Remarks on the moduli space of the $K3$ fibre}
\label{app:k3fibremodulispace}
In the main text we have been concerned with the calculation
of the periods $\theta_I = \int_{c_I} \Omega^{(2,0)}$
for the $K3$ fibre occurring in Calabi--Yau manifolds that are $K3$ fibrations.
We have  focused on two specific examples but our aim has been to illustrate the general
aspects of the construction. In particular, we have
stressed  the implication of the fibration structure
of the Calabi--Yau 3-fold
for its complex structure moduli space.
\par
In the present appendix we review some general features of $K3$ moduli spaces
and give some more details on issues that were just mentioned in the main text.
In particular, we give the relation between our determination
of the transcendental periods and the
embedding of complex structure moduli in the complete $K3$ moduli space, which
is well known to be a quaternionic homogeneous space quotiened by a
discrete group, see \cite{Aspinwall}.
\subsection{Local geometry of the complex structure moduli of $K3$}
As we emphasized in the main text (see the discussion in
section~\ref{ss:K3Pictransc}),
differently from the case of Calabi--Yau 3-folds where the moduli
space of K\"ahler class deformations and complex structure deformations are
independent factors in the full moduli space of the Calabi--Yau 3-fold,
the moduli space ${\cal M}$ for string compactifications on $K3$,
which includes complexified K\"ahler class
moduli and complex structure moduli,
is the quotient of a single quaternionic coset manifold with respect to a
global duality group \cite{Aspinwall}:
\begin{equation}
{\cal M} =
\frac{O(4,20)}{O(4) \times O(20) } /O(4,20;\ZZ)\ .
\label{k3modgen}
\end{equation}
Within ${\cal M}$, the submanifold ${\cal M}_{\rm c}$ of
complex structure deformations
also has a local coset manifold structure:
\begin{equation}
{\cal M}_{\rm c} =
\frac{O(3,19)^+}{(O(2) \times O(1,19))^+ } /O(3,19;\ZZ)^+\ .
\label{k3comstruc}
\end{equation}
Although the total number of $(1, 1)$-forms is always 20,
in every algebraic representation $\Sigma_{K3}$ of the $K3$ surface only a
subset of $n_{\rm tr}$ of them
can be traced back to a change in the coefficients of the defining polynomial.
The $(1, 1)$-forms in the Picard lattice generate deformations that
induce a change of the algebraic representation of the $K3$ surface.
The number $n_{\rm tr}$ is related to the rank of the Picard lattice,
$\rho\left(\Sigma_{K3}\right)$,
and to that of the transcendental one%
\footnote{Note, however, that none of the forms in the
transcendental lattice are of purely $(1,1)$-type. }:
 \begin{eqnarray}
 20=h^{(1,1)}&=& \rho(\Sigma_{K3})\, + \, n_{\rm tr}
 \nonumber\\
  \mbox{rank}\, Pic \left(
  \Sigma_{K3} \right)&=&\rho(\Sigma_{K3})\nonumber\\
  \mbox{rank} \, \Lambda^{\rm tr} \left(
  \Sigma_{K3} \right )&=&n_{\rm tr}+2\ .
 \label{algh11}
 \end{eqnarray}

If we use\footnote{We skip, however, some subtle points, see \cite{Aspinwall}.)}
the general ideas of mirror symmetry as it is
formulated, for instance, in the context of toric geometry (see e.g.\ \cite{hosono})
to each algebraic representation $\Sigma_{K3}$ we can associate a mirror
one $\Sigma^*_{K3}$ in which the number of algebraic and
non-algebraic deformations are interchanged: $n_{\rm tr} \leftrightarrow
\rho(\Sigma_{K3})$.
For instance, we can take the algebraic representation $X_4[1,1,1,1]$, for
which, as discussed in section~\ref{ss:K3Pictransc}, the Picard number
$\rho(\Sigma_{K3}) = 1$ and $n_{\rm
tr}=19$, and we can mode it by the action of a group
$G' = \ZZ_4^2$ of identifications
that acts on the homogenous coordinates
as in \eqn{ideqnK3},\eqn{identK3}.
The resulting manifold is the mirror algebraic representation
$ {X}^*_4[1,1,1,1]$, with $\rho(\Sigma_{K3}) = 19$ and $n_{\rm tr}=1$.
Indeed the space of parameters of the defining quartic polynomial $W$,
invariant under $G'$, is one dimensional, as described in section~\ref{ss:descrCY2}, see
\eqn{K31111}.
As discussed in previous sections, ${X}^*_4[1,1,1,1]$ is the $K3$ fibre for the
CY 3-fold $X^*_8[1,1,2,2,2]$ and it is
 the first of the two examples of algebraic representations
we have studied in this paper. The second example is
provided by the $2$-moduli algebraic representation in
\eqn{WW121146}.
In this case the identification group that restricts the available deformations
to those listed above is the $\ZZ_{6}$ group
acting on the homogeneous coordinates as
described in \eqn{neuqnK3} and \eqn{neuntK3}.
Using the methods of toric geometry one can regard \eqn{WW121146} as
the mirror manifold of the algebraic surface ${{X}}_{12}[1,1,4,6]$
with no quotiening.
\par
As already mentioned, our main interest has been in the period vector
$\vec\theta = (\theta_1,\ldots \theta_{n_{\rm tr}+2})$
containing the periods of the holomorphic $2$--form $\Omega^{(2,0)}$ along
a basis of transcendental cycles.
The period vector $\vec\theta$ has thus $2+n_{\rm tr}$ components, while
it depends on $n_{\rm tr}$ parameters, the parameters of the defining
polynomial, that we name `algebraic complex structure
deformations'.
As a consequence of \eqn{k3comstruc} and of the
constraint in \eqn{interfono},
which has signature $(2,n_{\rm tr})$, we can immediately
conclude that the moduli space of algebraic complex structure
deformations is given by
\begin{equation}
{\cal M}_{\rm c}^{\rm alg} =
\frac{O(2,n_{\rm tr} )}{O(2) \times O(n_{\rm tr})}/O(2,n_{\rm tr};\ZZ)~,
\label{algcomstruc}
\end{equation}
where
$O(2,n_{\rm tr};\ZZ)$ is named the duality group. It is obtained,
according to a general analysis
reviewed for instance in  \cite{fresoriabook} as the
semidirect product of the duality group of the potential $\Gamma_{W}$
with the monodromy group $\Gamma_{\rm M}$ we  derive from the Picard--Fuchs
equations.
Following a different way of reasoning, we could have anticipated
that  the $n_{\rm tr}+2$ periods  \eqn{transcper}
should satisfy some constraint
equation from the very fact that they depend only on $n_{\rm tr}$
moduli parameters.
The form of this constraint equation is fixed by the {\it a priori\/}
knowledge of the local geometry of the moduli space encoded in
\eqn{algcomstruc}. Indeed, as noted above, \eqn{interfono} is fully
equivalent to the statement that the moduli space of algebraic complex
structure variations is the coset manifold \eqn{algcomstruc}.
\subsection{The intersection matrix of the transcendental 2-cycles and the definition
of the duality group
$O(2,n_{\rm tr},\ZZ)$}
If we consider a generic $K3$ manifold, the complete second
homology group $H_2 (K3,\ZZ)$ with integer coefficients admits
a basis of 22 integral cycles $e_A$. Their intersection matrix is
\begin{equation}
C^{full}_{K3}\equiv e_A \, \cap \, e_B = \left( \matrix{ \matrix{
 \sigma_1 & 0 & 0 \cr
 0 & \sigma_1 & 0 \cr
 0 & 0 & \sigma_1 \cr} &  {\bf 0} \cr
{\bf 0} & \matrix{ -C(E_8) & {\bf 0} \cr {\bf 0} & -C(E_8) \cr }\cr } \right)
\ ,\label{e8e8}
\end{equation}
where $\sigma_1$ denotes the standard Pauli matrix
and $C(E_8)$ the Cartan matrix of the $E_8$ Lie algebra.
The above matrix is symmetric and its signature  is clearly $(3,19)$.
Canonical integral homology bases are defined up to transformations
$M$ such that
\begin{equation}
M\,\mbox{ is integral valued}~, \quad M \, C_{K3} \, M^T = C_{K3}~.
\label{so22z}
\end{equation}
The matrices satisfying \eqn{so22z} are the elements of the
relevant $O(3,19,\ZZ)$ group.
\par
When we introduce a specific algebraic representation $\Sigma_{K3}$
the integral homology lattice contains the transcendental
sublattice\footnote{We consider here
the transcendental lattice in the cycles rather than the forms,
the identification between those being made by the Poincar\'e duality.}
$\Lambda^{\rm tr} \subset H_2(K3,\ZZ)$,
and the moduli space of the algebraic deformations
is the following submanifold:
\begin{equation}
{\cal M}_{\rm c}^{\rm alg} =
\frac{O(2,n_{\rm tr})}{O(2) \times O(n_{\rm tr})}
\,/O( 2,n_{\rm tr},\ZZ) \, \subset \,
\frac{O \left( 3,19 \right)}{O(2) \times O\left(1,19\right)}\,
/O \left(3,19,\ZZ \right)
\label{subuman}
\end{equation}
of the complex structure moduli space.
The appropriate embedding \eqn{subuman} of coset manifolds is
determined by the lattice embedding
$\Lambda^{\rm tr} \subset H_2(K3,\ZZ)$.
\par
Now reconsider the constraint \eqn{interfono} for the case of a generic $K3$.
Allowing non-integral changes of bases ${\cal S}$, and naming
$\theta'_I = {\cal S}_I{}^J \, \theta^\prime_J$,
it can be reduced to the form
\begin{equation}
0  = \theta^\prime_I \, \theta^\prime_J \,\eta^{IJ}\qquad
\mbox{with}\qquad  \theta'_I = {\cal S}_I{}^J \,
\theta^\prime_J\qquad \mbox{and}\qquad
\eta = ({\cal S} \, {\cal I} \, {\cal S}^T)^{-1}\ ,
\label{frisby}
\end{equation}
where ${\cal I}$ is ${\cal I}_{IJ}$ with
$I,J=1,\dots , 2+n_{\rm tr}$,
and the inverse intersection matrix $\eta^{IJ}$ is
\begin{equation}
   \eta^{IJ}=
 \left(\matrix{{\matrix{a & 0 \cr 0 & b\cr}} & {  {\bf 0}}\cr {  {\bf 0}}
 & -{\bf h}^{ij} \cr} \right)\ ,
\label{etaint}
\end{equation}
$a,b>0$ being some positive real numbers and
the
$n_{\rm tr} \times n_{\rm tr}$
matrix ${\bf h}^{ij}$
being symmetric with positive-definite signature (i.e.\ positive
eigenvalues).
We recognize in \eqn{frisby} the basic constraint describing the
geometry of the K\"ahler homogeneous manifold
${\cal M}_{\rm c}^{\rm alg}$ of \eqn{subuman}
in Calabi--Visentini coordinates \cite{calvis}.
The standard solution to this constraint, often considered in the literature
\cite{PietroSKGlect,f0art,monodrcy} is given by
\begin{equation}
 {\vec \theta}= f(\mu)
  \, \left ( \matrix{ \frac{1}{2\sqrt{a}}\left(1+{\vec Y}^2 \right ) \cr
\frac{{\rm i}}{2\sqrt{b}}(1-{\vec Y}^2) \cr
 Y_i \cr }\right )   \qquad ; \qquad \begin{array}{l}
 i=1, \dots, n_{\rm tr}\\
 {\vec Y}^2 \, \equiv \, Y_i Y_j \,  h^{ij}
 \end{array}\ .
 \label{bellissima}
\end{equation}
 $f(\mu)$ is some overall holomorphic function  of the $n_{\rm tr}$
moduli coordinates  $\mu$ and henceforth, by inversion,
also of the functions $Y^i(\mu)$.
These latter can be utilized as a set of {\it special coordinates\/}
for the moduli manifold, in place of the original ones $\mu$.
The K\"ahler potential of
$O(2,n_{\rm tr})/\left(O(2)\times O(n_{\rm tr})\right)$
is given by
\begin{equation}
{\cal K}_{\rm tr} \, = \, \log \left[{\bar \theta}^\prime_I\theta^\prime_J
\, \eta^{IJ} \right]\ ,
\label{kalpot}
\end{equation}
showing that the overall holomorphic factor $f(\mu)$ is immaterial for the
determination of the K\"ahler metric since it corresponds to a
K\"ahler transformation.
\par
The matrix ${\cal G}$ in \eqn{matcalG}, that relates the algebraic and
transcendental cycles to the basis $e_A$ of $H_2(K3,\ZZ)$,
is integral. It transforms the
intersection matrix to
\begin{equation}
{\cal G}\, C_{K3} \, {\cal  G}^T = {\hat C}_{K3}=
\left( \matrix { {\cal I} & 0 \cr 0 & \star
\cr }\right) \ .
\end{equation}
The elements $M \in O(3,19,\ZZ)$, defined by  \eqn{so22z},
are mapped into matrices ${\hat M} = {\cal G}\, M \, {\cal G}^{-1}$
that satisfy
\begin{equation}
{\hat M}\, {\hat C}_{K3} \, {\hat M}^T = {\hat C}_{K3} \ .
\label{newso22}
\end{equation}
When ${\cal G}$ has determinant $\pm 1$ all the ${\hat M}$ matrices
are integral.
If $|\det{\cal G}|>1$, only a subset of them is integral.
\par
The definition of the duality group
$\Gamma_D=O(2,n_{\rm tr},\ZZ)$
which is relevant for our discussion is provided by identifying such a group
with the set of {\it integral\/}
$(n_{\rm tr}+2)\times (n_{\rm tr}+2)$
matrices $K$ satisfying the equation
\begin{equation}
{\cal I} = K\, {\cal I} \, K^T~.
\label{o2hzeta}
\end{equation}
We obtain an intersection $\gamma_D\cap O(3,19,\ZZ)$ by selecting
those matrices $M\in O(3,19,\ZZ)$ that, after the
transformation with ${\cal G}$ have the following form:
\begin{equation}
\hat M \equiv {\cal G}M{\cal G}^{-1}=
\left( \matrix { m & {\bf 0}\cr  {\bf 0}  & \bfone}
\right)~, \quad
\mbox{with $m$ integral and s.t.
${\cal I}= m {\cal I} m^T$}~.
\label{o2hino319}
\end{equation}
The monodromy matrices corresponding to the various algebraic
representations of the $K3$ surface belong to such an intersection.
Indeed, on one hand they must transform integral cycles into integral
cycles, so that they must be in $O(3,19,\ZZ)$. On the other hand, they
must transform integral transcendental cycles into integral
transcendental cycles, so that they must belong to
$O(2,n_{\rm tr},\ZZ)$.
\par
For our 2 examples, $X^*_{4}[1,1,1,1]$  and $X^*_{12}[1,1,4,6]$,
the transcendental lattices are of rank~3 and 4, respectively.
We can embed the transcendental lattices in the integral lattice spanned by the
first four entries in \eqn{e8e8}. Indeed, define the
matrices
\begin{equation}
{\cal G}_{(1)}=\pmatrix{1&0&0&0\cr 0&1&0&0\cr 1&-2&2&1\cr 0&0&-2&1}\ ;\qquad
{\cal G}_{(2)}=\pmatrix{1& -1& 0& 1\cr 0& 0& 1& -1\cr 0& 0& 0& 1\cr 0& -1& 0& 0}\ .
\end{equation}
These transform the upper $4\times 4$ block of the intersection matrix to
\begin{equation}
{\cal G}_{(1)} \pmatrix{\sigma_1&0\cr 0&\sigma_1}{\cal G}_{(1)}^T=\pmatrix{
{\cal I}_{(1)}& 0\cr 0&  -4}\ ;\qquad
{\cal G}_{(2)} \pmatrix{\sigma_1&0\cr 0&\sigma_1}{\cal G}_{(2)}^T={\cal I}_{(2)}\ ,
\end{equation}
where ${\cal I}_{(1)}$ and ${\cal I}_{(2)}$ are the intersection
matrices which we found in the main text
for the one-modulus and two-modulus $K3$ manifolds in an integral basis, given in
\eqn{intersectionmatrix1} and \eqn{SU3int}, respectively.
The Picard lattice consists of the 18 basis vectors not used in this
transformation, and in the first example it also has the fourth basis vector
after the transformation.
The transformations which we gave are not unique.
For instance, we have also found another embedding of ${\cal I}_{(2)} $ in the full
$K3$ intersection matrix, involving basis vectors corresponding to an $E_8$
Cartan matrix in \eqn{e8e8}. Note that in both cases the intersection matrix
has two positive eigenvalues.
\par
A difference between the two examples is the determinant of ${\cal G}$. For the
first example the determinant is 16, while for the second one it is 1.
Therefore in the second example the transcendental and algebraic lattices have
as direct sum  the full lattice $H^2(K3,\ZZ)$, while this is not the case for
the first example, where some elements of $H^2(K3,\ZZ)$ can not be written as a
linear combination of the basis vectors of both lattices with integer
coefficients.
\subsection{Embedding  the monodromy group into
$O(2,n_{\rm tr},\ZZ)$ and special features
of the cases
$n_{\rm tr}=1,2$}
Given an explicit algebraic representation $\Sigma_{K3}$,
the Picard--Fuchs approach  to be discussed in appendix~\ref{ss:permonK3}
leads to a system of linear partial differential equations in
$n_{\rm tr}$ variables that is satisfied
by each component of the $2+n_{\rm tr}$-dimensional period vector
$\vec\theta$.
Hence the period vector can be identified with a {\it basis of solutions\/}
to this system. The analytic continuation of a particular solution
around each singular point of the differential system
produces the monodromy group generators. Such monodromy
generators will be integer-valued pseudo-orthogonal matrices
when the solution basis is
brought to the integral basis where the intersection matrix is
${\cal I}$. Consequently, once a general solution of the
differential system has been obtained we have to look for changes of
bases that bring the monodromy matrices to be integral and such that
they leave some suitable intersection matrix ${\cal I}$ invariant.
The form of ${\cal I}$ is not determined by the Picard--Fuchs differential
system so that we have to resort to different sources of
information. In
the main text
we have  obtained  ${\cal I}$ by a direct
construction of the basis of transcendental cycles.
\par
Summarizing our discussion,
the Picard--Fuchs problem in the case of
the $K3$ fibre reduces to  a uniformization problem.
Differently from the case of Calabi--Yau 3-folds where even the
local geometry of the moduli space is not known {\it a priori\/} (apart from
the fact that it is of the {local special K\"ahler} type) so that the solution
of the Picard--Fuchs equations not only provides a basis of special
coordinates but also determines the form of the metric, in the $K3$ case the
moduli space is known {\it a priori}, but what we have to determine is the
relation between the  special coordinates $Y^i$ (obtained by solving the
algebraic constraint satisfied by the periods) and
the `Landau--Ginzburg' coordinates $\mu$ appearing explicitly in the
potential.

For
$n_{\rm tr} \ge 3$
there is nothing more that can be stated
in general. However, for
$n_{\rm tr} = 1$ and $n_{\rm tr} = 2$
there is some more {\it a~priori} information due to the accidental
isomorphisms
\begin{equation}
\label{iso1}
\frac{SO(2,1)}{SO(2) }\sim \frac{SL(2,\IR) }{SO(2)}
\sim\frac{SU(1,1) }{U(1)}
\end{equation}
and\footnote{The isomorphisms between $SL(2,\IR)$ and $SU(1,1)$ is simply
realized on doublets $(P_1,P_2)$ of the former and
$(U_1,U_2)$ of the latter by the Cayley transformation
$U_1 = P_1 + \ii P_2$ and $U_2 = P_1 - \ii P_2$.
}
\begin{equation}
\label{iso2}
\frac{SO(2,2)}{SO(2) \otimes SO(2)}\sim
\frac{SL(2,\IR)_1}{SO(2)} \otimes \frac{SL(2,\IR)_2}{SO(2)}\sim
\frac{SU(1,1)_1}{U(1)} \otimes \frac{SU(1,1)_2}{U(1)}~.
\end{equation}

In the case
$n_{\rm tr} = 1$, as in our first example of algebraic
$K3$ manifold $X^*_4[1,1,1,1]$,
we learn from this observation
that the period vector $\vec\theta$, which transforms as a triplet of
$SO(2,1)$ must be expressible in terms of two functions
$(U_1,U_2)$ in the doublet of $SU(1,1)$. Of course, the triplet of  $SO(2,1)$,
i.e.\ the $J=1$ representation, corresponds to a symmetric product of the
fundamental representation of $SU(1,1)$, the doublet, with itself.
Let us be explicit. In the case $n_{\rm tr}=1$, it is convenient
to choose in the inverse
intersection matrix $\eta$ of (\ref{etaint}) $a=b=1$ and $h=1$;
the period vector is given by \eqn{bellissima} with these specific choices.
Explicitly if the transformation of the doublet vector is
\begin{equation}
\label{k3t242bis}
\left(\begin{array}{c}U_1\\U_2\end{array}\right)
\rightarrow \left(\matrix{\alpha & \beta\cr \gamma & \delta}\right)
\left(\begin{array}{c}U_1\\U_2\end{array}\right) \quad ; \quad
\left(\matrix{\alpha & \beta\cr \gamma & \delta}\right) \, \in  \,
SU(1,1)
\end{equation}
the transformation of triplet ${\vec \theta}$ will be
\begin{equation}
\left(\begin{array}{c} \theta_1\\ \null \\ \theta_2\\ \null \\ \theta_3\end{array}\right)
\rightarrow
\left(\begin{array}{ccc}
{1\over 2} (\alpha^2 + \beta^2 + \gamma^2 + \delta^2) & {\ii\over 2}
(\alpha^2 - \beta^2 + \gamma^2 - \delta^2)
& \alpha \beta + \gamma \delta \\ \null & \null & \null \\
{-\ii\over 2}(\alpha^2 + \beta^2 - \gamma^2 - \delta^2) & {1\over 2}
(\alpha^2 - \beta^2 - \gamma^2 + \delta^2)
& -\ii (\alpha\beta - \gamma\delta) \\ \null & \null & \null \\
\alpha \gamma + \beta \delta & \ii(\alpha \gamma - \beta \delta) & \alpha \delta +
\beta \gamma\end{array}\right)
\left(\begin{array}{c} \theta_1\\ \null \\ \theta_2\\ \null \\
\theta_3\end{array}\right)\ .
\label{Abfmat}
\end{equation}
This amounts to expressing $\vec\theta$ in terms of $\vec U$ as follows:
\begin{equation}
\vec\theta =
f(\mu) \, \left ( \matrix{ \frac{1}{2}\left(1+{\vec Y}^2 \right ) \cr
 \frac{{\rm i}}{2}\left(1-{\vec Y}^2 \right )  \cr
 Y  \cr }\right ) = \left(\begin{array}{c} {1\over 2} (U_2^2 + U_1^2) \\
{\ii \over 2} (U_2^2 - U_1^2) \\ U_1 U_2\end{array}\right)~.
\label{predirelaz1}
\end{equation}
We have seen in the main text  how the above relation is
retrieved while computing the periods directly.
Indeed, the reader should compare the present discussion
with \eqn{ig0} and \eqn{tu4}. We can retrieve the same relation
by solving the third order Picard--Fuchs
equation. This will be done in appendix~\ref{ss:permonK3}.
In that process we will also discover the meaning of the functions
$(U_1,U_2)$. They form a basis of solutions for a second-order
hypergeometric equation to which we are able to reduce the original
third-order one. The monodromy of this hypergeometric equation will
be generated  by $2\times 2$ matrices belonging to the $SU(1,1)$ group.
Their $3\times 3$ image via the map \eqn{predirelaz1} consists
of integer-valued $O(2,1)$ matrices.
\par
Let us now turn our attention to the
$n_{\rm tr} = 2$ case, which applies to our second example, $X^*_{12}[1,1,4,6]$.
In this case the appropriate choices for the integers
$a,b$ and for the $2 \times 2$ matrix
${\bf h}_{ij}$ are
\begin{equation}
\mbox{2 modulus $K3$ } \quad : \quad  a=2, \, b=6, \, \quad
{\bf h}_{ij}=\left(\matrix{ 2 & -1\cr -1 & 2}\right)
=\mbox{ $A_2$ Cartan matrix.}
\label{hmat2}
\end{equation}
Indeed, consider the
 transcendental intersection matrix ${\cal I}_{(2)}$, i.e.\ \eqn{SU3int}.
If we change basis with the
matrix
\begin{equation}
{\cal N}= \left (
\matrix{ 1 & 1 & 1 & -1 \cr -1 & 1 & 3 & 3 \cr 1
   & 0 & 0 & 0 \cr 0 & 1 & 0 & 0 \cr  }
\right )\ ,    \label{cNpseuoort}
\end{equation}
the intersection form assumes the Calabi--Visentini form \eqn{etaint},
with the choices \eqn{hmat2}:
\begin{equation}
\eta^{(2)} \equiv {\cal N} \, {\cal I} \,{\cal N}^T = \left(
\begin{array}{rrrr}
2 & 0 & 0 & 0 \\
0 & 6 & 0 & 0 \\
0 & 0 & -2 & 1 \\
0 & 0 & 1 & -2
\end{array}
\right)~.\label{intortsect}
\end{equation}
In this case, however, the Calabi--Visentini basis cannot be an integral
basis for the transcendental lattice. Indeed,
the determinant of ${\cal N}$ is 6; therefore it does not
transform elementary cycles to a new basis of elementary cycles.
So in the basis \eqn{intortsect} the monodromy generators
are not necessarily integral.
In fact, one can show that there does not exist an integer basis
transformation matrix $M$ with integer inverse, i.e.\ with $\det M = \pm 1$,
which transforms the intersection matrix  ${\cal I}_{(2)}$ in
 \eqn{SU3int} to one of the form \eqn{etaint}.
\par
Though it is not an integral basis, the  Calabi--Visentini basis defined by
 \eqn{intortsect} is a perfectly appropriate
basis to discuss the solution of Picard--Fuchs equations and their
group-theoretical structure. Once that is done it
suffices to change back basis with
the matrix ${\cal N}$ and all the considerations we are about to make
are transferred to an integral basis.
\par
This being clarified, we realize the real pseudo-orthogonal group
 $O(2,2,\IR)$  as the set of real matrices $M$ satisfying the equation:
\begin{equation}
M  \, \eta^{(2)} \, M^T \, = \,  \eta^{(2)}
\label{intorteta}
\end{equation}
The integral subgroup $O(2,2,\ZZ)\subset O(2,2,\IR)$ is, however, given by
those $M$ that are integral in the integral basis in which the intersection
is ${\cal I}_{(2)}$: that is, ${\cal N}\, M \, {\cal N}^{-1}$ must be integral.
\par
Relying on the isomorphism \eqn{iso2}
we infer from the fact that the four-dimensional fundamental
representation of $O(2,2)$ is
given by the symmetric tensor product of the two fundamental representations
of $SL(2,\IR)_1$ and $SL(2,\IR)_2$ that the period vector $\vec\theta$
should be expressible in terms of two doublets $(P_1,P_2)$ and $(Q_1,Q_2)$
of the two factors.

The explicit group isomorphism is expressed by mapping
\begin{equation}
\label{primamappa}
\left( \matrix{ a_1 & b_1\cr c_1 & d_1\cr} \right) \,  \otimes
 \, \left( \matrix{ 1 & 0\cr 0 & 1\cr} \right)  \, \in   SL(2,\IR)_1
 \,  \otimes \, SL(2,\IR)_2
\end{equation}
into the $O(2,2,\IR)$ element
\begin{equation}
\label{pmbis}
\left (
\matrix{ {{{  a_1} + {  d_1}}\over 2} &
  {{{\sqrt{3}}\,\left( -{  b_1} + {  c_1} \right) }\over 2} &
  {{-{  a_1} - {\sqrt{3}}\,{  b_1} -
      {\sqrt{3}}\,{  c_1} + {  d_1}}\over 4} &
  {{{  a_1} - {  d_1}}\over 2} \cr
  {{{  b_1} - {  c_1}}\over {2\,{\sqrt{3}}}} &
  {{{  a_1} + {  d_1}}\over 2} &
  {{-3\,{  a_1} + {\sqrt{3}}\,{  b_1} +
      {\sqrt{3}}\,{  c_1} + 3\,{  d_1}}\over
    {12}} & {{-\left( {  b_1} + {  c_1} \right)
      }\over {2\,{\sqrt{3}}}} \cr
  -{{{  b_1} + {  c_1}}\over {{\sqrt{3}}}} &
  -{  a_1} + {  d_1} &
  {{3\,{  a_1} - {\sqrt{3}}\,{  b_1} +
      {\sqrt{3}}\,{  c_1} + 3\,{  d_1}}\over 6}
   & {{{  b_1} - {  c_1}}\over {{\sqrt{3}}}}
   \cr {{3\,{  a_1} - {\sqrt{3}}\,{  b_1} -
      {\sqrt{3}}\,{  c_1} - 3\,{  d_1}}\over 6}
   & {{-{  a_1} - {\sqrt{3}}\,{  b_1} -
      {\sqrt{3}}\,{  c_1} + {  d_1}}\over 2} &
  {{-{  b_1} + {  c_1}}\over {{\sqrt{3}}}} &
  {{3\,{  a_1} + {\sqrt{3}}\,{  b_1} -
      {\sqrt{3}}\,{  c_1} + 3\,{  d_1}}\over 6}
   \cr  }  \right )~,
\end{equation}
and
\begin{equation}
\label{secondamappa}
\left( \matrix{ 1 & 0\cr 0 & 1\cr} \right) \,  \otimes
 \, \left( \matrix{ a_2 & b_2\cr c_2 & d_2\cr} \right) \, \in   SL(2,\IR)_1
 \,  \otimes \, SL(2,\IR)_2
\end{equation}
into the $O(2,2,\IR)$ element
\begin{equation}
\label{smbis}
\left(
\matrix{ {{{ a_2} + { d_2}}\over 2} &
  {{{\sqrt{3}}\,\left( -{  b_2} + {  c_2} \right) }\over 2} &
  {{{  a_2} - {\sqrt{3}}\,{  b_2} -
      {\sqrt{3}}\,{  c_2} - {  d_2}}\over 4} &
  {{-{  a_2} + {  d_2}}\over 2} \cr
  {{{  b_2} - {  c_2}}\over {2\,{\sqrt{3}}}} &
  {{{ a_2} + { d_2}}\over 2} &
  {{-3\,{  a_2} - {\sqrt{3}}\,{  b_2} -
      {\sqrt{3}}\,{  c_2} + 3\,{  d_2}}\over
    {12}} & {{{  b_2} + {  c_2}}\over
    {2\,{\sqrt{3}}}} \cr
  -{{{  b_2} + {  c_2}}\over {{\sqrt{3}}}} &
  -{  a_2} + {  d_2} &
  {{3\,{  a_2} + {\sqrt{3}}\,{  b_2} -
      {\sqrt{3}}\,{  c_2} + 3\,{  d_2}}\over 6}
   & {{-{  b_2} + {  c_2}}\over {{\sqrt{3}}}}
   \cr {{-3\,{  a_2} - {\sqrt{3}}\,{  b_2} -
      {\sqrt{3}}\,{  c_2} + 3\,{  d_2}}\over 6}
   & {{-{  a_2} + {\sqrt{3}}\,{  b_2} +
      {\sqrt{3}}\,{  c_2} + {  d_2}}\over 2} &
  {{{  b_2} - {  c_2}}\over {{\sqrt{3}}}} &
  {{3\,{ a_2} - {\sqrt{3}}\,{ b_2} +
      {\sqrt{3}}\,{  c_2} + 3\,{  d_2}}\over 6}
   \cr  }\right )~.
\end{equation}
The two matrices above
commute, and  their product is a generic element of $O(2,2)$.
This embedding fixes the relation between the period vector
and the two doublet basis:
\begin{eqnarray}
 {\vec \theta}&=& f(\mu) \, \left ( \matrix{ \frac{1}{2\sqrt{2}}\left(1+2
 Y_1^2+ 2 Y_2^2 - 2 Y_1 Y_2\right ) \cr
 \frac{{\rm i}}{2\sqrt{6}}\left(1-2 Y_1^2- 2 \, Y_2^2
 + 2 Y_1 Y_2\right ) \cr
 Y_1 \cr
 Y_2 \cr }\right ) \nonumber\\
&= &
 \left(\matrix{ \frac{1}{\sqrt{2}} & 0 & 0 & 0 \cr 0 & \frac{1}{\sqrt{6}} & 0 & 0 \cr 0 & 0 &
  {\sqrt{{2\over 3}}} & 0 \cr 0 & 0 &
  {1\over {{\sqrt{6}}}} & {1\over {{\sqrt{2}}}} \cr  }\right) \,
 \left(\matrix{ - \frac{1}{2}( P_1 Q_2 + P_2  Q_1 ) \cr
 - \frac{1}{2}( P_1  Q_1 - P_2  Q_2 ) \cr
 \frac{1}{2}( P_1 Q_1 + P_2  Q_2 ) \cr
 - \frac{1}{2}( P_1  Q_2 - P_2  Q_1 ) \cr }\right )~.
 \label{tesoro}
\end{eqnarray}
Going to the $SU(1,1)_1 \otimes SU(1,1)_2$ basis of two doublets $(U_1,U_2)$
and $(V_1,V_2)$, a little algebra shows that
\eqn{tesoro} provides the identifications
\begin{eqnarray}
f &=& \frac{\rm i}{2}  \, U_1 \, V_1~, \nonumber\\
\left(\matrix {Y^1 \cr Y^2\cr}\right) &=& \frac{1}{f(\mu)}
\left(\matrix { {\sqrt{{2\over 3}}} & 0 \cr
{1\over {{\sqrt{6}}}} & {1\over {{\sqrt{2}}}} \cr  }\right) \,
\left(\matrix {  \frac{  1}{4} \, \left(
U_2 \, V_1  + \, U_1 \, V_2 \right)\cr
  \frac{\rm i}{4} \, \left(   \, U_2 \, V_1
-  \, U_1 \, V_2\right) \cr}\right)
\label{ispira}
\end{eqnarray}
\par
In the next section we shall also retrieve these
identifications from the solutions of the Picard--Fuchs equations
obtaining the functions $(U_1,U_2)$, $(V_1,V_2)$ in terms of the
solutions for a pair of second-order differential equations.

\section{Picard--Fuchs equations}
\label{ss:permonK3}
There are several ways to obtain the differential equations on the periods. One
way would be to consider first the derivatives of fundamental (2,0)-form
$\Omega^{(2,0)}$ with respect to the moduli. This can be expressed in terms of
the $(1,1)$ forms. Then the derivatives of the (1,1) and (0,2) forms can be expressed
again in a basis of all 2-forms. These differential equations can be combined
to give third-order differential equations on the fundamental $(2, 0)$ form. We
have
followed this procedure at first, and the results are the same as the
differential equations which we will present below.
\par
We will present a simpler method, which is inspired by the one used in toric
geometry \cite{hosono}. We will, however, not need the full machinery of toric
geometry and arrive at lower-order differential equations. A main part of the
method consists in keeping all moduli non-gauge fixed. That allows one to take
derivatives with respect to them which can be compared. The derivatives with
respect to these parameters are then expressed as derivatives with respect to
the invariant variables. To do so, we should first take a representation of the
(2,0) form which is gauge invariant. In the Griffiths representation of this
form, $W$ is gauge invariant, but the volume form $\omega$ is not invariant
under the rescalings (it is invariant under the shifts respecting the
weights). Therefore we should multiply first with a function of the moduli
which compensates for the scaling weights of $x_0x_3x_4x_5$.

\subsection{First example}
\subsubsection{The Picard--Fuchs equation}
In the first example the scaling weights of the volume form can be
compensated for by a factor $\psi_0$. Thus we consider
\begin{equation}
\hat \theta(\mu)=\int\hat \Omega^{(2,0)}\propto
\int\frac{\psi_0}{W^{(1)}(B',b_3,b_4,b_5)}\omega_{K3}\ ,
\end{equation}
where $W^{(1)}$ is given in \eqn{K3-I} and $\mu$ is the
truly invariant variable
\begin{equation}
\mu=\frac{\psi_0}{\left( B' b_3 b_4 b_5\right) ^{1/4}} \ ;\qquad
z=-\mu^{-4} \ ,  \label{defmu}
\end{equation}
for which the global identifications $\mu\sim \ii \mu\sim -\mu \sim -\ii\mu$
remain to be made.
It is now straightforward to obtain {\it Picard--Fuchs equation\/} by writing
\begin{equation}
4^4\frac{\partial}{\partial B'}
\frac{\partial}{\partial b_3}
\frac{\partial}{\partial b_4}
\frac{\partial}{\partial b_5}\hat\vartheta(\mu)=\left(
\psi_0\frac{\partial^4}{\partial \psi_0^4}\right)   \frac{1}{\psi_0}\hat\vartheta(\mu) \ .
\label{toricPF1}
\end{equation}
On functions of $\mu$ we have
\begin{eqnarray}
&&B'\frac{\partial}{\partial B'}=
b_3\frac{\partial}{\partial b_3}=
b_4\frac{\partial}{\partial b_4}=
b_5\frac{\partial}{\partial b_5}= -\frac{\mu}{4}\frac{\partial}{\partial \mu}
\nonumber\\
&&  \frac{\partial}{\partial \psi_0} =
\frac{1}{\left( B' b_3 b_4 b_5\right) ^{1/4}}\frac{\partial}{\partial \mu}\ .
\end{eqnarray}
Therefore \eqn{toricPF1} reduces to
\begin{equation}
\left( \mu\frac{\partial}{\partial \mu}\right) ^4\hat\vartheta=
\mu\frac{\partial^4}{\partial \mu^4}\frac{1}{\mu}\hat\vartheta\ .
\label{equaquart}
\end{equation}
If we introduce the rescaled period $\vartheta (\mu) = \hat\vartheta(\mu)/\mu$, it
can be immediately verified that the above \eqn{equaquart}
 can be rewritten as the operator $\mu{\partial\over\partial\mu}$ acting over
the third-order equation:
\begin{equation}
\label{k3t224a}
\biggl({\dop^3\over \dop \mu^3} - {6\mu^3\over 1 - \mu^4}
{\dop^2\over \dop \mu^2} - {7\mu^2 \over 1 - \mu^4}
{\dop\over\dop \mu} - {\mu\over 1 - \mu^4}\biggr) \vartheta (\mu) = 0\ .
\end{equation}
Hence we have obtained just in one simple stroke the third-order
equation that one can derive through the traditional method of considering
the first-order system satisfied by the periods.
\subsubsection{Solutions in terms of hypergeometric functions.}
Equation~(\ref{k3t224a}) is a Fuchsian equation, with regular singular
points located at infinity and in the zeros of the denominator,
i.e.\ for $\mu^4 = 1$. It admits obviously three linearly independent
solutions.
The monodromy generators corresponding to non-trivial paths encircling
the singular points have a matrix action on the chosen 3-vector of solutions
$\vec \vartheta (\mu)$:
$ \vec \vartheta \rightarrow M_P \vec \vartheta  $,
where $P$ is one of the singular points.
Moreover, the $K3$ defining potential is invariant under the $\ZZ_4$ symmetry
generated by $\mu \rightarrow \ii \mu$, that
acts on the solution vector as a $3\times 3$
matrix satisfying $A^4=\unity_{3\times 3}$.
\par
It was observed in \cite{LSW} that the solution of (\ref{k3t224a})
can be expressed in terms of the solution of a second-order
differential equation, due to the crucial fact that the so-called
$W_3$-invariant of the third-order equation vanishes. It turns thus
out that a generic solution of (\ref{k3t224a}) is of the form
\begin{equation}
\label{add1}
\vartheta (\mu) = {1\over \sqrt{1 - \mu^4}} \left({\dop Y\over\dop\mu}
\right)^{-1}
\hskip 0.1cm \left(a + b Y + c Y^2\right)~,
\end{equation}
where $a,b,c$ are constants and  $Y(\mu)$ satisfies the Schwarzian equation
\begin{equation}
\label{k3t237}
\sch Y \mu = {\mu^2\over 2} {\mu^4 + 11\over \left(1 - \mu^4\right)^2}~.
\end{equation}
This property of the PF equation is related geometrically to the fact
that any basis $\theta_I$ of transcendental periods of the $K3$
(i.e.\ any basis of solutions to the PF equation) satisfy the quadratic
constraint \eqn{interfono},
where ${\cal I}^{IJ}$ is the inverse intersection matrix of the
transcendental cycles, with signature $(2,1)$. As explained in
appendix~\ref{app:k3fibremodulispace}, this
relation basically tells us that the periods $\theta_I$
(over-)parametrize, in the guise of generalized Calabi--Visentini
coordinates, the moduli space of the transcendental cycles that,
as well known \cite{Aspinwall} has the form\footnote{$\Gamma_D$
is the duality group (formed by the monodromies and the symmetry of the
potential). It is integer-valued in an integral basis; we give
explicit matrices later.}
$\Gamma_D\backslash O(2,1)/O(2)$. Indeed the quadratic constraint (\ref{interfono})
implies that only one (and not two) of the ratios
of periods  are independent, in agreement with \eqn{add1}.
\par
In the $z$-variable \eqn{defmu},
using \cite{Bateman} 2.7.(11), equation~(\ref{k3t237}) becomes
\begin{equation}
\label{nmon1}
\sch Y z =
{1\over 2}{1\over z^2} + {3\over 8} {1\over (z+1)^2} - {13\over 32}
{1\over z(z+1)} = 2 I(0,{1\over 4},{1\over 2};-z)
\end{equation}
where we use the notation of \cite{Bateman}, equation~2.7(9), and $Y$ is a
Schwarzian
function $s(0,{1\over 4},{1\over 2};-z)$,
and can therefore be expressed
as $Y = {y_1\over y_2}$, where $y_{1,2}(z)$ are two linear independent
solutions of the Fuchsian equation
\begin{equation}
\label{nmon2}
{{\rm d}^2 y\over {\rm d} z^2} + I(0,{1\over 4},{1\over 2};-z)y = 0~.
\end{equation}
We also have
\begin{equation}
\label{nmon3}
y_{1,2} =(- z)^{1/2} (1 + z)^{1/4} U_{1,2}(z)\ ,
\label{pippovede}
\end{equation}
where $U_{1,2}$ are two linearly independent solutions of the
hypergeometric equation of parameters $\{{1\over 8},{3\over 8},1\}$.
Recalling that
the wronskian $y_1' y_2 - y_2' y_1$ is a constant since the first
derivative is absent in \eqn{nmon2},
it follows from \eqn{add1} and \eqn{nmon3}, that the rescaled period
$\hat\vartheta = \mu \vartheta $, has the generic form
\begin{equation}
\hat \vartheta(z) = C_{\alpha\beta} U_\alpha(z)\, U_\beta(z)~,\hskip 0.5cm
(\alpha,\beta=1,2)\ ,
\label{pippovisto}
\end{equation}
As promised, by means of \eqn{pippovisto} we have retrieved
the relation between the
triplet and the doublet representation anticipated
in \eqn{predirelaz1} via group-theory arguments. Indeed the
$3 \times 3$ $O(2,1)$-monodromy matrices of the third-order equation
are nothing else but the image, in the triplet representation, of the
$2 \times 2$ $SU(1, 1)$-monodromy matrices of the associated
hypergeometric equation. The construction of the third-order equation
solutions as tensor products of the hypergeometric equation solutions
is nothing else but a necessary consequence of this fact.
\subsection{Second example} \label{ss:PFex2}
\subsubsection{The Picard--Fuchs equation.}
In this case as invariant $(2,0)$ form we take the following one (with $W^{(2)}$ as in
\eqn{K3genpolyn2}):
\begin{equation}
\hat g(\nu_1,\nu_2)= \int
{B'}^{1\over 12}b_3^{1\over 12}b_4^{1\over 3}b_5^{1\over
2}\omega\frac{1}{W^{(2)}}   \ .   \label{inv20ex2}
\end{equation}
A pair of second-order differential equations satisfied by the
periods can be obtained through the  use of the following
relations:
\begin{eqnarray}
6^2\frac{\partial}{\partial\psi_1}  \frac{\partial}{\partial\psi_1}
\frac{1}{W^{(2)}}&=& 12^2 \frac{\partial}{\partial B'}
\frac{\partial}{\partial b_3} \frac{1}{W^{(2)}}  \label{toricPF21}\\
4\frac{\partial}{\partial\psi_4}  \frac{\partial}{\partial\psi_4}
\frac{1}{W^{(2)}}&= &6 \frac{\partial}{\partial \psi_1}
\frac{\partial}{\partial \psi_3} \frac{1}{W^{(2)}}
\ .  \label{toricPF22}
\end{eqnarray}
As alternatives for the second line in \eqn{toricPF22}
one can take various other relations but they all lead to the same final result.
Applying the identities  \eqn{toricPF22} on the invariant definition
\eqn{inv20ex2} of the period one gets the following two equations in
terms of the invariants \eqn{invnu12}:
\begin{eqnarray}
&&\left( \frac1{2}\nu_2^{1/2} \frac{\partial}{\partial \nu_2}\nu_2^{1/2}
\frac{\partial}{\partial\nu_2} \nu_1^{\frac{1}{12}}-
\left( \nu_1 \frac{\partial} {\partial\nu_1}
+\nu_2\frac{\partial}
{\partial\nu_2} \right) ^2 \nu_1^{\frac{1}{4}}\right) \hat g(\nu_1,\nu_2)
\nonumber\\
&&\left( 4\nu_1^{1/3}\frac{\partial}{\partial\nu_1}
\nu_1^{2/3}\frac{\partial}{\partial\nu_1} -
\nu_2^{1/2} \frac{\partial}{\partial\nu_2}
\nu_2^{1/2} \frac{\partial}{\partial\nu_2}
\right) \hat g(\nu_1,\nu_2)\ .\label{pfex2}
\end{eqnarray}
For instance, to obtain  the second of the above
equations from the second of the identities \eqn{toricPF22}
it suffices to fix the gauge \eqn{gaugefibr2}.
\subsubsection{Lian--Yau solution.}
Our goal is that of writing the general solution of such a system.
This can be done by using some results already existing in the literature,
in particular those obtained by Lian and Yau in \cite{lianyau}.
They use two variables $x$ and $z$, related to our $\nu_1$ and $\nu_2$ by
\begin{equation}
\nu_1^{-1}= ( 864\,x)^2 \, z \ ;\qquad \nu_2= 4 \nu_1 \, (-1 + 864\,x)\ .    \label{nuasfxz}
\end{equation}
and the above equations~\eqn{pfex2} reduce to:
\begin{equation}
L_2 \hat \vartheta  =0\ ;\qquad L_1\hat\vartheta  =0   \ ,\label{linyaueq}
\end{equation}
where
\begin{eqnarray}
 L_1 &\equiv& x\frac{d}{dx} \left(x\frac{d}{dx} -2\,
 z\frac{d}{dz}\right)-12 \, x\, \left(6\,x\frac{d}{dx} +5\right)\,
 \left(6\,x\frac{d}{dx} +1 \right ) \nonumber\\
 L_2 &\equiv& \left(z\frac{d}{dz}\right)^2 - z\, \left( 2\, z\frac{d}{dz}
 - x\frac{d}{dx} +1 \right)\left( 2 z\frac{d}{dz}-
 x\frac{d}{dx}\right)\ .
\end{eqnarray}
and \begin{equation}
\hat \vartheta= \nu_1^{1/12}\hat g
\end{equation}
which in the gauge \eqn{usualgauge2} takes the form (using as in the
first example $\hat \Omega^{(2,0)}=\psi_0\Omega^{(2,0)}$)
\begin{equation}
\hat \vartheta=  \int \frac{\psi_0\, \omega_{K3}}{W^{(2)}} \propto \int\hat \Omega^{(2,0)}  \ .
\label{hatvarthetaex2}
\end{equation}
\par
To obtain the solutions of Lian and Yau it suffices to
introduce the following differential operator:
\begin{equation}
L_3(u) \equiv \frac{d}{du}\left(u\frac{d}{du}\right)
-  \left(u\frac{d}{du} +\frac{5}{6}\right)\,
 \left(u\frac{d}{du} +\frac{1}{6} \right )=
 u \, \left(1 - u \right) \frac{d^2}{du^2} +\left(1-2 \, u
\right)\, \frac{d}{du}  - \frac{5}{36}
\end{equation}
and verify that a general solution to the system \eqn{linyaueq} of partial
differential equations is of the  following form
 \begin{equation}
{\hat \vartheta} \left(x,z \right)  =  \left(a_{12} , a_{21}, a_{11}, a_{22} \right) \,
\cdot \, \left(\matrix {\xi_1 ({ r}) \, \xi_2({ s}) \cr
\xi_2 ({ r}) \, \xi_1({ s}) \cr   \xi_1 ({ r}) \, \xi_1({ s}) \cr
 \xi_2 ({ r}) \, \xi_2({ s})
 \cr }\right )
 \label{dirproduc}
\end{equation}
where $\left( a_{12} , a_{21}, a_{11}, a_{22} \right)$ is a set of
arbitrary constants and $\xi_1(u)$ and $\xi_2(u)$
constitute a basis of solutions for the second-order equation
\begin{equation}
\label{lin2equ}
 L_3(u) \, \xi(u) = 0 \ .  \label{hygeomeq}
\end{equation}
The variables $r$ and $s$ are a pair of algebraic functions of the
original variables $x,z$ determined as {\bf any one} of the {\bf four} branches of the
pair of algebraic equations:
\begin{eqnarray}
 r + s -2\, r\, s - 432\,x=0 \ ;\qquad&\mbox{or}&\qquad
 (1-2r)(1-2s)=2\sqrt{\frac{\nu_2}{\nu_1}}\nonumber\\
r\, s\,( 1- r) ( 1-s)- (432\,x)^2 z =0\ ;\qquad&\mbox{or}&\qquad
4\,r\, s\,( 1- r) ( 1-s)=(4\,\nu_1)^{-1}\ .
\label{alegeq}
\end{eqnarray}
Equation~\eqn{hygeomeq} is a hypergeometric equation of parameters $
a=\frac{1}{6}$, $b=\frac{5}{6}$, $c=1$,
which admits as regular solution in the neighbourhood
of $u=0$ the hypergeometric series:
\begin{equation}
\xi(u)= {\xi }_0(u)= \null_2 F_1\left(\frac{1}{6},\frac{5}{6},1 ;
u\right) \ .
\label{regperio}
\end{equation}
A complete set of linear independent solutions of the hypergeometric
equation~\eqn{hygeomeq}  for large values of $u$ is given by
\begin{eqnarray}
 {\xi }_1(u) &=&  B_1 \,  u^{-1/6} \, \null_2 F_1\left(\frac{1}{6},\frac{1}{6},\frac{1}{3};
\frac{1}{u}\right)\nonumber\\
 {\xi }_2(u) &=&  B_2 \,  u^{-5/6} \, \null_2 F_1\left(\frac{5}{6},\frac{5}{6},\frac{5}{3};
\frac{1}{u}\right)   \ ,
\label{hygeobase}
\end{eqnarray}
where $B_1$ and $B_2$ are arbitrary constants.
\par
\subsubsection{Solution of the algebraic equations.}\label{ss:solalgeq}
With a further change of variable:
\begin{equation}
r=\frac{a+1}{2} \quad ; \quad  s=\frac{b+1}{2}\ ,
\label{furthvar}
r=\frac{a+1}{2} \quad ; \quad  s=\frac{b+1}{2}\ ,
\end{equation}
and setting
\begin{equation}
A \equiv \left( \sqrt{\nu_2}+\ft{1}{2}\sqrt{\nu_1}\right)^2 - 1\ ;\qquad
B \equiv \left( \sqrt{\nu_2}-\ft{1}{2}\sqrt{\nu_1}\right)^2 - 1\ .
\label{singpoint}
\end{equation}
the system of equations~\eqn{alegeq} goes into the form
\begin{eqnarray}
(a^2-1)\, (b^2-1) \, A &=& (a+b)^2 \nonumber\\
(a^2-1)\, (b^2-1) \, B &=& (a-b)^2  \ .
\label{vernewal}
\end{eqnarray}
Then with a few elementary manipulations one can obtain the explicit
solutions of the algebraic system \eqn{vernewal}:
\begin{equation}
a=\frac{\sqrt{A}+\sqrt{B}}{\sqrt{A+1}+\sqrt{B+1}}  \ ;\qquad
b=\frac{\sqrt{A}-\sqrt{B}}{\sqrt{A+1}+\sqrt{B+1}}\ ,
\end{equation}
where the sign of $\sqrt{A}$, $\sqrt{B}$, $\sqrt{A+1}$ and $\sqrt{B+1}$
can be taken as arbitrary but the same in the two expressions. In other
words one can take one solution, and
the other possible branches of the solution can be easily obtained
using the symmetries of the algebraic system:
\begin{equation}
r \, \leftrightarrow \, s \quad ; \quad r \, \leftrightarrow \, (1-r) \quad ; \quad
\quad s \, \leftrightarrow \, (1-s)\ .
\label{symaleq}
\end{equation}
\par
In the usual gauge, the invariants are given by \eqn{invnu12usgauge},
and
\begin{equation}
A=\frac{1}{B'}\left( (\psi_1+\psi_0^6)^2 -B'\right) \ ;\qquad
B=\frac{1}{B'}\left( \psi_1^2 -B'\right) \ ,
\end{equation}
and one branch of the solution can be written as
\begin{eqnarray}
r&=& \frac{1}{2} \left(1+ \frac{\sqrt{(\psi_1+\psi_0^6)^2-B'}-
\sqrt{\psi_1^2 -B'} }{\psi_0^6} \right) \nonumber\\
s&=& \frac{1}{2} \left(1+ \frac{\sqrt{(\psi_1+\psi_0^6)^2-B'}+
\sqrt{\psi_1^2 -B'} }{\psi_0^6} \right) \ .
\label{solbran}
\end{eqnarray}
In this way we have reached an explicit solution of the Picard--Fuchs
differential system. It suffices to choose the algebraic branch \eqn{solbran} and to
use the basis of solutions \eqn{hygeobase} for the $\xi_i(r)$ and $\xi_i(s)$ functions
appearing in \eqn{dirproduc}.
\par
The monodromy of this period vector has been obtained in the main text by
considering the elliptic fibration structure or the two modulus $K3$ surface
$X^*_6[1,1,4,6]$ (compare with section~\ref{ss:pmiex2}). There we have
related  the differential
equation \eqn{hygeomeq} to the Picard--Fuchs equation of the fibre torus.
\par
In the present appendix the derivation of the direct product
structure \eqn{dirproduc} of the solutions was our main goal. Indeed,
we wanted to verify the group-theory prediction encoded in
\eqn{tesoro}. Once again this factorization of the solutions is a
consequence of the accidental factorization of the group
$SO(2,2)\equiv SU(1,1) \times SU(1,1)$. The two $SU(1,1)$ factors
contain the monodromy matrices of the two hypergeometric equations
in \eqn{lin2equ}.

\section{The $\ZZ_8$ basis of integer cycles in example~1}
\label{ss:z8c}

Apart from the strategy used in the main text, there is also the
possibility to construct CY cycles similar to the construction
already used for the $K3$ fibre, using a fundamental cycle and its
analytic continuation. This will lead us also to monodromy matrices
and a formula for the CY intersection matrix from the intersection
matrix ${\cal I}$ and monodromy ${\cal M}_\infty$ of the $K3$ fibre.
In this way we make contact with \cite{candmirror1}.

\begin{figure}
\begin{center}
\null\hskip -1pt
\epsfxsize=5cm
\epsfysize=4.7cm
\epsffile{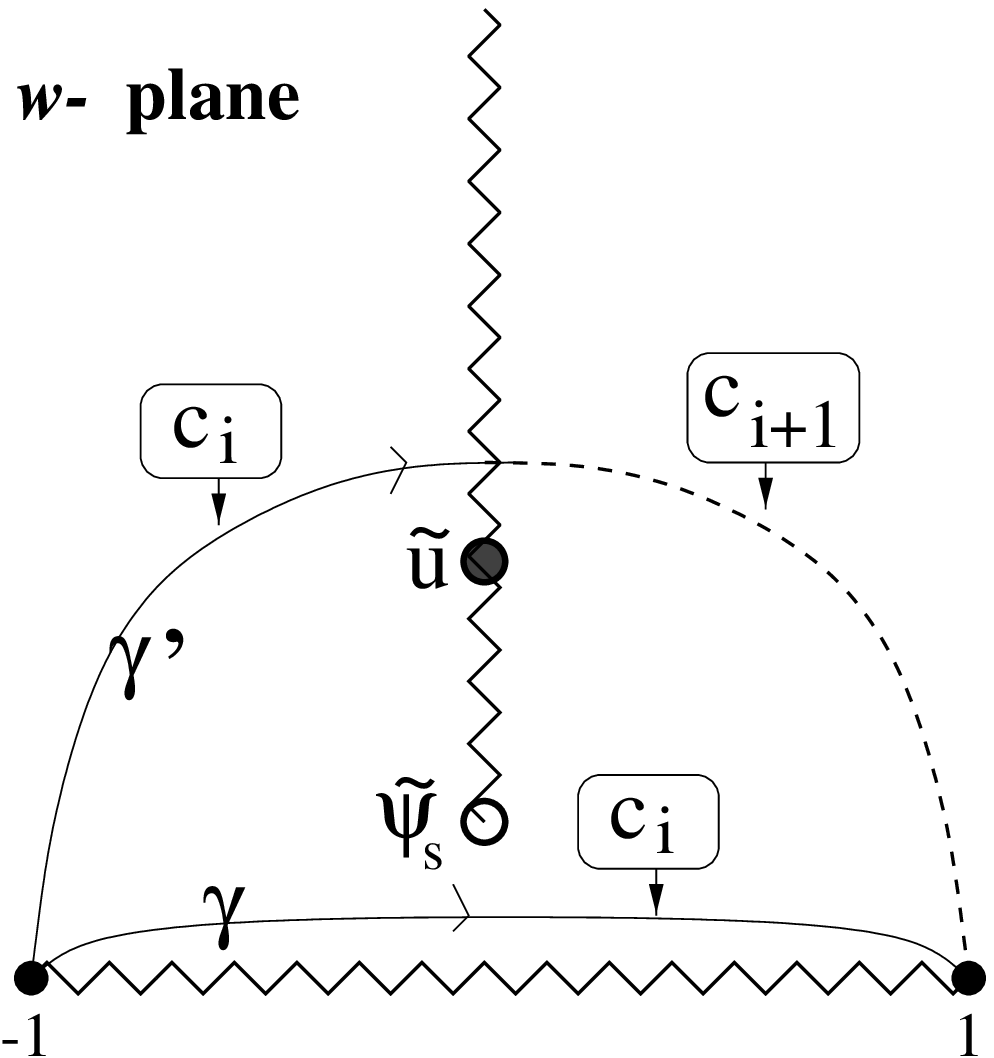}
\end{center}
\vskip -1.1cm
\mycaptionl{A basis $C_\sigma$
($\sigma=0,\ldots 7$) of CY 3-cycles, of which 6 are independent,
is obtained by fibring (as indicated by the boxes) a basis
$c_i$ ($i=0,\ldots 3$) of $K3$ 2-cycles, of which 3 are independent, over
the path $\gamma$ (getting $C_{2i}$) or $\gamma'$ (getting $C_{2i+1}$)
in the base. }
\label{bigfigb}
\end{figure}

We will define a
fundamental cycle $C_0$ (and a corresponding period $\hat\varpi_0$)
for a certain region in moduli space
($\psi_0$ large). This period cannot be defined globally and smoothly; as
we have done before
we choose cuts in the moduli space and analytically continue the period
to a different region,
$\psi_0^4$ and $\tilde\psi_s$ small and in the upper half-plane.
Then we use the $\ZZ_8$ symmetry
that acts on the moduli
(see definitions \eqn{deftildepsiu}) as
\begin{equation}
\label{sat1}
(\tilde\psi_s,\,\tilde u) \to
({\rm e}^{\ii\pi}\tilde\psi_s,\,{\rm e}^{\ii\pi}\tilde u)\ ,
\end{equation}
 and define a whole set of periods
 $\hat\varpi_\sigma\stackrel{\ZZ_8}{\to}
\hat\varpi_{\sigma+1}$, $\sigma=0,\ldots 7$.
Only six of these periods are independent, since $\sum_{\sigma\,
{\rm even}}\hat\varpi_\sigma = \sum_{\sigma\,{\rm odd}}\hat\varpi_\sigma =0$.
Finally, all the periods $\hat\varpi_\sigma$ are analytically
continued to the whole moduli space and their
non-trivial monodromies around the various singularities are
analysed. In the following we describe this construction in some detail.
\paragraph{The fundamental cycle.}
An integer cycle $C_0$ can be  defined, for $\psi_0$ large,
in complete analogy  to the $K3$ fundamental cycle of
 (\ref{fredag6}):
\begin{eqnarray}
\label{cyfredag6}
C_0&=& \left\{ (x_1,x_2,x_3,x_4,x_5) | x_5 =
\mbox{const.}, |x_1|^2=|x_2|^2=|x_3|=\delta, \right.\nonumber \\
&& \left. \mbox{$x_4$ the solution of $W^{(1)}(x)=0$ that tends
to $0$ as $\psi_0 \to \infty$} \right\}/(2|G|) ~,
\end{eqnarray}
where $G$ is the discrete identification group for the CY \eqn{GisZKGp}.
Here we divide by $(2|G|)$ in order to get an elementary cycle. Indeed,
the subset of transformations \eqn{identif2}
which leaves invariant the specified set of variables $\{x_i\}$ contains
not only the group $G=\ZZ_4^3$ (i.e.\ arbitrary $m_1$, $m_2$ and $m_3=m_0$ in \eqn{identif2}), but
contains also  the projective transformation $(x_1,x_2)\to-(x_1,x_2)$.
Changing variables as in  (\ref{x1x2x0z}) from $x_1$, $x_2$ to
$\zeta$, $x_0$,
the torus $|x_1|^2=|x_2|^2=\delta$ covers eight times the torus
$\delta |\zeta| = |x_0|=\delta$. With the notations of
\eqn{fredag6} and \eqn{tirsdag11}, such that
$b_0=(\gamma_0\times\gamma_3)/|G'|$, and denoting the circle around
$\zeta=0$ as $C$, we have
\begin{equation}
C_0 = \frac{2|G|}{8|G'|} \, C_0 =
\frac{1}{8|G'|} \gamma_1\times\gamma_2\times\gamma_3 =
\frac{1}{|G'|} C\times \gamma_0\times\gamma_3 = C\times b_0~.
\end{equation}
 Using \eqn{tirsdag11} we obtain
\cite{candmirror1}
\begin{eqnarray}
\label{funcy}
\hat\varpi_0 &\equiv & \int_{C_0} \hat\Omega^{(3,0)}
= \int_{C\times b_0}\hat\Omega^{(2,0)}\frac{d\zeta}{2\pi\ii\,\zeta}\nonumber\\
&=&{1\over 2 \pi\ii} \int_C {\dd\zeta\over\zeta}
\hat\vartheta_0\left(z(\zeta)\right)
= {1\over \pi} \int_{\gamma} {\dd w\,\hat\vartheta_0\left(z(w)\right)
\over\sqrt{1 - w^2}} ~.
\end{eqnarray}
Note that in terms of the variable $w$, equation~(\ref{defw}), the circle $C$ is
mapped
into a contour surrounding the
cut from $w=-1$ to $w=1$ associated to the factor $1/\sqrt{1 - w^2}$,
i.e.\ twice the path $\gamma$ `above the cut'
shown in figure~\ref{bigfigb} to  describe the `even' cycles.
\paragraph{Analytically continued cycles.}
Now consider the fundamental period for
$\psi_0,\tilde\psi_s$ small and in the upper
half-plane, i.e.\ the situation depicted in figure~\ref{bigfigb}.
\begin{figure}
\begin{center}
\null\hskip -1pt
\epsfxsize=11cm
\epsffile{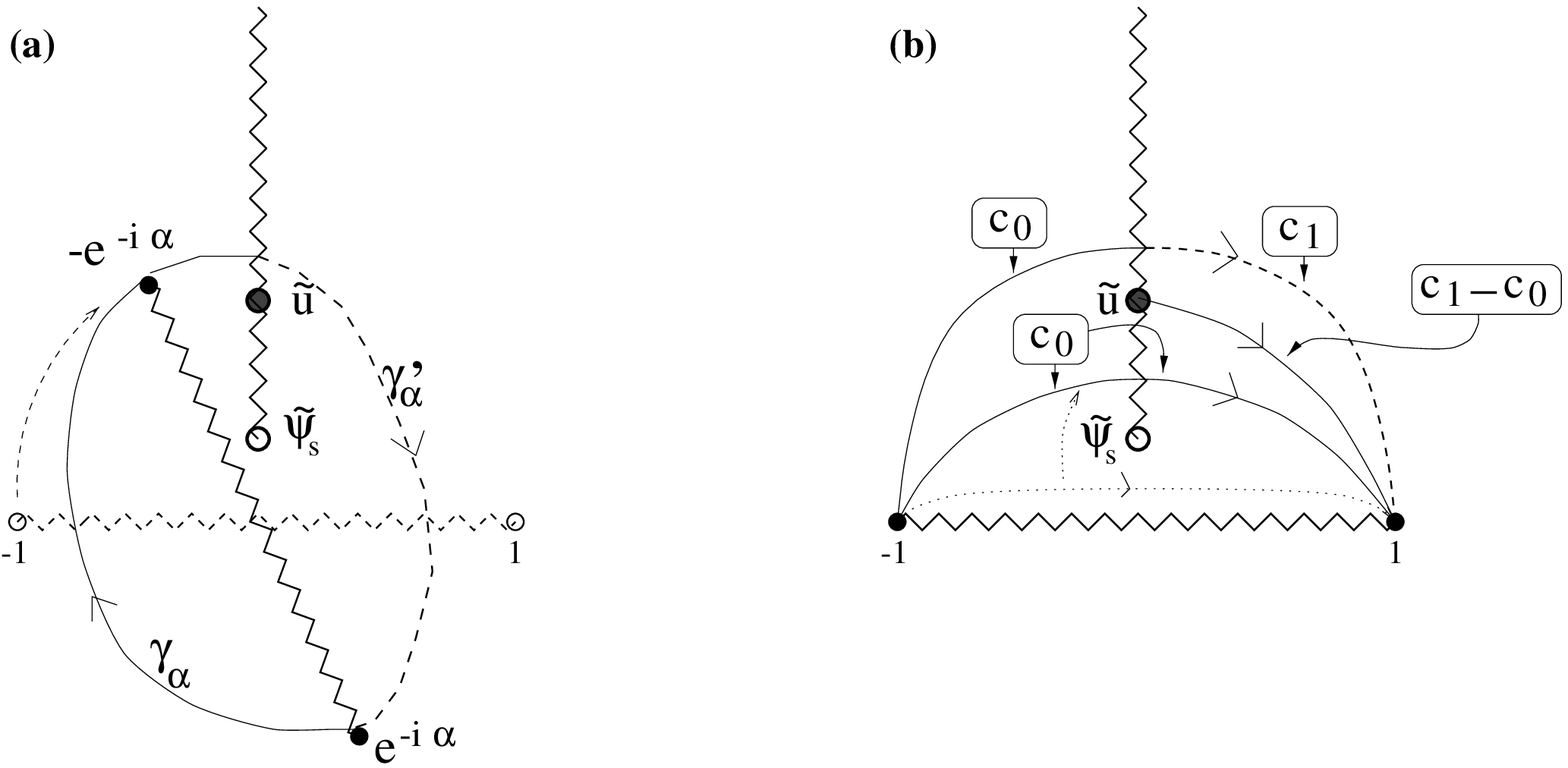}
\vskip -0.1cm
\end{center}
\mycaptionl{{\bf (a)} The rotation $(\tilde\psi_s,\tilde u)\to
({\rm e}^{\ii\alpha}\tilde\psi_s, {\rm e}^{\ii\alpha}\tilde u)$ is
discussed changing variables in the base to $w' = {\rm
e}^{-\ii\alpha}w$. The paths $\gamma,\gamma'$ become
$\gamma_\alpha,\gamma_\alpha'$; notice that with respect to figure~\ref{bigfigb}
we draw $\gamma$ under the square-root cut, changing its orientation.
When $\alpha=\pi$, we have $\gamma_\pi=\gamma'$ and $\gamma_\pi' =
\gamma$; moreover the cycles fibred over $\gamma'$
have undergone a $\ZZ_4$ rotation.
{\bf (b)} The vanishing cycle $V_v$ at the conifold: consider
$C_0\sim \gamma\times c_0;$
the path $\gamma$ can be lifted past the singular point
$\tilde\psi_s$, as the $K3$ cycle $c_0$ is regular there. Then $C_0$ is clearly
equal to $C_1 - V_v$, where $V_v$ is the fibration of $c_1 - c_0 \equiv
c_0'$, the vanishing cycle in the $K3$, on an open line from $1$ to
$\tilde u$.}
\label{fig:alpharot}
\end{figure}

Under a rotation $(\tilde\psi_s,\tilde u)\to
({\rm e}^{\ii\alpha}\tilde\psi_s, {\rm e}^{\ii\alpha}\tilde u)$
the modulus of the $K3$ fibre
transforms, according to  (\ref{fredag2}), as
$z(w)\to z({\rm e}^{-\ii\alpha}w)$. Changing integration variable
in  (\ref{funcy}), $w'={\rm e}^{-\ii\alpha}w$, we obtain
$\hat\varpi_0\to{1\over\pi}\int_{\gamma^{(\alpha)}}{\dd w'\over
\sqrt{1 - {\rm e}^{2\ii\alpha}w'^2}}
\hat\vartheta_0\left(z(w')\right)$.
As explicitly shown in figure~\ref{fig:alpharot}a,
under a rotation of $\pi$,
we have $\gamma\times c_0\to \gamma'\times c_0 \to \gamma\times
c_1\to\ldots$, i.e.\ $C_0\to C_1\to C_2\to\ldots$.
Thus we have generated the  basis of cycles $C_\sigma$ described in
figure~\ref{bigfigb}, i.e.\ $C_\sigma\stackrel{\ZZ_8}{\to}C_{\sigma+1}$.
The corresponding periods $\hat\varpi_\sigma$ are naturally divided
into `even' and `odd' periods
\begin{eqnarray}
  \label{i2}
  \hat\varpi_{2 k} & = & {1\over \pi}\int_{\gamma} {\dd w\over \sqrt{1 -
  w^2}}\,\hat\vartheta_k\left(z(w)\right)~,\nonumber\\
  \hat\varpi_{2 k + 1} & = & {1\over \pi}\int_{\gamma'} {\dd w\over \sqrt{1 -
  w^2}}\,\hat\vartheta_k\left(z(w)\right)~,
\end{eqnarray}
where $\hat\vartheta_k$, $j=0,\ldots,3$ is the $\ZZ_4$-basis of $K3$
periods described previously. We will choose as a basis of independent periods
$\{\hat\varpi_\Sigma\} = (\hat\varpi_0,\hat\varpi_2,\hat\varpi_4,
\hat\varpi_1,\hat\varpi_3,\hat\varpi_5)$.
\paragraph{Intersection matrix.}
As discussed in section~\ref{ss:stratK3per}, to obtain the intersection of
two CY 3-cycles realized as fibrations of $K3$ 2-cycles
 we have to add  (with the appropriate sign) the
intersections of the 2-cycles in the fibre above
the intersection points in the base.

Let us apply this to the basis $\{C_\Sigma\} =
(C_0,C_2,C_4,C_1,C_3,C_5)$ of CY cycles
(associated to the basis $\hat\varpi$ of periods) we are considering. These
cycles are described as fibrations (see figure~\ref{bigfigb}
and  (\ref{i2})): $C_{2I} = \gamma\times c_I$,
$C_{2I+1}=\gamma'\times c_I$. Using  (\ref{fibint}) it is immediate
to see that $C_{2I}\cdot C_{2J} = C_{2I+1}\cdot C_{2J+1} = 0$, since
$\gamma\cap\gamma =\gamma'\cap\gamma' = 0$.
The interesting intersections are
\begin{equation}
C_{2I} \cdot C_{2J+1} = (+1) c_I \cdot c_J +(-1) c_J \cdot ({\cal M}_\infty
c)_J
= ({\cal I} - {\cal I}{\cal M}_\infty^T)_{IJ}\ ,
\label{CYint}
\end{equation}
where ${\cal I}_{IJ} = c_I\cdot c_J$ is the intersection matrix
(\ref{intersectionmatrix1})
and ${\cal M}_\infty$ the $\ZZ_4$ action (\ref{Minf}) in the $K3$ fibre.
Thus we obtain $C_\Lambda\cdot C_\Sigma = I_{\Lambda\Sigma}$, with
\begin{equation}
I = \left(\matrix{0 &{\cal I}(\unity -{\cal M}_\infty^T) \cr
-(\unity -{\cal M}_\infty){\cal I} & 0}\right)\ ;\qquad
{\cal I}(\unity -{\cal M}_\infty^T)=\left(
\begin{array}{rrr}
-1 & 3 & -3 \\
1 & -1 & 3 \\
-3 & 1 & -1
\end{array}
\right)~.
\label{defD}
\end{equation}
Observe that the determinant of this intersection matrix is $16^2$,
while the determinant of the $K3$ intersection matrix \eqn{intersectionmatrix1}
is $-4$. We expect that fibring the $K3$ over paths in $\zeta$ with
intersection matrix 1 should also lead to a construction of CY
cycles, and then the determinant of the intersection matrix would be
$4^2$. If this is the case, then the basis $\hat\varpi$ of integer cycles,
constructed here, is not an elementary basis. Indeed, nothing
guarantees that differences between the analytic continued fundamental cycle,
and the fundamental cycle itself are elementary (we just know that
they are integer). In fact, below we shall construct a more
elementary basis of CY cycles, removing the factor 16 in the
intersection matrix.
\paragraph{Monodromies.}
First of all the monodromy around $\psi_0=0$ corresponds to
the action of the $\ZZ_8$ symmetry on the $\varpi$ basis,
 $\hat\varpi\to {\cal A} \hat\varpi$, with
\begin{equation}
\label{pad1}
{\cal A} = \left(\matrix{0 & \unity\cr {\cal M}_\infty & 0}\right)~.
\end{equation}

Other singular points of the CY manifold are located at
$\tilde u=\pm 1$ and $\tilde\psi_s= \pm 1$. Since the singularities
for $+1$ are related by the $\ZZ_8$ symmetry to the ones for $-1$, we
concentrate on the $+1$ representatives. First we study
the conifold singularity at $\tilde u =1$.

In the main text we found at this point the `vanishing cycle' $V_v$,
equation~\eqn{nbas1}. We now re-express it in terms of the basis $C_\Lambda$.
In figure~\ref{fig:alpharot}{(\bf (b)} it is explained how the $C_0$ cycle is deformed
to a $C_1$ cycle. In that procedure, it does not change by lifting it past the
singularity at $w=\tilde\psi_s$, because the transported $K3$ cycle $c_0$ has no
monodromy at that point, see \eqn{M0}. But lifting it through the singularity
at $w=\tilde u$, one finds that at the right of the cut appears ${\cal M}_\infty c_0$,
and the difference $C_1-C_0$ is therefore a transport of
\begin{equation}
c_0- {\cal M}_\infty c_0=c_0-c_1=-c'_0  \label{C0-C1}
\end{equation}
 over the line from $w=1$ to $w=\tilde u$. Thus we have
$-{\cal V}_v = \hat\varpi_1 -\hat \varpi_0$.
Knowing this expression for the singular cycle, we can obtain the
monodromy around the CY conifold singularity $\tilde u=1$,
denoted as $M_{\tilde u}$, through the Picard--Lefshetz formula \eqn{Picard-LefshetzK3CY}
with $\nu=C_1-C_0$.
Around the point $\tilde\psi_s= 1$
there is one more non-trivial monodromy, whose matrix action
we denote by $M_s$.  In order to compute $M_s$ it turns out that
it is easier\footnote{The computation is simply done by
manipulating the pictures of the cycles in figure~\ref{bigfiga};
we first apply $M_{\tilde u}$ by letting $\tilde u$
run around $1$ in the
$w$-plane, and then $M_s$ by doing the same thing with $\tilde\psi_s$.
Then by deforming the paths, we write the resulting integral
in terms of our basis periods.} to evaluate first
the combined monodromy $M_{\tilde u} M_s$, and then obtain
$M_s$ multiplying by $M_{\tilde u}^{-1}$. The final result is
\begin{equation}
M_{\tilde u} =
\pmatrix{2& 0& 0& -1& 0& 0\cr -1& 1& 0& 1& 0& 0\cr 3& 0& 1& -3& 0& 0
\cr 1& 0& 0& 0& 0& 0\cr -3& 0& 0& 3& 1& 0\cr 3& 0& 0& -3& 0& 1}~,
\hskip 0.3cm
M_s = \pmatrix{
 1& 0& 0& 0& 0& 0\cr
 0& 2& 1& 0& -1& -1\cr
 -3& -4& -2& 3& 4& 3\cr
 0& -1& -1& 1& 1& 1\cr
 3& 4& 3& -3& -3& -3\cr
 -3& -3& -2& 3& 3& 3} \ . \label{monodrex1CY}
\end{equation}
All the monodromy matrices derived here agree with those given in
\cite{candmirror1}.
\paragraph{Comparison of bases.}
To translate results we want to compare the basis here with that
in the main text. The integrals ${\cal T}_I$ in the main text
over the circle correspond to the even periods $\hat\varpi_{2I}$, see
\eqn{i2}, and the change of basis \eqn{bas2}. The relation of ${\cal
V}_v$ to the $\hat\varpi$ was determined above. The same method can
be used for obtaining expressions of ${\cal V}_{1,2}$. In fact, we
can do all three at once.
The three $K3$ cycles can be decomposed in one which has no
monodromy at $w=\tilde u$, and two others which have no monodromy at
$w=\tilde\psi_s$. Therefore in a similar way as above we can make the
difference between the CY cycles $C_{2I+1}$ and $C_{2I}$.
The former consist of a transport over a $\gamma'$ of a
 $K3$ cycle. This $K3$ cycle, say $c$  at the left hand of the cut
in figure~\ref{fig:alpharot}{\bf (b)}, will become ${\cal M}_\infty c$ at the right hand side.
If we pull the path $\gamma'$ through the two singular points, a
cycle gets an extra part at one of the two singular points (if we
write it in the basis of the cycles such that they only have
monodromy in one of the points), and the difference between
$C_{2I+1}-C_{2I}$ is at the end the
transport over a path from $w=1$ to either $w=\tilde u$ or
$w=\tilde\psi_s$ of the cycle $(\unity -{\cal M}_\infty )c_I$. We
still have to express that in the $c'$ basis, using $F$ \eqn{bas2},
and obtain
\begin{equation}
\hat \varpi_{2I+1}-\hat \varpi_{2I}=(\unity -{\cal M}_\infty )
F^{-1}{\cal V}_I\ .
\end{equation}
The full transformation matrix is then
\begin{equation}
\pmatrix{{\cal V}_I\cr {\cal T}_I}=\pmatrix{-F(\unity -{\cal M}_\infty )^{-1}
&F(\unity -{\cal M}_\infty )^{-1}\cr F&0}\pmatrix{\hat \varpi_{2I}\cr \hat \varpi_{2I+1}}\ .
\label{cbZ8VT}
\end{equation}
Note that the matrix
\begin{equation}
F(\unity -{\cal M}_\infty )^{-1}=\pmatrix{-1&0&0\cr \ft32&0&-\ft12\cr \ft34&\ft12&\ft14}
\end{equation}
is non-integer. Its determinant is $-\ft14$. As the basis $v=\{{\cal V},{\cal T}\}$ is build from
integer cycles, this shows that the previous basis was not elementary, as we
expected due to the determinant of the intersection matrix \eqn{defD}. The transformation
\eqn{cbZ8VT} thus eliminates the extra factor $4^2$ in the determinant of $I$.

One can then calculate the intersection matrix $q$ in the basis $v$ using
\eqn{defD}, \eqn{cbIM}, and obtains \eqn{explicitICYex1} with a
compact expression for $q_{vv}$:
\begin{equation}
q_{vv}={\cal I}'(\unity -{\cal M}'_\infty )^{-1\,T}-(\unity -{\cal M}'_\infty )^{-1}
{\cal I}'\ .
\end{equation}
\def\Journal#1#2#3#4{{#1} {\bf #2} (#4) #3}
\def\NPB{{ Nucl. Phys.} {\bf B}}
\def\PLB{{ Phys. Lett.}  {\bf B}}
\def\PRL{Phys. Rev. Lett.}
\def\PRD{{ Phys. Rev.} {\bf B}}
\def\ZPC{{ Z. Phys.} {\bf B}}
\def\IJMPA{{ Int. J. Mod. Phys.}A}
\def\CMP{ Comm. Math. Phys.}
\def\CQG{ Class. Quantum Grav.}


\begin{thebibliography}{99}
\bibitem{KKLMV} S. Kachru, A. Klemm, W. Lerche, P. Mayr, C. Vafa, Nucl. Phys. {\bf B459}
(1996) 537; hep-th/9508155.
\bibitem{KLMVW} A. Klemm, W. Lerche, P. Mayr, C. Vafa, N. Warner,
Nucl. Phys. {\bf B477} (1996) 746; hep-th/9604034.
\bibitem{revLerche}
 W. Lerche, in {\em Gauge
Theories, Applied Supersymmetry, Quantum Gravity}, eds. B. de Wit et
al., Leuven Notes in Mathematical Physics, B6, 1996, p. 53;
Nucl. Phys. Proc. Suppl. 55B (1997) 83-117; Fortsch. Phys. 45 (1997);
hep-th/9611190.
\bibitem{Klemmreview} A. Klemm,
in "1996 summer school in High Energy Physics and cosmology",
Trieste 1996, World Scientific, p. 120; hep-th/9705131.
\bibitem{DWVP}  B. de Wit, P.G. Lauwers, R. Philippe, Su S.-Q.
and A. Van Proeyen, \Journal{\PLB}{134}{37}{1984};\\
B. de Wit and A. Van Proeyen, \Journal{\NPB}{245}{89}{1984}.
\bibitem{PKTN2} G. Sierra and P.K. Townsend, in {\em
Supersymmetry and Supergravity 1983}, ed. B. Milewski (World
Scientific, Singapore, 1983), p. 396;\\
S. J. Gates, \Journal{\NPB}{238}{349}{1984}.
\bibitem{SeiWit} N.~Seiberg and E.~Witten, Nucl. Phys. {\bf
B426} (1994) 19; {\bf B431} (1994) 484.
\bibitem{ItalianN2}
L. Andrianopoli, M. Bertolini, A. Ceresole, R. D'Auria,
S. Ferrara, P. Fr\`e and T. Magri, Journal of geometry and Physics
{\bf 23} (1997) 111; hep-th/9605032.
\bibitem{Seiberg88} N. Seiberg, Nucl. Phys. {\bf B303} (1988) 286.
\bibitem{revKlemmTheisen} A. Klemm and S. Theisen, in {\em Gauge
Theories, Applied Supersymmetry, Quantum Gravity}, eds. B. de Wit et
al., Leuven Notes in Mathematical Physics, B6, 1996, p. 27.
\bibitem{2ndqms}
S.~Kachru and C.~Vafa, Nucl. Phys. {\bf B450} (1995) 69,
hep-th/9505105; \\
S.\ Ferrara, J.\ A.\ Harvey, A.\ Strominger and C.\ Vafa,
Phys. Lett. {\bf B361} (1995) 59, hep-th/9505162.
\bibitem{Polchinski} J. Polchinski, Phys. Rev. Lett. 75 (1995) 4724;
hep-th/9507158.
\bibitem{GE} S. Katz, A. Klemm, C. Vafa,
Nucl. Phys. {\bf B497} (1997) 173; hep-th/9609239.
\bibitem{GE-review} S. Katz, P. Mayr, C. Vafa, Adv. Theor. Math. Phys. {\bf 1}
(1998) 53, hep-th/9706110.
\bibitem{M-sol} E. Witten, Nucl. Phys. {\bf B500} (1997) 3, hep-th/9703166.
\bibitem{KLM}
A.Klemm, W.Lerche and P.Mayr,
Phys. Lett. {\bf B357} (1995) 313, hep-th/9506112.
\bibitem{AspinwallLouis}
P.S. Aspinwall, and J. Louis, Phys. Lett. {\bf B369} (1996) 233, hep-th/9510234.
\bibitem{PietroSKGlect} P. Fr\`e, Nucl. Phys. Proc. Suppl. {\bf 45BC}
(1996) 59.
\bibitem{whatskg} B. Craps, F. Roose, W. Troost and A. Van Proeyen,
Nucl. Phys. {\bf B503} (1997) 565; hep-th/9703082.
\bibitem{GrifHar}  P. Griffiths and J. Harris, {\em Principles of Algebraic Geometry},
Wiley-Interscience, 1978.
\bibitem{Aspinwall} P. Aspinwall, {\em $K3$ surfaces and String Duality}, hep-th/9611137
\bibitem{PFeqns}
S.\ Ferrara and J.\ Louis, \Journal{\PLB}{278}{240}{1992};\\
A.\ Ceresole, R.\ D'Auria, S.\ Ferrara, W.\ Lerche and J. Louis,
\Journal\IJMPA{8}{79}{1993}.
\bibitem{candmirror1} P. Candelas, X. de la Ossa, A. Font, S. Katz and
D. Morrison, Nucl. Phys. {\bf B416} (1994) 481;
hep-th/9308083.
\bibitem{CdAF} L. Castellani, R. D' Auria and S. Ferrara, Phys. Lett.
{\bf B241} (1990) 57; Class. Quantum Grav.
{\bf 7} (1990) 1767,\\
R.~D'Auria, S.~Ferrara and P.~Fr\`e, Nucl.~Phys. {\bf B359}
(1991) 705.
\bibitem{special} A.~Strominger, \Journal{\CMP}{133}{163}{1990}.
\bibitem{f0art} A. Ceresole, R. D'Auria, S. Ferrara and A. Van
Proeyen, \Journal\NPB {444} {92}{1995}, hep-th/9502072.
\bibitem{symplconf} P. Claus, K. Van Hoof and A. Van Proeyen, in
preparation.
\bibitem{FerStro} S.~Ferrara and A.~Strominger, in {\em Strings
'89}, eds. R.~Arnowitt, R.~Bryan, M.J.~Duff, D.V.~Nanopoulos and
C.N.~Pope (World Scientific, 1989), p.~245.
\bibitem{griffiths} P. Griffiths, Ann. Math. {\bf 90} (1969) 460, 496.
\bibitem{LSW}
W. Lerche, D. Smit and N. Warner, Nucl. Phys. {\bf B372} (1992) 87.
\bibitem{Morrison} D.R. Morrison, in {\it Essays on Mirror
Manifolds}, ed. S.T. Yau, International Press Hong Kong, 1992.
\bibitem{hosono}
V. Batyrev, Duke Math. Journal {\bf 69} (1993) 349; Journal Alg.
Geom. {\bf 3} (1994) 493;\\
P.S. Aspinwall, B.R. Greene and  D.R. Morrison,
{\em The Monomial-Divisor Mirror Map}, Internat. Math. Res. Notices (1993), 319-337,
alg-geom/9309007; \\
S. Hosono, A. Klemm, S. Theisen, S.T. Yau, Comm.
Math. Phys. 167 (1995) 301;\\
S. Hosono, A. Klemm and S. Theisen,
Commun. Math.Phys. {\bf 167} (1995) 301, hep-th/9308122;
{\em `Lectures on Mirror Symmetry'},
In {\em Helsinki 1993, Proceedings, Integrable models and strings}, 235-280,
hep-th/9403096.
\bibitem{lianyau} B.~ H.~Lian and S.~T.~Yau, {\it Mirror Maps,
    Modular Relations and Hypergeometric Series I and II}, hep-th/9507151
    and hep-th/9507153.
\bibitem{Bateman} A. \ Erdelyi, F. \ Obershettinger, W. \ Magnus
and  F.G. Tricomi,  Higher Transcendental Functions, McGraw Hill,
New York (1953).
\bibitem{Arnold}{V.I. Arnold: Singularity Theory, selected papers. Cambridge
University Press 1981.}
\bibitem{fresoriabook} P. Fr\`e and P. Soriani, {\it The $N=2$
Wonderland, from Calabi--Yau manifolds to topological
field--theories}, World Scientific, 1995.
\bibitem{calvis} E.\ Calabi and E.\  Visentini,
Ann. of Math. 71 (1960) 472.
\bibitem{monodrcy}
M. Bill\'o, A. Ceresole, R. D'Auria, S. Ferrara, P. Fr\`e, T. Regge,
P. Soriani and A. Van Proeyen, Class. Quantum Grav. {\bf 13} (1996) 831;
hep-th/9506075.
\end{thebibliography}
\end{document}